\documentclass{aa}
\usepackage[varg]{txfonts}
\usepackage{epsfig}
\usepackage{natbib}
\bibpunct{(}{)}{;}{a}{}{,} 

\usepackage{graphicx}
\usepackage[citecolor=blue,colorlinks=true,linkcolor=blue,urlcolor=blue]{hyperref}

\usepackage{siunitx}
\usepackage{subfig}
\usepackage{bm}

\begin{document}

\title{Searching for X-ray counterparts of unassociated \textit{Fermi}-LAT sources and rotation-powered pulsars with \textit{SRG}/eROSITA}

\author{Martin G.~F.~Mayer\thanks{\email{mgf.mayer@fau.de}}\inst{1,2} \and Werner Becker\inst{1,3} }
\institute{Max-Planck Institut f\"ur extraterrestrische Physik, Giessenbachstrasse, 85741 Garching, Germany \and Dr. Karl Remeis-Sternwarte and Erlangen Centre for Astroparticle Physics, Friedrich-Alexander Universität Erlangen-Nürnberg, Sternwartstrasse 7, 96049 Bamberg, Germany \and
Max-Planck Institut f\"ur Radioastronomie, Auf dem H\"ugel 69, 53121 Bonn, Germany}

\date{Received on Nov. 15th, 2023 / Accepted for publication on Feb. 6th 2024 / arXiv:2401.17295}

\abstract
{The latest source catalog of the {\it Fermi}-LAT telescope contains more than 7000 $\gamma$-ray sources at GeV energies, with the two dominant source classes thought to be blazars and rotation-powered pulsars. Despite continuous follow-up efforts, around 2600 sources have no known multiwavelength association.} 
{Our target is the identification of possible (young and recycled) pulsar candidates in the sample of unassociated $\gamma$-ray sources via their characteristic X-ray and $\gamma$-ray emission. To achieve this, we cross-match the {\it Fermi}-LAT catalog with the catalog of X-ray sources in the western Galactic hemisphere from the first four all-sky surveys of eROSITA on the {\it SRG} (Spektrum-Roentgen-Gamma) mission. We complement this by identifying X-ray counterparts of known pulsars detected at $\gamma$-ray and radio energies in the eROSITA data.}  
{We use a Bayesian cross-matching scheme to construct a probabilistic catalog of possible pulsar-type X-ray counterparts to {\it Fermi}-LAT sources. Our method combines the overlap of X-ray and $\gamma$-ray source positions with a probabilistic classification (into pulsar and blazar candidates) of each source based on its $\gamma$-ray properties and a prediction on the X-ray flux of pulsar- or blazar-type counterparts. Finally, an optical and infrared counterpart search is performed to exclude coronally emitting stars and active galactic nuclei from our catalog.} 
{We provide a catalog of our prior $\gamma$-ray-based classifications of all 2600 unassociated sources in the {\it Fermi}-LAT catalog, with around equal numbers of pulsar and blazar candidates. Our final list of candidate X-ray counterparts to suspected new high-energy pulsars, cleaned for spurious detections and sources with obvious non-pulsar counterparts, contains around 900 X-ray sources, the vast majority of which lies in the 95\% $\gamma$-ray error ellipse. We predict between 30 and 40 new pulsars among our top 200 candidates, with around equal numbers of young and recycled pulsars. This candidate list may serve as input to future follow-up campaigns, looking directly for pulsations or for the orbital modulation of possible binary companions, where it may allow for a drastic reduction in the number of candidate locations to search. We furthermore detect the X-ray counterparts of 15 known rotation-powered pulsars, which were not seen in X-rays before.} 
{}


\titlerunning{Unassociated {\it Fermi}-LAT Sources seen with eROSITA}
\maketitle

\section{Introduction}
The {\it Fermi} Large Area Telescope (LAT) is by far the most sensitive $\gamma$-ray telescope in the GeV band ever flown \citep{Atwood09}. Since its launch in 2008, it has continuously monitored the $\gamma$-ray sky and detected thousands of new sources. The latest source catalog \citep[4FGL-DR4;][]{4FGLDR4} is based on 14 years of data and supersedes three previous iterations of the fourth {\it Fermi}-LAT source catalog \citep{4FGL, 4FGL_DR2, 4FGLDR3}. It contains 7195 $\gamma$-ray emitters from $50\,\si{MeV}$ to $1\,\si{TeV}$, distributed over the whole sky. 
The two most prominent contributing source classes are blazar-type active galactic nuclei (AGN) and rotation-powered pulsars (PSRs), of which around 3900 and 320 can be found in the catalog, respectively. 
However, a large number of sources, around 2600, does not yet have a convincing association at lower energies, providing considerable discovery potential 
in the {\it Fermi}-LAT catalogs. 
In particular, the distribution of unassociated {\it Fermi} sources in galactic latitude suggests a likely Galactic origin for a significant fraction of the 2600 unidentified sources. 
While a significant fraction of these low-latitude sources may originate from mismodelled diffuse emission \citep{4FGLDR3}, it is a promising effort to search for new pulsars in individual unassociated {\it Fermi} sources, as numerous previous studies have shown \citep[e.g.,][]{Au22, Bruzewski23, Hare22, Braglia20}. 

The sample of known $\gamma$-ray pulsars consists of approximately equal numbers of young, energetic (i.e.~non-recycled) pulsars and millisecond  (i.e.~recycled) pulsars, all of whose emission is powered by rotation rather than accretion \citep{3PC}. 
A considerable fraction of the young energetic pulsars is found to be radio-quiet \citep{2PC, 3PC}, so that their $\gamma$-radiation is often necessary for their detection.
On the other hand, a large fraction of MSPs is found in binary systems with an optical companion, due to their formation requiring the spin-up by accretion in a low-mass X-ray binary \citep{Alpar82, Radhakrishnan82}.
An especially interesting class of MSP binaries is given by black-widow and redback systems \citep[see][]{Roberts13, Hui19}, which are characterized by a strong interaction between the pulsar and its companion and short orbital periods. In these ``spider'' systems, strong particle winds from the pulsar continually ablate their companion star \citep{Fruchter88, Strader19}, often leading to enhanced X-ray emission from intra-binary shocks \citep{Gentile14,Koljonen23}. 

It is well established that rotation-powered pulsars typically emit X-rays at a level of $L_{X}/\dot{E} \sim 10^{-5}-10^{-3}$ of their spin-down power \citep{Becker97, Shibata16}, with frequent contributions from thermal emission from the neutron star surface and nonthermal magnetospheric emission. 
Typical GeV $\gamma$-ray efficiencies are significantly larger, at \mbox{$L_{X}/\dot{E} \sim 10^{-2}$ -- 1} \citep{2PC}. 
For {\it Fermi}-detected pulsars in particular, the GeV $\gamma$-ray and X-ray fluxes are correlated with large scatter, likely due to a variety of emission mechanisms, viewing geometries, and beaming fractions \citep{Marelli11, Marelli15}.
Typical flux ratios (or equivalently luminosity ratios) lie in the range $F_{\gamma}/F_{X} \sim 100 - 10\,000$, with young radio-quiet pulsars being relatively X-ray fainter than radio-loud ones and MSPs \citep{2PC}.     
In contrast to radio, X-ray, or $\gamma$-ray energies, pulsars exhibit comparatively weak emission at optical or infrared wavelengths \citep{2PC, Mignani11}. 

The biggest obstacle in identifying counterparts to {\it Fermi}-LAT sources is their inherently large positional error. 
This makes a direct search for counterparts, for instance in the optical, challenging, as the expected number of positional chance coincidences is enormous. 
Therefore, a frequently employed approach in the search for possible associations for {\it Fermi} sources is the usage of X-ray data, as a large fraction of blazars and rotation-powered pulsars are expected to show detectable X-ray emission. The positions of X-ray sources are usually accurate on an arcsecond-level, facilitating associations with counterparts at radio or optical wavelengths.
An excellent example of this is the systematic follow-up project of unassociated {\it Fermi} sources with {\it Swift}-XRT by \citet{StrohFalcone}\footnote{\url{https://www.swift.psu.edu/unassociated/}}. Within this program, X-ray observations of {\it Fermi} error ellipses have been routinely carried out to search for potential counterparts, leading to many newly detected pulsars. 
While snapshot surveys such as that by \citet{StrohFalcone} provide immense discovery potential, they are limited by the field of view of their telescope, which prevents the reliable estimation of the probability of chance coincidences. 

The eROSITA telescope aboard the {\it Spektrum-Roentgen-Gamma} observatory \citep[{\it SRG};][]{Predehl21, Sunyaev21} provides the first all-sky survey in X-rays since the {\it ROSAT} mission \citep{Truemper82} around 30 years prior. Within its bandpass of $0.2-10 \,\si{keV}$, the eROSITA all-sky survey (eRASS) is providing the deepest coverage of the X-ray sky to date, with an expected final (i.e., after the eight planned surveys) all-sky point-source sensitivity in the $0.2-2.3 \,\si{keV}$ band around 25 times deeper than the ROSAT all-sky survey \citep{Predehl21,Merloni12}. 
At the time of writing, four out of eight planned all-sky surveys have been completed, with the cumulative data set (named eRASS:4) reaching a typical point source sensitivity of around $1-2\times 10^{-14}\,\si{erg.s^{-1}.cm^{-2}}$ in the main band.  
This paper employs the eRASS:4 data of the western Galactic hemisphere, i.e.~in the range $180^{\circ} < l < 360^{\circ}$. This data set provides the deepest contiguous picture of half the X-ray sky, including the regions around 1400 unassociated {\it Fermi} sources, with almost three million detected X-ray sources.

The main goal of this work is to identify the X-ray counterparts of (candidate) rotation-powered pulsars in eRASS:4 data. 
On one hand, this is achieved via a probabilistic cross-match of eROSITA X-ray sources to unassociated {\it Fermi} sources, which are likely to be high-energy pulsars. This is achieved by combining a purely positional cross-match with a prior classification of unassociated 4FGL sources based on their $\gamma$-ray properties, and constraints on the expected X-ray flux of a potential match.
With this effort, we aim to provide useful lists of promising targets for radio, optical, X-ray, or $\gamma$-ray follow-up by quantifying the likelihood of an association with a given X-ray source using a Bayesian approach \citep{Budavari08, Salvato18}. 
On the other hand, we present the set of detected X-ray counterparts of known rotation-powered pulsars in eRASS:4, emphasizing those sources whose X-ray emission was observed for the first time to extend the sample of X-ray-detected pulsars. 

Our paper is organized as follows: We give an overview of the data products we used in Sect.~\ref{Data}, 
and we describe the individual components of our method for constructing probabilistic cross-match catalogs in Sect.~\ref{Strategy}. In Sect.~\ref{PSRs_X}, we present the sample of X-ray counterparts to previously identified pulsars to verify our capabilities of detecting X-ray emission from pulsars. In Sect.~\ref{Results}, we present our list of candidate pulsar matches to unassociated $\gamma$-ray sources and discuss their implications and follow-up prospects. Finally, we summarize our work in Sect.~\ref{Summary}. 

\section{Input Data from \textit{Fermi}-LAT and eROSITA\label{Data}}
The target of our analysis is the set of all unidentified sources in the {\it Fermi}-LAT 14-year source catalog (4FGL-DR4, in the following 4FGL), which contains 7195 GeV $\gamma$-ray sources \citep{4FGLDR4}. From this catalog, we selected all unidentified sources, including those with low-probability associations, meaning all sources labeled in the catalog with {\tt CLASS1 == `', `unk'}. 
In order to characterize the properties of the typical $\gamma$-emitting source classes, we also defined two samples of associated sources, a pulsar sample ({\tt CLASS1 == `psr', `msp'}), and a sample of blazar-type AGN ({\tt CLASS1 == `bll',`fsrq',`bcu'}). This results in a total of 2577 unassociated sources, of which 1371 lie within the western Galactic hemisphere.  
Among the complete set of associated sources, 320 are of pulsar- and 3934 of blazar-type according to our definition, of which 163 and 1922 are located in the western Galactic hemisphere, respectively.

The primary eROSITA data product we use is the latest internal eRASS:4 single-band source catalog.\footnote{This corresponds to the internal release version {\tt all\_s4\_SourceCat1B\_221031\_poscorr\_mpe\_photom.fits}.} This catalog contains almost three million X-ray sources with a $0.2-2.3$ keV detection likelihood \citep{Brunner22} above 5. The catalog columns important for this work are the sky positions ({\tt RA\_CORR}, {\tt DEC\_CORR}), and positional error ({\tt RADEC\_ERR}) of the sources, as well as the estimated X-ray flux and its error ({\tt ML\_FLUX\_1}, {\tt ML\_FLUX\_ERR\_1}). 
Furthermore, for the purpose of visual inspection of X-ray sources, we used the available eRASS:4 imaging data in the regions of all {\it Fermi} sources in the latest processing version {\tt c020}. We manually rebinned the available images from the archive to create simple count images with a pixel size of $12\arcsec$ for each region of interest. The energy bands we used for the image creation were the three standard eROSITA bands $0.2-0.6$, $0.6-2.3$, and $2.3-5.0$ keV.

\section{Cross-matching strategy \label{Strategy}}
Typically, the extent of 4FGL error ellipses is very large (the median $95\%$ semi-major axis being $5.8\arcmin$ for unassociated sources), and the density of X-ray sources in the all-sky survey is quite high (at an average source density per sky area of $140\,\si{deg^{-2}}$ in eRASS:4). Hence, one expects to observe many chance coincidences between 4FGL and eROSITA sources in a naive positional cross-match, with a median of three X-ray sources expected per 4FGL error ellipse. We, therefore, followed a more complex scheme, ranking potential associations between the two catalogs in a quantitative, probabilistic manner. 
Apart from a positional cross-match (Sect.~\ref{Positional}), our approach included a prior classification of 4FGL sources based on their $\gamma$-ray properties (Sect.~\ref{Prior}), obtaining constraints on the expected X-ray flux of a pulsar/blazar-type match (Sect.~\ref{FluxRatio}), and the visual inspection of the resulting matches and possible multiwavelength counterparts (Sect.~\ref{Counterpart}). Finally, we tested our method using the known associations of 4FGL sources (Sect.~\ref{Calibration}). 

\begin{figure}[t!]
\centering
\includegraphics[width=1.0\linewidth]{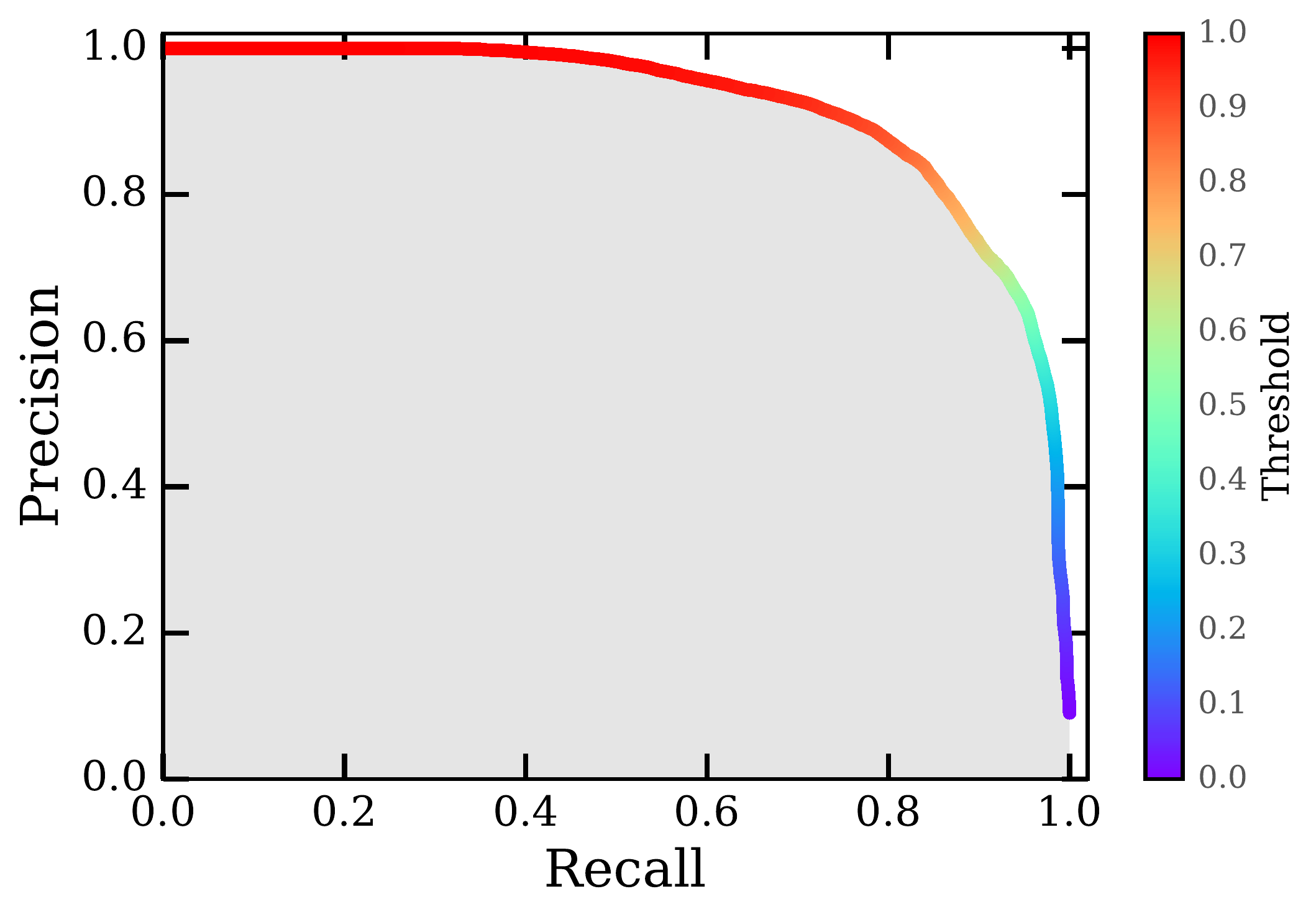} 
\caption{Precision-recall curve for the binary classifier separating pulsar- and blazar-candidate $\gamma$-ray sources. This graph compares the recall of the classifier, i.e. the fraction of pulsars in the test sample that are correctly identified, to its precision, i.e. the fraction of true pulsars among all objects selected as pulsars, dependent on the decision threshold.}
\label{ROC}
\end{figure}

\begin{figure}[t!]
\centering
\includegraphics[width=1.0\linewidth]{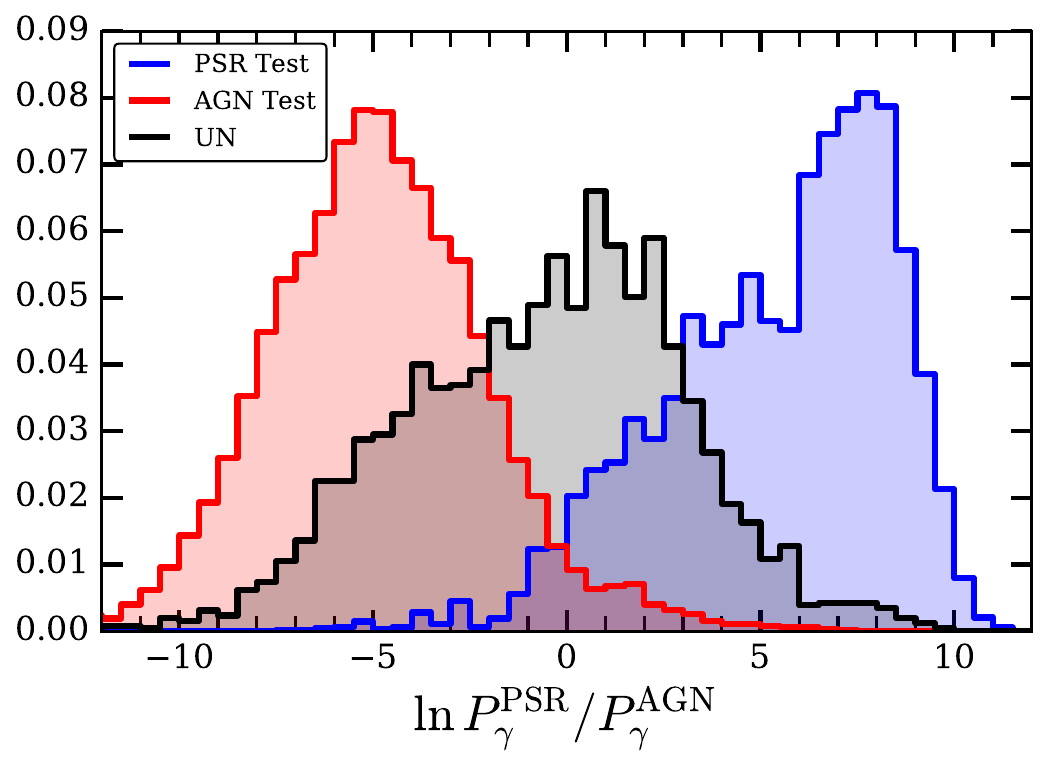} 
\caption{Distribution of binary classification probabilities $P_{\gamma}^{\rm PSR}$ for the targeted unassociated 4FGL sources. This is compared to the probabilities evaluated on PSR- and AGN-type ``out-of-box'' test samples during cross-validation, meaning targets the classifier did not see during training. The fractions of misclassified AGN and pulsars in the test sample are $4.2\%$ and $4.9\%$, respectively. 
}
\label{MachineLearning}
\end{figure}

\subsection{Classification of $\gamma$-ray sources 
\label{Prior}}

\begin{table*}
    \centering
    \caption{Optimal set of features and associated importance for the classifier separating pulsar-like and blazar-like $\gamma$-ray sources.}
    \begin{tabular}{ccc}
    \hline\hline
        Feature & Importance & Description \\ \hline
        {\tt SigCombined} & $0.276$ & $3 \log {\tt LP\_SigCurv}\tablefootmark{a} - \log {\tt Variability\_Index}\tablefootmark{a}$ \\
        {\tt ModSigCurv} & $0.176$ & $3\log {\tt LP\_SigCurv}\tablefootmark{a} - 2\log {\tt Signif\_Avg}\tablefootmark{a}$\\
        {\tt LP\_beta}\tablefootmark{a} & $0.128$ & \\
        {\tt LP\_SigCurv}\tablefootmark{a} & $0.109$ & \\[0.3ex]
        {\tt K24} & $0.079$ & $\frac{2 F(1.0-3.0\,\si{GeV}) \,-\, F(0.3-1.0\,\si{GeV}) \,- \,F(3-30\,\si{GeV})}{2 F(1.0-3.0\,\si{GeV}) \,+\, F(0.3-1.0\,\si{GeV}) \,+\, F(3-30\,\si{GeV})}$ \\[0.3ex]
        {\tt Frac\_Variability}\tablefootmark{a} & $0.051$ & \\[0.3ex]
        {\tt HR34} & $0.038$ &  $\frac{F(3-30\,\si{GeV}) \,-\, F(1.0-3.0\,\si{GeV}) }{F(3-30\,\si{GeV}) \,+\, F(1.0-3.0\,\si{GeV}) }$\\[0.3ex]
        {\tt SymLat} & $0.033$ & Symmetric Galactic latitude, i.e. $\lvert b\rvert$\\
        {\tt EFluxErr} & $0.032$ & $\log {\tt Unc\_Energy\_Flux100}\tablefootmark{a} - 0.4\log {\tt Energy\_Flux100}\tablefootmark{a}$\\
        {\tt Variability\_Index}\tablefootmark{a} & $0.031$ & \\[0.3ex]
        {\tt HR45} & $0.024$ & $\frac{F(30-1000\,\si{GeV}) \,-\, F(3-30\,\si{GeV}) }{F(30-1000\,\si{GeV}) \,+\, F(3-30\,\si{GeV}) }$ \\[0.3ex]
        {\tt PL\_Index}\tablefootmark{a} & $0.021$ & \\
        \hline
    \end{tabular}
    \tablefoot{See Appendix \ref{ML_Tuning} for the reasoning behind the definition of the compound features.
    \tablefoottext{a}{Catalog column \citep[see][]{4FGLDR3}}}
    \label{ML_Features}
\end{table*}

\begin{figure*}
\centering
\includegraphics[width=1.0\linewidth]{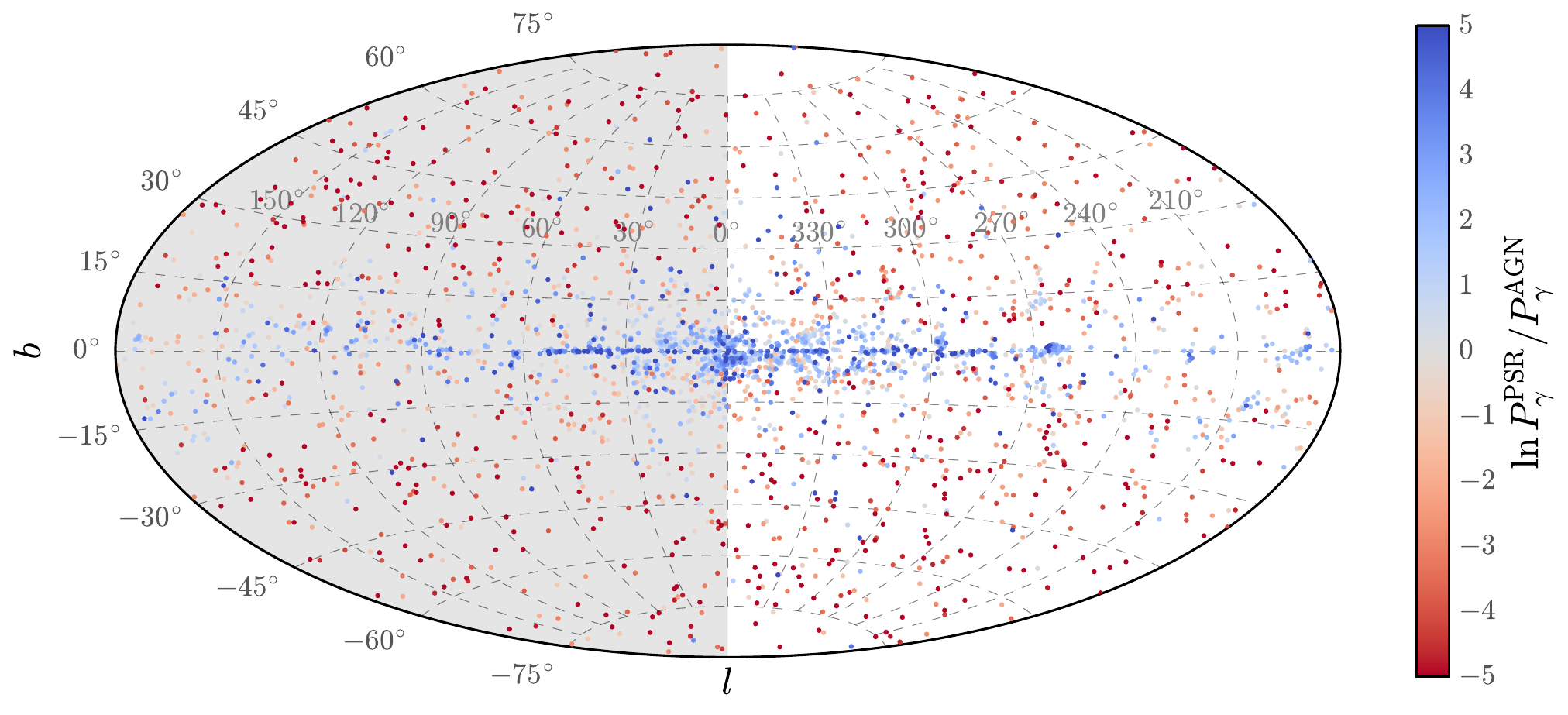} 
\caption{Distribution of binary classification probabilities $P_{\gamma}^{\rm PSR}$, indicated by the color scale of the markers, of unassociated 4FGL sources across the Galaxy. The grey shading marks the eastern Galactic hemisphere, for which no eROSITA X-ray data are available. }
\label{GalDist_Prior}
\end{figure*}

It has been shown numerous times \citep{Salvetti16, SazParkinson16, Kaur19,  Kerby21, Kerby21b, Germani21, Bhat22, Coronado-Blazquez22}, that the properties of $\gamma$-ray sources in the {\it Fermi}-LAT source catalogs can be used to predict their source type, and in particular to differentiate between pulsar-like and blazar-like sources. Typically, the primary differences between the two classes were found to be a stronger spectral downward curvature 
for pulsars, and stronger variability for blazars. 
Here, we performed a machine-learning-based classification of unassociated 4FGL-DR4 sources, with the main target of estimating a prior probability $P_{\gamma}^{\rm PSR}$ for a particular source to be a pulsar, rather than an AGN. 
While this neglects the presence of further source classes among the sample of unassociated 4FGL sources, we believe this approach is sufficient to separate pulsar candidates from the most prominent expected background population, as only $8\%$ of associated 4FGL sources are assigned to such further classes. 

The type of classifier we used for this task was a random forest classifier, as was frequently done before \citep{Salvetti16, SazParkinson16, Bhat22}. In order to tackle the strongly imbalanced nature of the training set (with a factor of ten more blazars than pulsars), we decided to use a {\tt BalancedRandomForest} classifier \citep{Chen04}, as implemented in the {\tt imbalanced-learn} package \citep{Lemaitre17}. Briefly, this tool combats the class imbalance by undersampling the majority class, rather than artificially creating synthetic data points to oversample the minority class \citep[e.g. SMOTE,][]{SMOTE}. 

The process we followed to tune our classifier is explained in detail in Appendix \ref{ML_Tuning}. Importantly, we attempted to avoid introducing biases against faint sources, by excluding absolute fluxes from our list of candidate features and by constructing flux-independent versions of certain features (e.g., curvature significance, flux error). 
%
In order to determine the optimal set of features for our classifier, we performed recursive feature elimination \citep{Luo20}, with the target of maximizing its average precision (see Appendix \ref{ML_Tuning}), meaning the ability of the classifier to construct a pure and complete sample of pulsar candidates from a test sample. 
The optimal set of features, along with their associated importance, is displayed in Table \ref{ML_Features}, and the corresponding precision-recall curve, determined via repeated five-fold cross-validation, is depicted in Fig.~\ref{ROC}. Our classifier achieved an average precision of $0.920$, 
and a ROC AUC\footnote{The receiver operating characteristic area-under-curve (ROC AUC) specifies the skill of a binary classifier by measuring the integral of the curve relating the fractions of true and false positives.} score of $0.989$, 
which is at a similar level as similar classifiers based on earlier source catalogs. For comparison, the work of \citet{SazParkinson16} reports a ROC AUC of $0.993$.

The result of our classifier is a binary probability $P_{\gamma}^{\rm PSR}$ for each unassociated source, quantifying the likelihood of it being a pulsar rather than an AGN based on its $\gamma$-ray properties. The distribution of this quantity is compared to that obtained for pulsar and AGN test samples in Fig.~\ref{MachineLearning}.
Overall, our classifier yields a relatively large fraction of sources with $P_{\gamma}^{\rm PSR} > 0.5$, namely around $45\%$, 
compared to other works \citep[e.g.,][]{Bhat22, Coronado-Blazquez22}. This is likely caused by our approach of using a balanced classifier, which implicitly assumes that both classes are a priori equally likely, as well by our attempt to not bias ourselves against faint sources (see Appendix \ref{ML_Tuning}).
Fig.~\ref{GalDist_Prior} shows the distribution of pulsar or blazar candidates identified by our classifier across the Galaxy. As expected, a large fraction of sources in the Galactic plane are classified as likely pulsars, whereas the majority of sources off the plane are found to be likely AGN. Nonetheless, there is a noteworthy presence of high-latitude pulsar candidates, which are particularly promising for identification due to less crowded fields at any wavelength. 

In order to classify whether individual pulsar candidates are likely to be young or MSPs, we analogously tuned and trained a second classifier on the corresponding pulsar subsamples in the 4FGL-DR4 catalog (141 young pulsars vs. 179 MSPs). This classifier (see Table \ref{ML_Features_Youngold} for the used features) returns a probability $P_{\gamma}^{\rm YNG}$ for a given source to be young (rather than recycled), conditional on its pulsar nature. This allows us to identify, for instance, likely MSP candidates in our sample, which could constitute promising sources for optical follow-up. Our algorithm achieved a ROC AUC score of $0.964$ 
on the test sample during 5-fold cross-validation, preforming similarly to previous attempts \citep{SazParkinson16}.  

All results of our $\gamma$-ray based prior classification efforts are provided in electronic form at CDS. In the supplied table, for each unassociated 4FGL source, we give the raw parameters used by our classifiers, and the resulting binary classification probabilities $P_{\gamma}^{\rm PSR}$ and $P_{\gamma}^{\rm YNG}$, as defined in this section.   

\subsection{Positional cross-match \label{Positional}}
For the probabilistic quantification of positional agreement between two sources in the eRASS:4 and 4FGL-DR4 catalogs, we follow an approach equivalent to that outlined by \citet{Budavari08} and implemented in {\tt NWAY} \citep{Salvato18}:

The prior probability for the hypothesis $H$ that a randomly selected eROSITA source is the correct counterpart of a given 4FGL source is given by  
\begin{equation}
    P(H) = \frac{c}{\nu \, \Omega}, 
\end{equation}
where $\nu$ describes the sky density of sources in the eROSITA catalog, $\Omega$ refers to the overlapping sky area covered by the two catalogs, and $c$ is the prior completeness factor, describing the fraction of {\it Fermi} sources for which we expect to find an X-ray counterpart \citep{Salvato18}. 
Given the observed data $D$, the likelihood for a match of two sources with a positional separation $\psi$, and (Gaussian) positional errors $\sigma_1$ and $\sigma_2$ is given by \citep[see eqs. 16 \& 28 in][]{Budavari08}:
\begin{equation} 
    P(D|H) = \left( \frac{\Omega}{4\pi} \right) \times \frac{2}{\sigma^2_1 + \sigma^2_2} \exp \left( - \frac{\psi^2}{2\left(\sigma^2_1 + \sigma^2_2\right)} \right). 
\end{equation}
Hence, following from Bayes' theorem, the posterior probability of a given match is proportional to
\begin{equation}   
   P(H|D) \propto \frac{c}{2 \pi \nu \left(\sigma^2_1 + \sigma^2_2 \right)} \exp \left( - \frac{\psi^2}{2\left(\sigma^2_1 + \sigma^2_2\right)} \right).
\end{equation}

\begin{figure}[t!]
\centering
\includegraphics[width=1.0\linewidth]{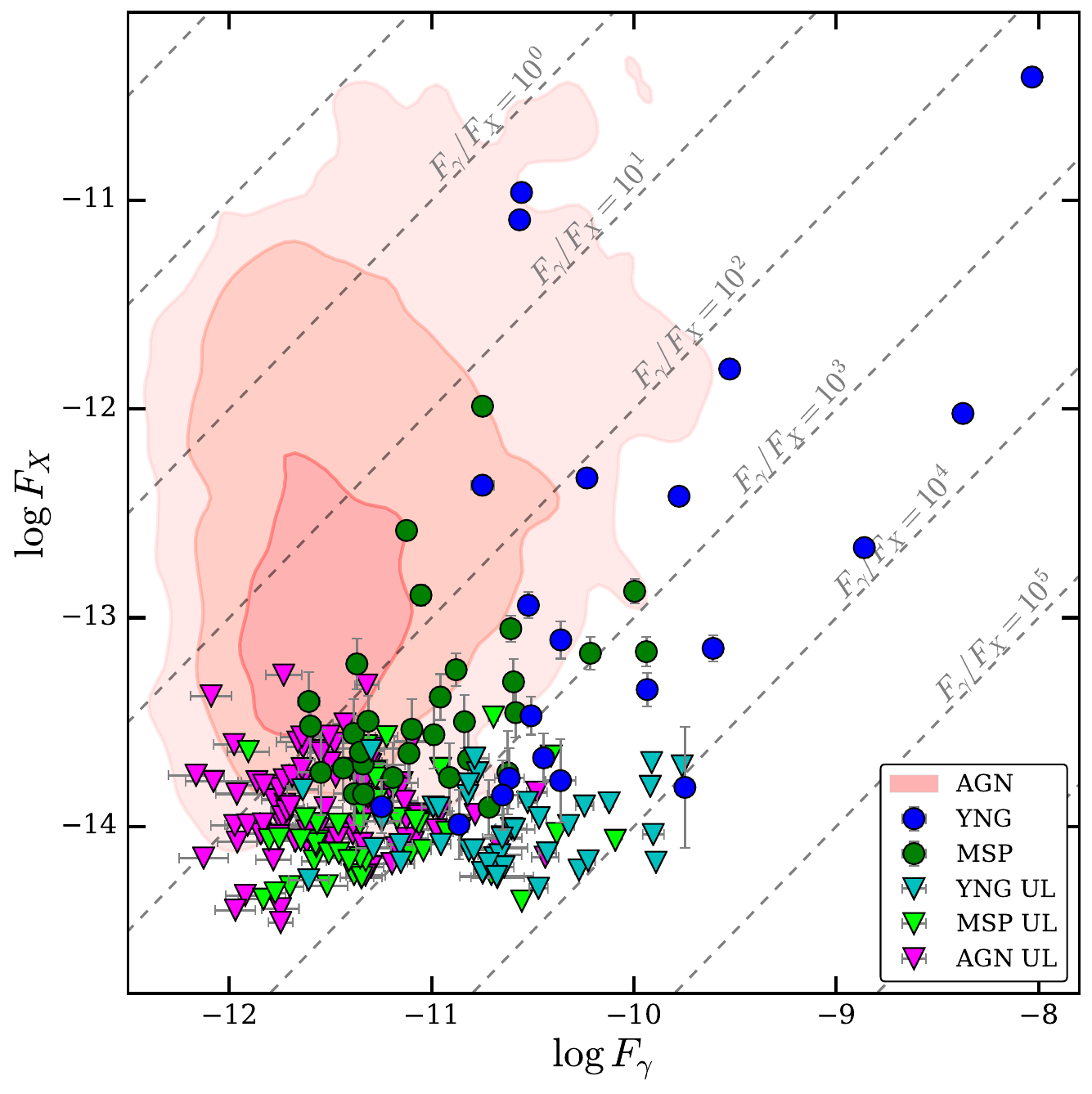} 
\caption{Distribution of X-ray ($0.2-2.3\,\si{keV}$) and $\gamma$-ray ($100 \,\si{MeV}-100\,\si{GeV}$) fluxes of young pulsars (blue circles) and MSPs (green). 
One-sigma X-ray upper limits are indicated with triangles in cyan (young pulsars), lime green (MSPs), and light red (AGN), respectively.
In order to provide a clean plot, the distribution of detected AGN is illustrated by the red filled contours, marking its one, two, and three-sigma outlines. }
\label{XGFluxes}
\end{figure}

\begin{figure}[t!]
\centering
\includegraphics[width=1.0\linewidth]{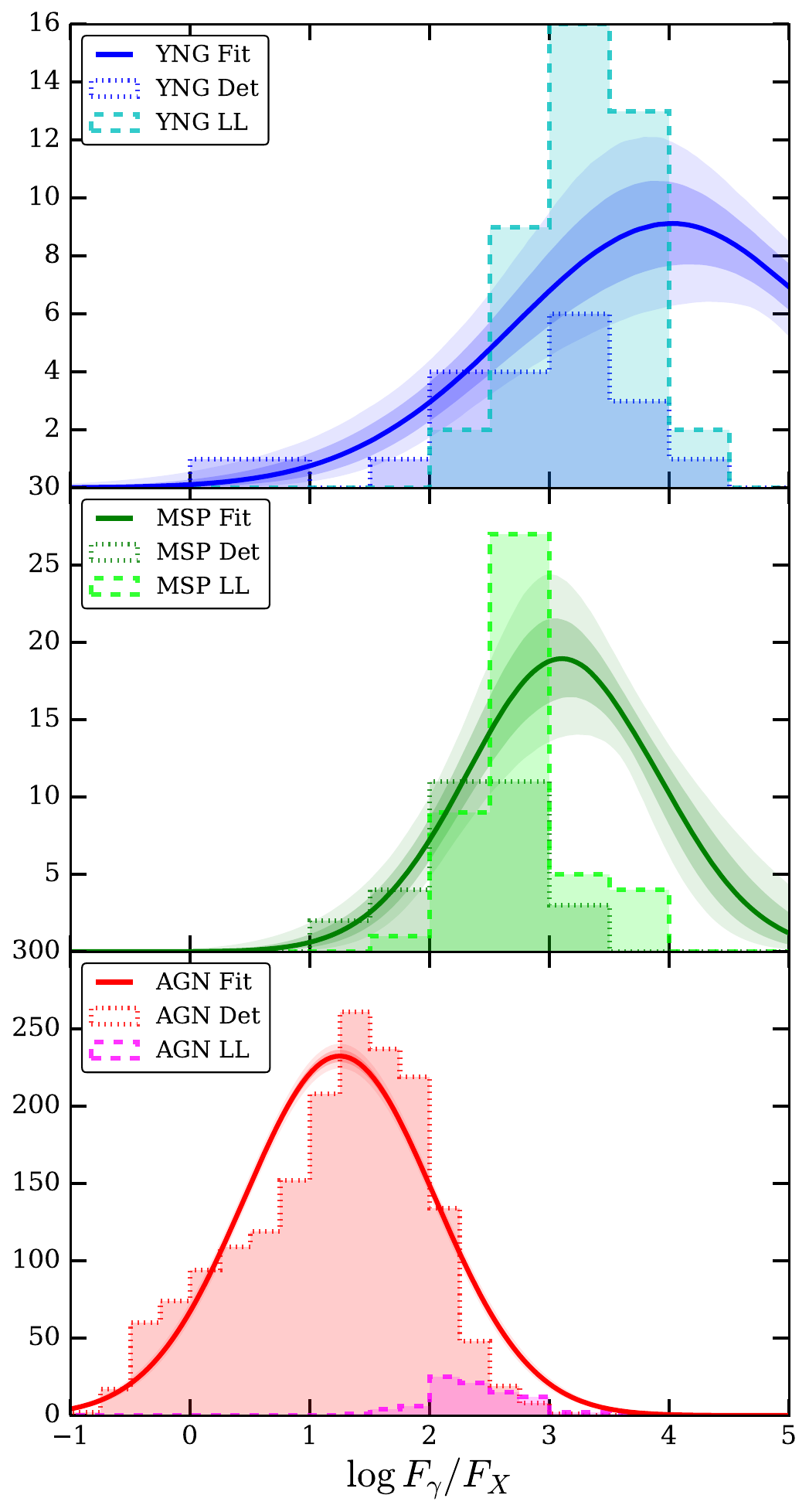} 
\caption{Distribution of the $\gamma$-to-X-ray flux ratio of 4FGL young pulsars (top), MSPs (middle), and AGN (bottom). In both panels, the distribution of the flux ratio of detected sources (dotted histogram) and corresponding one-sigma lower limits  (dashed histogram) are indicated, together with a best-fitting log-normal distribution (solid line).}
\label{FluxRatios}
\end{figure}

In practice, 
given that the {\it Fermi}-LAT source positions are specified with error ellipses, we used the full covariance matrix $\boldsymbol{\Sigma}$ to model the astrometric error accurately. Furthermore, given the considerable variation of the X-ray source density on the sky, we replaced the constant sky density of sources with a location-dependent one $\nu = \nu(\mathbf{x})$, which we estimated locally within a two-degree radius of each {\it Fermi} source.   
This yields the following expression for the astrometric component of our posterior:
\begin{eqnarray}  \label{PositionalPost}
     P(H|D) \, &\propto& \, \frac{c}{2 \pi \nu(\mathbf{x}) \sqrt{| \boldsymbol{\Sigma} |}} \exp \left( - \frac{1}{2} \boldsymbol{\psi}^{\intercal} \boldsymbol{\Sigma}^{-1} \boldsymbol{\psi} \right) ,\textrm{where} \\
\boldsymbol{\Sigma} \,&=&\, \left(\begin{array}{cc}
\sigma_{1,x}^2 + \sigma_{2}^2 & \rho_1 \sigma_{1,x} \sigma_{1,y} \\
\rho_1 \sigma_{1,x} \sigma_{1,y} &  \sigma_{1,y}^2 + \sigma_{2}^2 \\
\end{array}\right). \nonumber 
\end{eqnarray}
Here, we have decomposed the separation vector $\boldsymbol{\psi}$ into its $x$ and $y$ components. The positional covariance matrix components for 4FGL sources $(\rho_1,\sigma_{1,x},\sigma_{1,y})$ were determined from the given $95\%$ error ellipses (rescaled by a factor $0.409$ to obtain a ``one-sigma''-ellipse) following \citet{Pineau17}. The quantity $\sigma_2$ corresponds to the one-sigma positional error of the eRASS:4 sources, determined from their {\tt RADEC\_ERR} following the scheme of \citet{Merloni23}.

\subsection{Including X-ray and $\gamma$-ray flux information \label{FluxRatio}}
The majority of all sources in eRASS:4 are detected with only a few counts, the median value of background-subtracted counts being around 10 in the $0.2 - 2.3 \,\si{keV}$ band. 
Therefore, we are unable to aid our classification with detailed analyses of spectral or timing properties of X-ray sources.  
However, it is well known that rotation-powered pulsars and blazars exhibit vastly different $\gamma$-ray to X-ray flux ratios, with blazars being much X-ray brighter in a relative sense \citep{Kaur19, Kerby21b}. Furthermore, phenomenological differences exist also between different types of pulsars, with radio-quiet young pulsars    
tending to be X-ray fainter than radio-loud ones and MSPs \citep{Marelli11,Marelli15, 2PC}. Therefore, depending on the pulsar type (young or MSP), the expected X-ray flux of its potential counterpart may vary drastically. 

Hence, with the goal of predicting possible X-ray fluxes for blazars, young pulsars, and MSPs, we constrained the expected flux ratio $F_{\gamma}/F_{X}$ for each class based on the associated 4FGL sources and their X-ray counterparts in our catalog. We cross-matched all counterpart coordinates of associated sources with the eRASS:4 catalog using a $15\arcsec$ matching radius. For each match, we recorded the X-ray (pseudo-)flux in the $0.2-2.3\,\si{keV}$ band ({\tt ML\_FLUX\_1}\footnote{The source fluxes given in the catalog were calculated from their average count rate over the two-year duration of the survey, assuming an absorbed power-law spectrum, with a photon index $\Gamma=2$ and absorption column $N_{\rm H} = 3\times10^{20}\,\si{cm^{-2}}$.}), and $\gamma$-ray flux in the $100 \,\si{MeV}-100\,\si{GeV}$ band. In case of no matching source, we used the internal eROSITA upper limit server \citep[see][]{Tubin23} to determine the counts and background level within a $30\arcsec$ aperture\footnote{This aperture was chosen by \citet{Tubin23} to be similar to the size of the point spread function, optimizing the signal-to-noise of a hypothetical weak source.} at the source position. Cases in which a nearby overlapping source was present  were excluded from this analysis, as no reliable upper limits could be extracted here. 
Overall, we obtained 21, 31, and 1762 usable X-ray fluxes and 42, 46, and 89 upper limits for young pulsars, MSPs, and $\gamma$-ray-bright blazars, respectively. The distribution of X-ray and $\gamma$-ray flux information for the associations is shown in Fig.~\ref{XGFluxes}, and the distribution of X-ray to $\gamma$-ray flux ratios in Fig.~\ref{FluxRatios}. In addition, we give a list of all X-ray detected counterparts of 4FGL pulsars in the appendix in Table \ref{XDetPulsarTable}.

\begin{table}
    \centering
    \caption{Results of fitting flux ratios $F_{\gamma}/F_{X}$ with a log-normal distribution $\mathcal{N}(\mu_{t}, \sigma_{t})$ for different source classes $t$. }
    \begin{tabular}{ccc}
    \hline\hline
        $t$ & $\mu_{t}$ & $\sigma_{t}$ \\ \hline
        YNG & $3.98\pm0.23$ & $1.28\pm 0.19$ \\
        MSP & $3.09\pm0.12$ & $0.77\pm 0.11$ \\
        AGN & $1.249\pm0.019$ & $0.793\pm 0.014$ \\
        \hline
    \end{tabular}
    \tablefoot{Errors are given at a one-sigma level. }
    \label{FR_Results}
\end{table}

In order to make predictions for fluxes of potential matches of unassociated 4FGL sources, we constructed a simple model of the flux ratio distribution for each source class. We chose to use a log-normal distribution with arbitrary offset and scale for this purpose $\log\,F_{\gamma}/F_{X} \sim \mathcal{N}(\mu, \sigma)$. 
To determine the best-fit parameters of the distribution, we employed an unbinned likelihood approach for the X-ray detected sources, where the likelihood of a given set of parameters $(\mu, \sigma)$ is given by  
\begin{equation}
    \log\,\mathcal{L}(\mu, \sigma) = \sum_{i} \log\, \mathcal{N}(\log\,F_{\gamma,i}/F_{X,i}\,|\,\mu, \sigma),
\end{equation}
where the sum runs over all detected sources $i$.
For undetected sources, meaning sources with X-ray upper limits only, we used the full Poissonian likelihood of the intrinsic X-ray flux \citep[see][]{Tubin23} together with the observed $\gamma$-ray flux, to estimate the likelihood of possible flux ratios for a given source. By combining the likelihoods of X-ray detected and undetected sources, we obtained an estimate of the intrinsic flux ratio distribution of each source class, which is unbiased by the eRASS:4 sensitivity limit.
For each class, Fig.~\ref{FluxRatios} displays the distribution of observed flux ratios $F_{\gamma}/F_{X}$ and their lower limits for X-ray nondetections, together with the best-fitting log-normal distribution, whose parameters are given in Table \ref{FR_Results}. 

Our analysis qualitatively confirms several previously established findings. This includes the much brighter appearance of blazars compared to pulsars in X-rays, indicated by the much lower flux ratios. In Fig.~\ref{FluxRatios}, this can be observed in the much larger fraction of detected AGN compared to young pulsars and MSPs, and also in the evident difference between the peaks of the distributions of detected sources.   
Furthermore, young pulsars clearly show a much larger spread in $F_{\gamma}/F_{X}$ than MSPs. This may be caused by a significant contribution of very X-ray faint radio-quiet pulsars to the population of young $\gamma$-ray pulsars \citep{2PC}. 
We note, however, that a significant discrepancy may exist between the cataloged X-ray fluxes, which are based on count rates and a simple energy conversion factor \citep{Merloni23}, and the intrinsic source fluxes, especially in the case of very soft or hard spectra, or strong foreground absorption.   
In addition, since the main eROSITA energy band is very soft, there is likely a significant contribution of thermal X-ray emission from the neutron star surface, while, for instance, \citet{Marelli11} only considered its nonthermal component, which may lead to a relative decrease in $F_{\gamma}/F_{X}$ here. 
Furthermore, the nonthermal X-ray and $\gamma$-ray emission of pulsars is likely to be nonisotropic, as significant beaming is expected \citep{Becker09}. Since the spatial origin of magnetospheric X-ray and $\gamma$-ray emission may be different, different beaming fractions and viewing geometries may affect the two regimes, with variations between young and recycled pulsars being possible \citep{Marelli11}. 
All these factors may introduce offsets and scatter into the observed $F_{\gamma}/F_{X}$ relation, compared to what may be the relation of their intrinsic magnetospheric luminosities.

\begin{figure}[t!]
\centering
\includegraphics[width=1.0\linewidth]{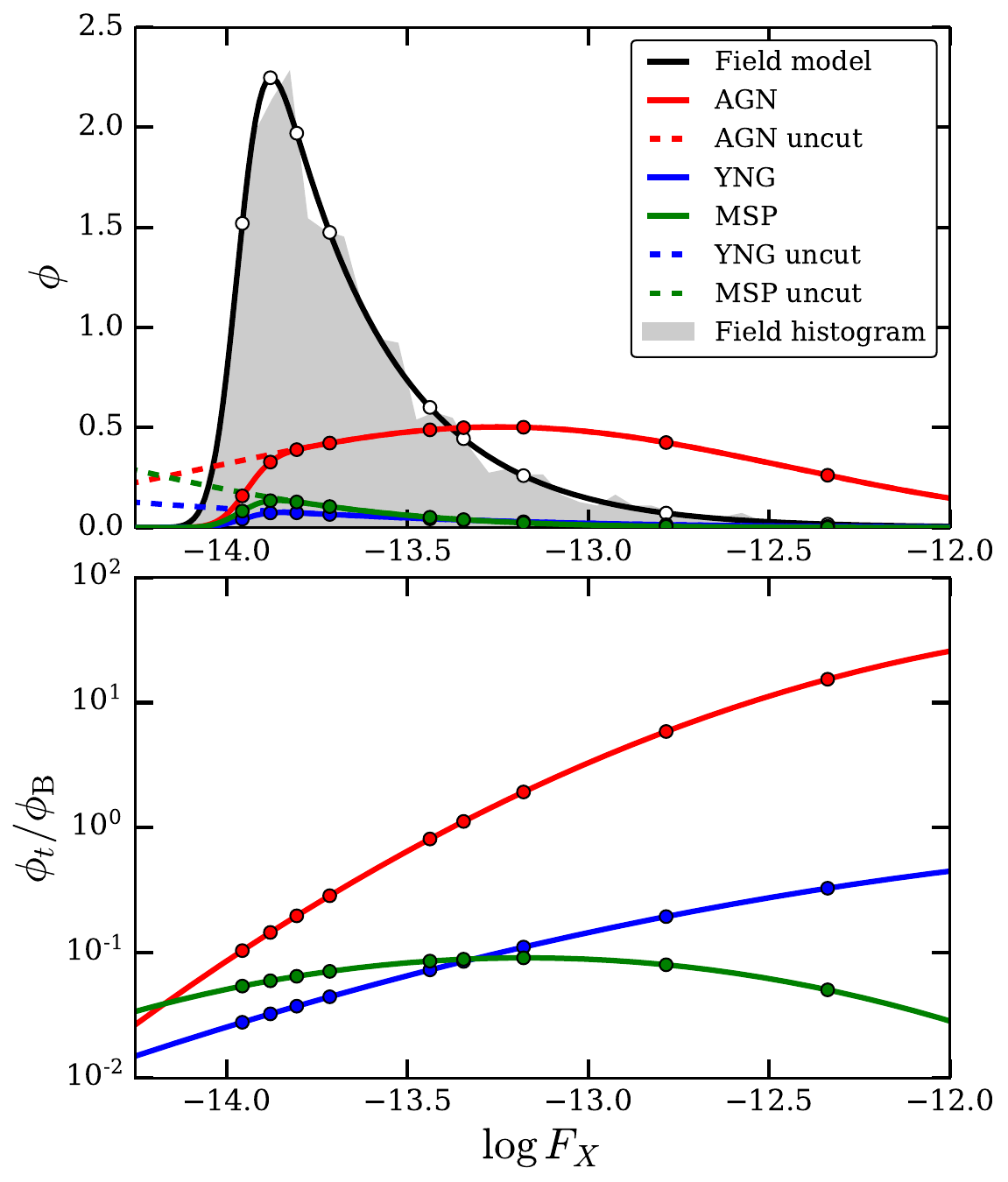} 
\caption{
Example for the computation of the expected X-ray flux for a specific $\gamma$-ray source, 4FGL J0159.8$-$2234. The top panel indicates the normalized X-ray flux distributions of field sources (black), and for the assumption of a young pulsar, MSP, or AGN nature. The bottom panel depicts the corresponding ratio of probabilities. The markers indicate the fluxes of the detected X-ray sources around the {\it Fermi} source. For reference, the $\gamma$-ray flux of the source is $\log F_{\gamma}=-12.00$, and the limiting X-ray flux is $\lambda = -13.95$. 
}
\label{Example_FluxRatioComp}
\end{figure}

\begin{figure*}[t!]
\centering
\includegraphics[width=18cm]{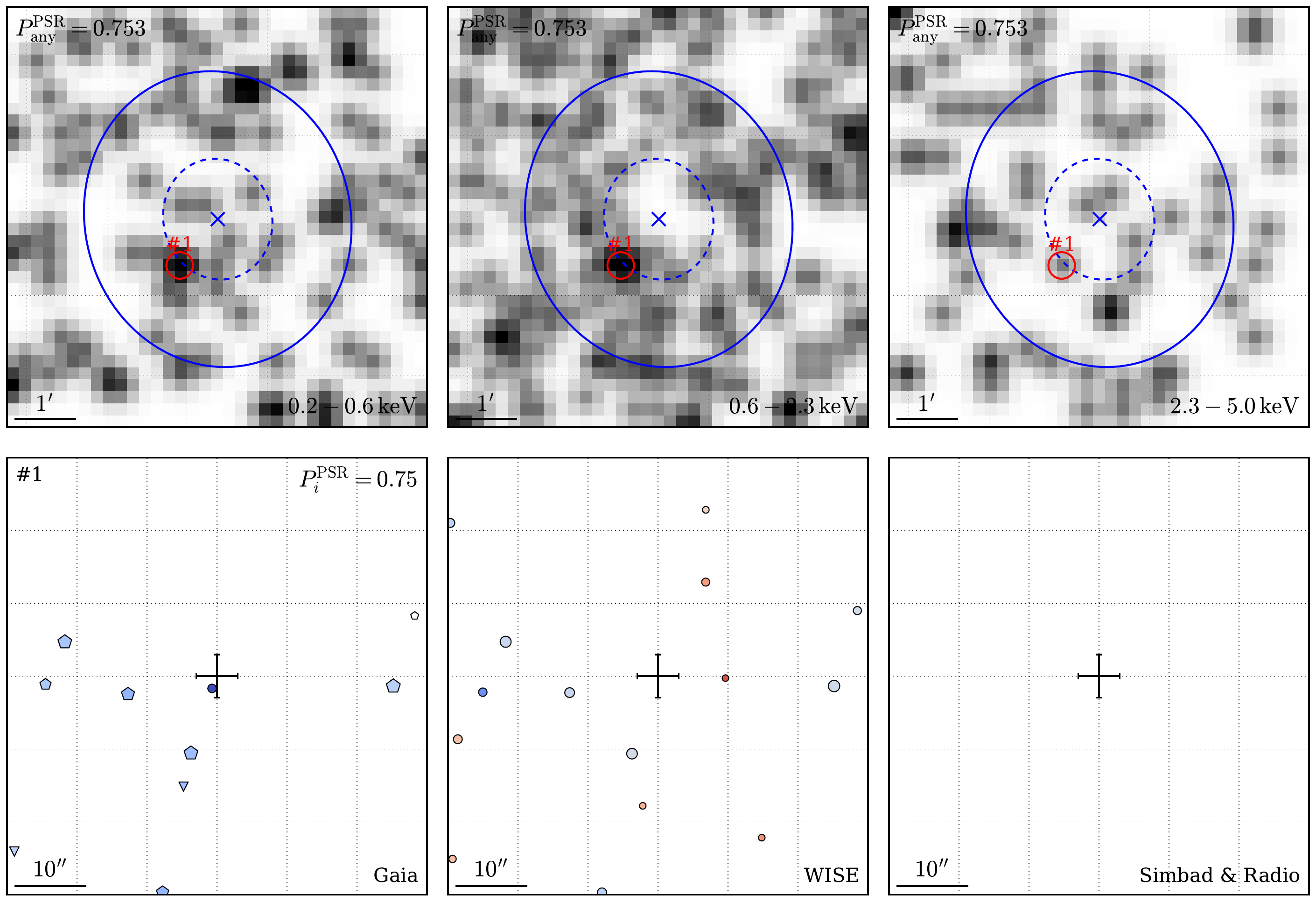} 
\caption{Example diagnostic plot for visual inspection. The top panels display eRASS:4 images of the position of the source \mbox{4FGL J1544.2$-$2554} in the energy bands $0.2-0.6$, $0.6-2.3$, and $2.3-5.0\,\si{keV}$, overlaid with the $1\sigma$ (blue dashed line) and $95\%$ error ellipse of the 4FGL source (blue solid line), and the position of the detected X-ray source (red). The pixel size is $12\arcsec$.  
The bottom panels show the local distribution of {\it Gaia} DR3 (left), CatWISE2020 (center), as well as SIMBAD entries and radio sources (right) around the X-ray source position (error bars). In the left and center panels, the size of the markers is proportional to the source brightness, and their color reflects the photometric colors (with bluer markers indicating bluer colors). In the left panel, pentagonal (triangular) markers correspond to sources with statistically significant (insignificant) proper motion or parallax, circular markers indicate the lack of a complete astrometric solution for a source. 
}
\label{VisInsp}
\end{figure*}

In order to evaluate whether a given candidate match is a likely X-ray counterpart to an unassociated 4FGL source, it is necessary to compare our expected class-dependent flux distribution with the flux distribution of field X-ray sources in the region of interest. To achieve this, for each $\gamma$-ray source, we constructed a histogram of X-ray source fluxes within a $2^{\circ}$ radius and fitted a simple empirical model for the background flux distribution $\phi_{B}$ of the type:
\begin{equation} \label{FR_Back}
    \phi_{B}(\log F_{X}\,|\,A,\Gamma, \lambda, \theta) = A\,(F_{X})^{1+\Gamma}\, \Phi(\log F_{X}\,|\, \lambda, \theta),
\end{equation}
where $\Phi$ denotes a cumulative normal distribution, which models the sensitivity-limited cutoff \citep[as in][]{Ge21}. Our fit yields constraints on the characteristic slope of the local X-ray flux distribution $\Gamma$, on the logarithm of the limiting X-ray flux at which half the sources are detected $\lambda$, and on the characteristic width of this cutoff $\theta$. The constant $A$ normalizes the probability distribution to unity.  
In a few, mostly Galactic, cases, we found that a single power law could not describe the intrinsic distribution of X-ray fluxes well. Hence, for each region, we systematically tested whether adding a second power law or a log-normal component could improve the fit, keeping the model with the lowest Akaike information criterion \citep{Akaike74} in each case.  

By applying this sensitivity cutoff to the expected class-dependent X-ray flux distribution for a given $\gamma$-ray flux $F_{\gamma}$, we can determine the expected probability distribution function $\phi_{t}$ of observing an X-ray counterpart to a source of class $t$ at a given flux $F_{X}$:
\begin{equation} \label{FR_Assoc}
    \phi_{t}(\log F_{X}\,|\,\log F_{\gamma}) = \Phi(\log F_{X}\,|\, \lambda, \theta) \,\mathcal{N}(\log F_{X}/F_{\gamma}\,|\,\mu_{t}, \sigma_{t}).
\end{equation}
Hence, by evaluating the quotient $\phi_{t}(\log F_{X}\,|\,\log F_{\gamma})/\phi_{B}(\log F_{X})$ for any potential match, we obtain a factor modifying its likelihood \citep[see][]{Budavari08, Salvato18}, under the assumption of a certain source class. Furthermore, integrating the function $\phi_{t}$ over all possible X-ray fluxes yields an estimate of the prior completeness, that is, the probability of a given $\gamma$-ray source having a detected X-ray counterpart (see Sect.~\ref{Combine}).
The approach of comparing the expected flux distributions for counterparts and field sources is illustrated for an example 4FGL source in Fig.~\ref{Example_FluxRatioComp}. 

This step of constraining the expected flux of the X-ray counterpart to a given {\it Fermi} source is vital to avoid overestimating the likelihood of having detected a pulsar-type match. X-ray counterparts to $\gamma$-ray pulsars frequently have a small chance of having a flux detectable by eRASS:4, in particular for fainter $\gamma$-ray sources ($F_{\gamma} \lesssim 3\times10^{-12}\,\si{erg.s^{-1}.cm^{-2}}$). The converse is true for blazars, which are typically expected to stand out as a particularly bright X-ray source compared to the field. 

To test whether rudimentary spectral information of our X-ray sources could be used to aid our classification, we also inspected the band-to-band hardness ratios for the energy ranges $0.2-0.5$, $0.5-1.0$, $1.0-2.0$, and $2.0-5.0\,\si{keV}$ for our different source types. 
We did find some global differences between the populations, with pulsar counterparts exhibiting larger, and $\gamma$-ray blazars smaller, scatter than field X-ray sources unrelated to the 4FGL catalog. However, due to the weak nature of these trends in our sample, we were unable to construct a quantitatively reliable predictor estimating the likelihood of a particular source being a pulsar- or AGN-type counterpart to a 4FGL source. 

\subsection{Computation of posterior match probabilities \label{Combine}}
To ``piece together'' the information obtained in the previous sections, we calculated a combined posterior probability for a given X-ray source to be the correct match to an unidentified {\it Fermi} source \emph{and} for that source to be of a certain class.
Considering all possibly matching X-ray sources and object types, the normalized posterior probability for an individual match is given by:
\begin{eqnarray} \label{FullPosterior}
    P_{i}^{t} &=& \frac{\rho_{i}^{t}}{\bar{c} + \sum_{i}\sum_{t} \rho_{i}^{t}}, \,\text{with} \\
    \rho_{i}^{t} &=& P_{\gamma}^{t} \frac{\phi_{t}(\log F_{X}\,|\,\log F_{\gamma})}{\phi_{\rm B}(\log F_{X})} 
    \frac{\exp \left( - \frac{1}{2} \boldsymbol{\psi}^{\intercal} \boldsymbol{\Sigma}^{-1} \boldsymbol{\psi} \right) }{2 \pi \nu(\mathbf{x}) \sqrt{| \boldsymbol{\Sigma} |}} \nonumber \\
    \bar{c} &=& 1 - \sum_{t} P_{\gamma}^{t} \int \phi_{\rm t}(\log F_{X}\,|\,\log F_{\gamma}) \,\mathrm{d}\log F_{X},\nonumber 
\end{eqnarray}
where the index $i$ runs over all possibly associated X-ray sources, and $t=\mathrm{\{YNG, MSP, AGN\}}$ indicates the considered object types. Here, $\rho_{i}^{t}$ can be understood as an unnormalized match probability, taking into account the $\gamma$-ray properties (via the source-type prior $P_{\gamma}^{t}$), X-ray flux (via the expected flux distribution for an association $\phi_{t}$ and the background $\phi_{\rm B}$), and the astrometry of the involved sources. The quantity $\bar{c}$ measures the ``prior incompleteness'' \citep{Salvato18} for a given $\gamma$-ray source, that is, the prior probability for its true counterpart to be missing, given its expected X-ray flux and the sensitivity limit of the catalog. 
For a given $\gamma$-ray source, the probability of having any match of a given type $t$ is given by $P_{\rm any}^{t} = \sum_{i} P_{i}^{t}$, and of having any match of any type by $P_{\rm any} = \sum_{i}\sum_{t} P_{i}^{t}$.

After having computed the ``pulsar-match'' probability $P_{i}^{\rm PSR} = P_{i}^{\rm YNG} + P_{i}^{\rm MSP} $ for each possible X-ray association of a 4FGL source, we compiled our final candidate list for counterparts to possible $\gamma$-ray pulsars by applying the following cuts:    
\begin{itemize}
    \item Location within ``three sigma'' from the $\gamma$-ray position, that is within the error ellipse containing $99.7\%$ of the probability mass (around 1.4 times the radius of the $95\%$ ellipse in the 4FGL catalog). 
    \item Minimum pulsar-match probability $P_{i}^{\rm PSR} \geq 0.02$. 
\end{itemize}
This left us with around 1100 candidate matches, 200 of which have a probability $P_{i}^{\rm PSR} \geq 0.1$. 

\begin{figure*}[t!]
\centering
\includegraphics[width=9cm]{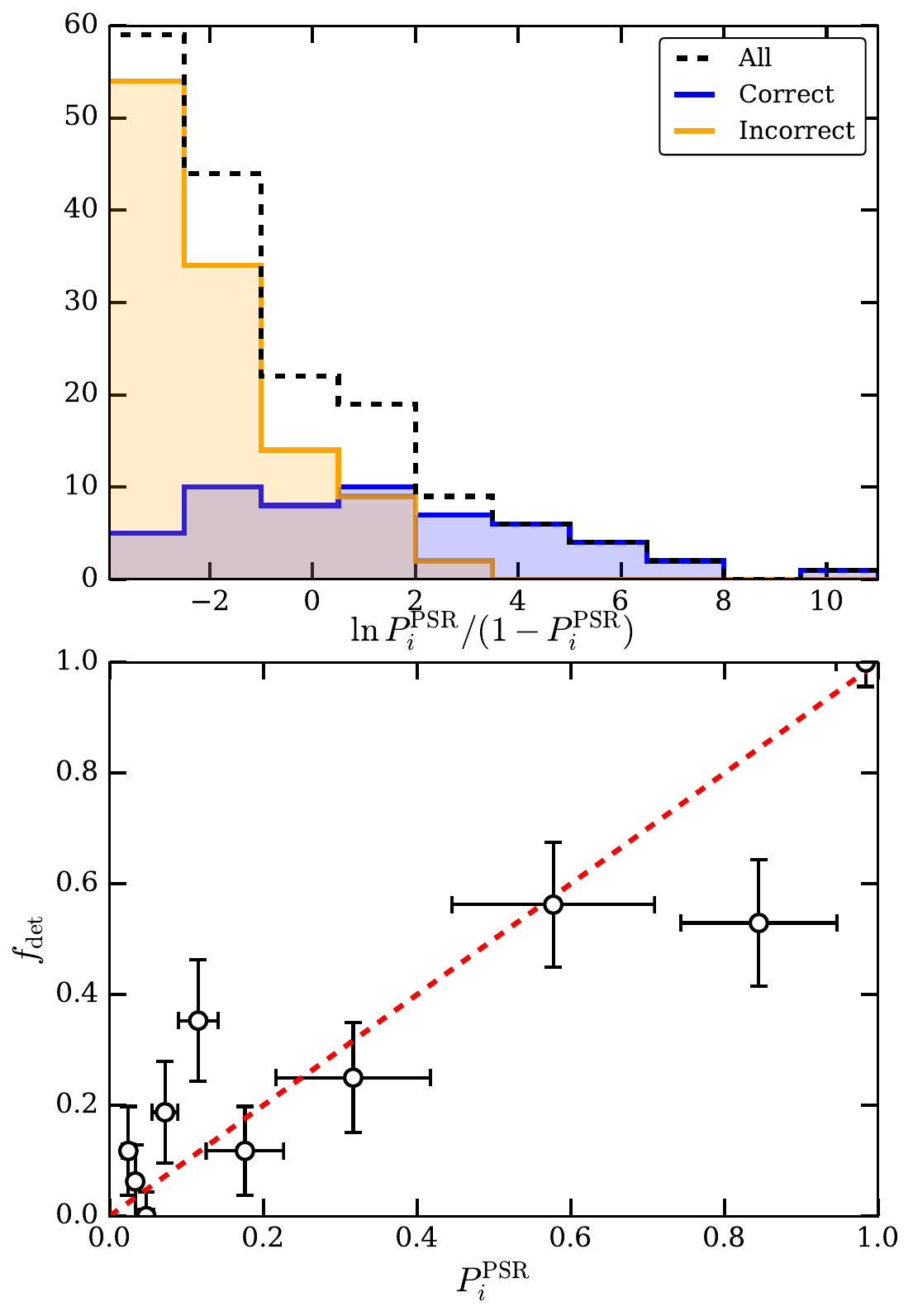} 
\includegraphics[width=9cm]{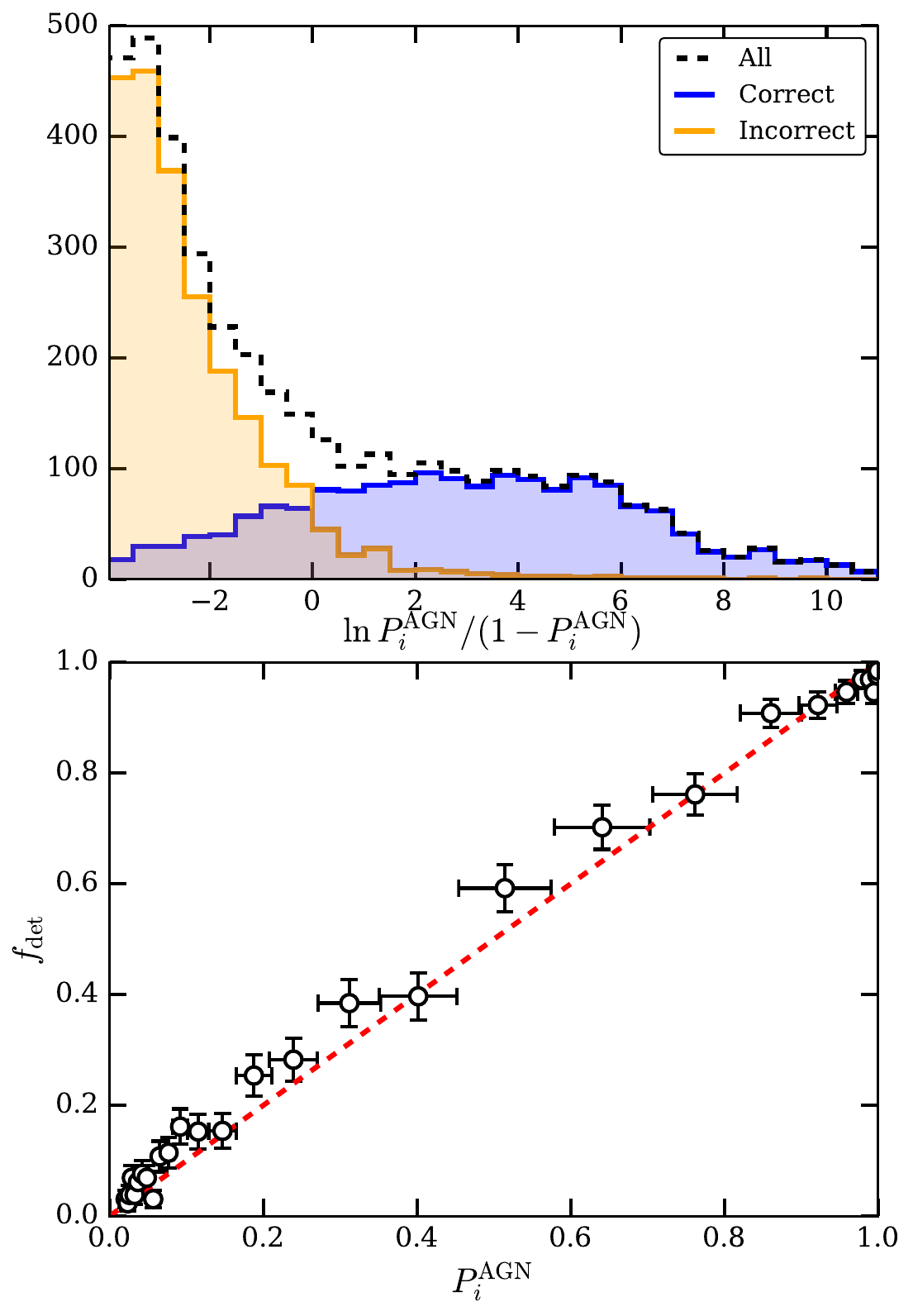} 
\caption{Verification of match probabilities. The top panels indicate the distribution of match probabilities $P_{i}^{\rm PSR}$/$P_{i}^{\rm AGN}$ (left/right) for associated 4FGL sources. The blue (orange) histograms indicate the match probabilities assigned to sources known to be the correct (incorrect) counterparts of the given 4FGL source, with the black dashed line indicating the sum of the two distributions. 
The lower panels show calibration curves for the two classes, plotting the observed fraction of correctly detected counterparts $f_{\rm det}$, for sources binned by predicted match probability $P_{i}^{\rm PSR}$/$P_{i}^{\rm AGN}$. The red line indicates the $1:1$ relation expected for an ideal, unbiased classifier.
}
\label{ProbCalibration}
\end{figure*}

\begin{table*}[t!]
 \begin{center}
 \caption[]{Observational properties of the ATNF pulsars with newly discovered X-ray counterparts in eRASS:4 data.} 
\resizebox{18.4cm}{!}{
 \begin{tabular}{c c c c c c c c c}\hline\hline\\[-1ex]
      PSR        & $\gamma$-ray counterpart&  $P$  &   $\dot{P}$  &     $d$   &   Age     &     $\dot{E}$  &     Rate        &      $\mathcal{L}_{\rm det}$     \\
      {}         &         {}           &     s     &     $\si{s.s^{-1}}$      &    kpc  &   yr     &    $\si{erg.s^{-1}}$   &      $\si{ct.s^{-1}}$      &       {}     \\[1ex]\hline\\[-0.5ex]
$\;$ J0837$-$2454  &       {}           &  0.62941 & $3.49\times10^{-13}$    &  0.20    & $2.86\times10^{4}$ & $5.50\times10^{34}$    &    0.0125    &    61.6     \\
$\;$ J0711$-$6830  &       {}           &  0.00549 & $1.49\times10^{-20}$    &  0.11  & $5.84\times10^{9}$ & $3.61\times10^{33}$     &    0.0056     &     16.2     \\
$\;$ J1902$-$5105  & 4FGL J1902.0$-$5105 &  0.00174 & $9.20\times10^{-21}$    &  1.65   & $3.00\times10^{9}$ & $6.87\times10^{34}$    &    0.0194     &     5.3      \\
$\;$ J1045$-$4509  &       {}           &  0.00747 & $1.76\times10^{-20}$    &  0.34  & $6.70\times10^{9}$ &  $1.70\times10^{33}$    &    0.0129     &     7.4      \\
$\;$ J1125$-$6014  & 4FGL J1126.4$-$6011 &  0.00263 & $3.73\times10^{-21}$    &  0.99   & $1.12\times10^{10}$ & $8.09\times10^{33}$   &    0.0151     &     11.1     \\
$\;$ J1312$+$0051  & 4FGL J1312.7+0050 &  0.00423 & $1.75\times10^{-20}$    &  1.47   & $3.82\times10^{9}$ & $9.15\times10^{33}$      &    0.0406     &     31.6     \\  
$\;$ J0509$+$0856  &       {}           &  0.00406 & $4.41\times10^{-21}$    &  0.82  & $1.46\times10^{10}$ & $2.61\times10^{33}$    &    0.0150     &     7.4      \\ 
$\;$ J0514$-$4408  & 4FGL J0514.6$-$4408 &  0.32027 & $2.04\times10^{-15}$    &  0.97  & $2.48\times10^{6}$ & $2.45\times10^{33}$    &    0.0134     &     17.3     \\
$\;$ J1207$-$5050  & 4FGL J1207.4$-$5050 &  0.00484 & $6.06\times10^{-21}$    &  1.27   & $1.27\times10^{10}$ & $2.11\times10^{33}$   &    0.0174     &     14.1     \\
$\;$ J0952$-$0607  & 4FGL J0952.1$-$0607 &  0.00141 & $4.77\times10^{-21}$    &  7.60   & $4.69\times10^{9}$ & $6.67\times10^{34}$    &    0.0325     &     12.3     \\
$\;$ J0610$-$2100  & 4FGL J0610.2$-$2100 &  0.00386 & $1.23\times10^{-20}$    &  3.26  & $4.96\times10^{9}$ & $8.45\times10^{33}$    &    0.0241     &     15.6     \\
$\;$ J1439$-$5501  & 4FGL J1440.2$-$5505 &  0.02864 & $1.42\times10^{-19}$    &  0.66   & $3.20\times10^{9}$  & $2.38\times10^{32}$   &    0.0180     &     13.3     \\
     B1036$-$45  &       {}           &  0.66199 & $1.25\times10^{-15}$    &  0.92  & $8.38\times10^{6}$ & $1.71\times10^{32}$       &    0.0167     &     11.6     \\
$\;$ J1405$-$5641  &       {}           &  0.61757 & $1.20\times10^{-15}$    & 12.38  & $8.17\times10^{6}$ & $2.01\times10^{32}$     &    0.0133     &     6.6      \\
$\;$ J1057$-$5851  & 4FGL J1056.9$-$5852 &  0.62037 & $1.01\times10^{-13}$    &   {}     & $9.77\times10^{4}$ & $1.66\times10^{34}$  &    0.0099     &     8.4      \\[-2ex]
\multicolumn{9}{c}{\rule[0mm]{0mm}{0mm}}\\
\hline\\[-3ex]
\end{tabular}
}
\tablefoot{
 The individual columns are as follows:
 PSR: Pulsar name,  $\gamma$-ray counterpart: Name of the source in the 4FGL catalog.
 The pulsar parameters $P$ (rotation period), $\dot{P}$ (period derivative), $d$ (distance), (spin-down) age, and $\dot{E}$ (spin-down power) are taken from the ATNF pulsar database \citep{ATNF}. Rate is the count rate in the $0.2-2.3$ keV band. 
 $\mathcal{L}_{\rm det}$ is the detection likelihood  for the $0.2-2.3$ keV band  \citep{Brunner22}. 
 The list is sorted in descending order of $\dot{E}/ 4\pi d^2$. 
 \label{PSR_counterpart}}
 \end{center}
 \end{table*}

\begin{table*}[t!]
    \centering
    \caption{List of top 50 candidate pulsar-type matches to 4FGL sources.}
    \resizebox{18.4cm}{!}{
    \begin{tabular}{cccccclcccccl}
    \hline\hline
        \#&$\alpha_{X}$&$\delta_{X}$&$\sigma_{X}$&$\mathcal{L}_{\rm det}$&$\mathcal{L}_{\rm ext}$&\tt Source\_Name\_4FGL&$P_{\gamma}^{\rm PSR}$&$P_{i}^{\rm PSR}$&$P_{i}^{\rm YNG}$&$P_{i}^{\rm MSP}$&$P_{i}^{\rm AGN}$&Comment\\ 
        &$\rm deg$&$\rm deg$&$\rm arcsec$&&&&&&&&&\\ \hline%
        1&$236.06418$&$-25.92528$&$3.0$&$19.8$&$0.0$&\tt 4FGL J1544.2-2554&$0.997$&$0.753$&$0.000$&$0.753$&$0.001$&\\%
        2&$197.23184$&$-62.41172$&$1.6$&$180$&$0.0$&\tt 4FGL J1309.1-6223&$0.997$&$0.637$&$0.630$&$0.007$&$0.003$&\\%
        3&$253.31237$&$-43.82513$&$4.2$&$5.2$&$0.0$&\tt 4FGL J1653.2-4349&$0.989$&$0.563$&$0.562$&$0.000$&$0.000$&\\%
        4&$143.50470$&$-62.56495$&$2.5$&$38$&$0.0$&\tt 4FGL J0933.8-6232&$1.000$&$0.540$&$0.032$&$0.509$&$0.000$&\it (a)\\%
        5&$244.20254$&$-53.69458$&$2.3$&$48$&$0.0$&\tt 4FGL J1616.6-5341&$1.000$&$0.480$&$0.176$&$0.303$&$0.000$&\it (b)\\%
        6&$252.34131$&$-44.73401$&$1.8$&$112$&$0.0$&\tt 4FGL J1649.3-4441&$0.990$&$0.477$&$0.460$&$0.017$&$0.009$&\it (b)\\%
        7&$118.04167$&$-29.50531$&$1.6$&$217$&$0.0$&\tt 4FGL J0752.0-2931&$0.999$&$0.452$&$0.436$&$0.016$&$0.003$&\it (c)\\%
        8&$263.86181$&$-29.75370$&$7.7$&$18.0$&$6.1$&\tt 4FGL J1735.4-2944&$0.913$&$0.451$&$0.427$&$0.024$&$0.291$&\\%
        9&$201.34839$&$-54.23968$&$3.8$&$10.3$&$0.0$&\tt 4FGL J1325.3-5413&$0.968$&$0.448$&$0.018$&$0.430$&$0.005$&\it (d)\\%
        10&$266.41332$&$-36.42802$&$3.0$&$37$&$0.0$&\tt 4FGL J1745.6-3626&$0.751$&$0.448$&$0.023$&$0.425$&$0.267$&\\%
        11&$236.33279$&$-45.90416$&$2.2$&$77$&$0.0$&\tt 4FGL J1545.2-4553&$0.911$&$0.361$&$0.024$&$0.337$&$0.131$&\\%
        12&$260.52566$&$-32.08855$&$4.2$&$5.4$&$0.0$&\tt 4FGL J1722.1-3205&$0.971$&$0.349$&$0.033$&$0.316$&$0.003$&\\%
        13&$209.23867$&$-61.38800$&$1.8$&$130$&$0.0$&\tt 4FGL J1357.3-6123&$0.952$&$0.348$&$0.337$&$0.011$&$0.037$&\it (b, e)\\%
        14&$236.81672$&$-48.00443$&$3.5$&$16.1$&$0.0$&\tt 4FGL J1547.4-4802&$0.944$&$0.348$&$0.035$&$0.313$&$0.018$&\\%
        15&$183.49466$&$-44.25681$&$4.0$&$10.8$&$0.0$&\tt 4FGL J1213.9-4416&$0.945$&$0.335$&$0.002$&$0.332$&$0.007$&\\%
        16&$214.66390$&$-61.19200$&$5.2$&$10.3$&$0.0$&\tt 4FGL J1418.7-6110&$0.917$&$0.326$&$0.318$&$0.007$&$0.005$&\\%
        17&$260.15801$&$-26.87634$&$2.5$&$33$&$0.0$&\tt 4FGL J1720.6-2653c&$0.400$&$0.324$&$0.031$&$0.293$&$0.539$&\\%
        18&$245.57301$&$-72.05278$&$4.1$&$13.2$&$0.0$&\tt 4FGL J1622.2-7202&$0.997$&$0.323$&$0.000$&$0.323$&$0.001$&\\%
        19&$253.57303$&$-49.11719$&$3.3$&$12.6$&$0.0$&\tt 4FGL J1654.2-4907c&$0.287$&$0.319$&$0.046$&$0.273$&$0.306$&\\%
        20&$246.53852$&$-49.29462$&$2.6$&$28.3$&$0.0$&\tt 4FGL J1626.0-4917c&$0.483$&$0.311$&$0.293$&$0.018$&$0.248$&\\%
        21&$225.02645$&$-58.77219$&$5.1$&$7.5$&$0.0$&\tt 4FGL J1500.1-5846&$0.509$&$0.306$&$0.301$&$0.005$&$0.068$&\\%
        22&$171.97380$&$-62.02326$&$1.5$&$163$&$0.0$&\tt 4FGL J1127.9-6158&$0.998$&$0.303$&$0.291$&$0.012$&$0.001$&\it (f)\\%
        23&$126.56484$&$-50.90072$&$4.2$&$8.9$&$0.0$&\tt 4FGL J0826.1-5053&$0.986$&$0.301$&$0.002$&$0.298$&$0.001$&\\%
        24&$92.18263$&$20.59542$&$4.0$&$39$&$0.0$&\tt 4FGL J0608.8+2034c&$0.969$&$0.299$&$0.286$&$0.013$&$0.018$&\\%
        25&$249.93840$&$-46.68515$&$4.9$&$10.2$&$0.0$&\tt 4FGL J1639.8-4642c&$0.957$&$0.280$&$0.279$&$0.001$&$0.001$&\it (e)\\%
        26&$118.73457$&$-39.88257$&$4.4$&$7.4$&$0.0$&\tt 4FGL J0754.9-3953&$1.000$&$0.279$&$0.003$&$0.276$&$0.000$&\\%
        27&$252.85143$&$-44.36273$&$1.1$&$1220$&$0.0$&\tt 4FGL J1650.9-4420c&$0.915$&$0.279$&$0.270$&$0.009$&$0.258$&\it (b, e)\\%
        28&$254.16530$&$-48.22062$&$2.6$&$39$&$0.0$&\tt 4FGL J1656.9-4814&$0.722$&$0.270$&$0.077$&$0.194$&$0.110$&\\%
        29&$194.20780$&$-63.66546$&$4.2$&$6.8$&$0.0$&\tt 4FGL J1257.0-6339&$0.959$&$0.265$&$0.155$&$0.111$&$0.001$&\\%
        30&$221.32135$&$-60.00743$&$3.1$&$18.7$&$0.0$&\tt 4FGL J1445.1-5958c&$0.816$&$0.259$&$0.254$&$0.005$&$0.027$&\\%
        31&$244.81362$&$-50.78953$&$3.2$&$22.0$&$0.0$&\tt 4FGL J1619.3-5047&$0.982$&$0.258$&$0.253$&$0.005$&$0.002$&\\%
        32&$259.37013$&$-44.03974$&$2.4$&$62$&$0.0$&\tt 4FGL J1717.6-4404&$0.760$&$0.253$&$0.095$&$0.158$&$0.238$&\\%
        33&$258.72454$&$-33.38508$&$3.6$&$10.9$&$0.0$&\tt 4FGL J1714.9-3324&$0.999$&$0.252$&$0.046$&$0.206$&$0.000$&\\%
        34&$264.39070$&$-33.54471$&$2.5$&$22.3$&$0.0$&\tt 4FGL J1737.3-3332&$0.942$&$0.250$&$0.212$&$0.038$&$0.007$&\it (b)\\%
        35&$198.12297$&$-62.57599$&$2.4$&$41$&$0.0$&\tt 4FGL J1312.6-6231c&$0.987$&$0.245$&$0.200$&$0.045$&$0.004$&\it (g)\\%
        36&$246.65399$&$-48.96485$&$3.5$&$18.8$&$0.0$&\tt 4FGL J1626.5-4858c&$0.882$&$0.237$&$0.231$&$0.006$&$0.013$&\it (b)\\%
        37&$252.33402$&$-45.23386$&$4.2$&$15.8$&$0.0$&\tt 4FGL J1649.2-4513c&$0.964$&$0.232$&$0.221$&$0.011$&$0.002$&\\%
        38&$148.43061$&$-15.16169$&$5.3$&$5.2$&$0.0$&\tt 4FGL J0953.6-1509&$1.000$&$0.229$&$0.000$&$0.228$&$0.000$&\\%
        39&$251.25342$&$-41.38724$&$2.7$&$27.5$&$0.0$&\tt 4FGL J1645.1-4123c&$0.947$&$0.228$&$0.022$&$0.206$&$0.031$&\\%
        40&$254.48273$&$-46.92163$&$1.9$&$96$&$0.0$&\tt 4FGL J1657.7-4656c&$0.527$&$0.226$&$0.102$&$0.123$&$0.433$&\it (b)\\%
        41&$243.00184$&$-51.42570$&$3.7$&$17.5$&$0.0$&\tt 4FGL J1611.9-5125c&$0.957$&$0.224$&$0.222$&$0.002$&$0.001$&\it (b)\\%
        42&$86.16917$&$22.63266$&$2.8$&$44$&$0.0$&\tt 4FGL J0544.4+2238&$0.688$&$0.222$&$0.180$&$0.042$&$0.103$&\\%
        43&$228.44197$&$-15.34978$&$4.9$&$9.2$&$0.0$&\tt 4FGL J1513.7-1519&$0.989$&$0.219$&$0.002$&$0.217$&$0.004$&\\%
        44&$246.72777$&$-42.87342$&$2.9$&$25.0$&$0.0$&\tt 4FGL J1626.6-4251&$0.908$&$0.217$&$0.025$&$0.192$&$0.016$&\it (b)\\%
        45&$205.67808$&$-57.48130$&$4.0$&$17.2$&$0.0$&\tt 4FGL J1342.6-5730&$0.473$&$0.216$&$0.064$&$0.152$&$0.281$&\\%
        46&$275.72745$&$-47.31907$&$5.3$&$5.2$&$0.0$&\tt 4FGL J1822.9-4718&$0.961$&$0.212$&$0.000$&$0.212$&$0.008$&\\%
        47&$213.09350$&$-60.30956$&$2.4$&$37$&$0.0$&\tt 4FGL J1412.2-6018&$0.427$&$0.212$&$0.191$&$0.021$&$0.625$&\\%
        48&$254.33750$&$-39.28219$&$3.9$&$13.1$&$0.0$&\tt 4FGL J1657.4-3917c&$0.630$&$0.211$&$0.083$&$0.128$&$0.059$&\\%
        49&$198.15828$&$-62.95837$&$3.1$&$18.7$&$0.0$&\tt 4FGL J1312.3-6257&$0.606$&$0.209$&$0.201$&$0.008$&$0.025$&\\%
        50&$262.51595$&$-34.32575$&$3.1$&$22.1$&$0.0$&\tt 4FGL J1730.1-3422&$0.930$&$0.208$&$0.193$&$0.014$&$0.002$&\\%
        \hline
    \end{tabular}
    }
    \tablefoot{The full list of sources is available online at CDS.  
    The columns $\alpha_{X}$, $\delta_{X}$, $\sigma_{X}$ give the position and uncertainty of the matched X-ray source; $\mathcal{L}_{\rm det}$ and $\mathcal{L}_{\rm ext}$ specify its detection and extent likelihood \citep{Brunner22}. $P_{\gamma}^{\rm PSR}$ gives the prior probability of the respective $\gamma$-ray source to correspond to a pulsar (Sect. \ref{Prior}). 
    Finally, $P_{i}^{\rm PSR}$ ($P_{i}^{\rm AGN}$) indicates the Bayesian probability for a particular X-ray source to be the correct match {\it and} for the source's nature to be a pulsar (blazar), respectively. $P_{i}^{\rm YNG}$ and $P_{i}^{\rm MSP}$ show the same for the young pulsar and MSP subtypes. The entries have been sorted by descending $P_{i}^{\rm PSR}$.   
    \\ \tablefoottext{a}{Identified as likely MSP in \citet{Dai17};} \tablefoottext{b}{Possible stellar counterpart;} \tablefoottext{c}{Star-rich region;} \tablefoottext{d}{Possible binary counterpart;} \tablefoottext{e}{Classified as likely blazar by \citet{Kerby21};} \tablefoottext{f}{Possible Be star counterpart;} \tablefoottext{g}{Classified as likely pulsar by \citet{Kerby21}}.}
    \label{GoodMatches}
\end{table*}

\subsection{Visual inspection and multiwavelength counterparts} \label{Counterpart}
As a final step of our procedure, we performed a visual inspection of all potentially matching X-ray sources, fulfilling the criteria specified in the previous section. To achieve this, we created diagnostic plots for each matched 4FGL source, overlaying the X-ray and $\gamma$-ray source position(s) on local eRASS:4 X-ray images in the energy bands $0.2-0.6$, $0.6-2.3$, and $2.3-5.0\,\si{keV}$. The target of this effort was to recognize spurious X-ray sources, which were falsely detected, for instance, due to an enhanced level of diffuse emission in Galactic extended sources, or in the wings of the point spread function of very bright point sources \citep[see also][]{Merloni23}.
While such spurious emission is often classified as extended in the detection pipeline, we refrained from generally excluding extended sources from our cross-matching effort, as young pulsars can potentially exhibit X-ray-bright pulsar wind nebulae, which may show a significant angular size. 

In addition to examining X-ray images, we also found it helpful to check whether our potentially matching X-ray sources could be classified securely as non-pulsar via their multiwavelength counterparts. The weak optical and infrared emission of pulsars contrasts the properties of stars and AGN, the two most common X-ray-emitting source classes in the all-sky survey \citep{Salvato22}, which tend to be characterized by bright optical \citep{Schneider22} and mid-infrared \citep{Salvato18} emission, respectively. This can be exploited to classify the potential counterpart of an X-ray source \citep[e.g.,][]{Tranin22}. 
Therefore, we overplotted the positions of our X-ray sources with the distribution of sources in the {\it Gaia} DR3 \citep{Gaia, GaiaDR3} and CatWISE2020 catalogs \citep{CatWISE2020}, to check for evident non-pulsar counterparts. In this way, purely coronal X-ray emitters are identifiable as comparatively bright optical sources, with significant proper motion or parallax in the {\it Gaia} catalog, whereas AGN tend to appear as mid-infrared sources with typical {\it WISE} color $\rm W1-W2 > 0$ \citep{Salvato18}, without significant proper motion or parallax in their possible {\it Gaia} counterpart. 
For completeness, we also queried the SIMBAD database \citep{Wenger00} to check for existing identifications of astronomical sources in the vicinity. Furthermore, we queried a subset \citep[SUMSS, TGSS, FIRST, NVSS, VLSSR, WENSS, GLEAM, LOTSS, WISH;][]{SUMSS, TGSS, FIRST, NVSS, VLSSR, WENSS, GLEAM, LOTSS, WISH} of the radio source catalogs used by \citet{Bruzewski23} to check whether our sources exhibit radio emission at frequencies $\lesssim1\,\si{GHz}$, which may be expected for both blazars and pulsars. 

We note that there are two fundamental caveats to this approach: 
Since, due to their formation, most MSPs are expected to be located in binary systems, it is likely that a large fraction of X-ray-emitting rotation-powered pulsars in fact has an optical counterpart detectable by {\it Gaia}. However, since coronal X-ray emitters typically emit only a fraction of their luminosity at X-ray energies \citep{Pizzolato03, Wright11}, we expect such cases of optically detected binary companions to be much optically fainter than a coronal X-ray emitter would be. Hence, even for convincing positional matches, we only excluded sources with optical counterparts with a {\it Gaia} magnitude of $m_{\rm G} \lesssim 15$. For comparison, typically, redback companions have optical magnitudes on the order of $m_{\rm G} \sim 20$ \citep{Clark23}. 
For potential AGN counterparts, we did not apply a limiting magnitude, but required a convincing extragalactic nature (i.e. no significant proper motion or parallax in {\it Gaia}), and a clear mid-infrared counterpart.  
Second, for most sources, the X-ray astrometric errors and counterpart source densities are so high that a secure statement on the presence of a multiwavelength counterpart is impossible to make. 
Hence, the above criteria were applied rather conservatively, and sources were only excluded if the multiwavelength association was deemed very likely from a positional standpoint. In practice, this implies positional agreement within $1-2$ sigma and sufficiently small error bars for only a single source to be within $\sim 3$ sigma of the X-ray source.  

The diagnostic plots displaying X-ray images of all sources and multiwavelength catalogs around their positions are available on Zenodo\footnote{\url{https://zenodo.org/doi/10.5281/zenodo.10634392}}. For illustration, Fig.~\ref{VisInsp} displays an example plot, showing the region of the source 4FGL J1544.2$-$2554.  
In that particular example, an optical counterpart appears to exist for the matched X-ray source. While it is much too faint to explain the observed X-ray source with coronal emission, it could be a viable binary counterpart to a possible X-ray bright MSP. 

\begin{figure*}[t!]
\centering
\includegraphics[width=18cm]{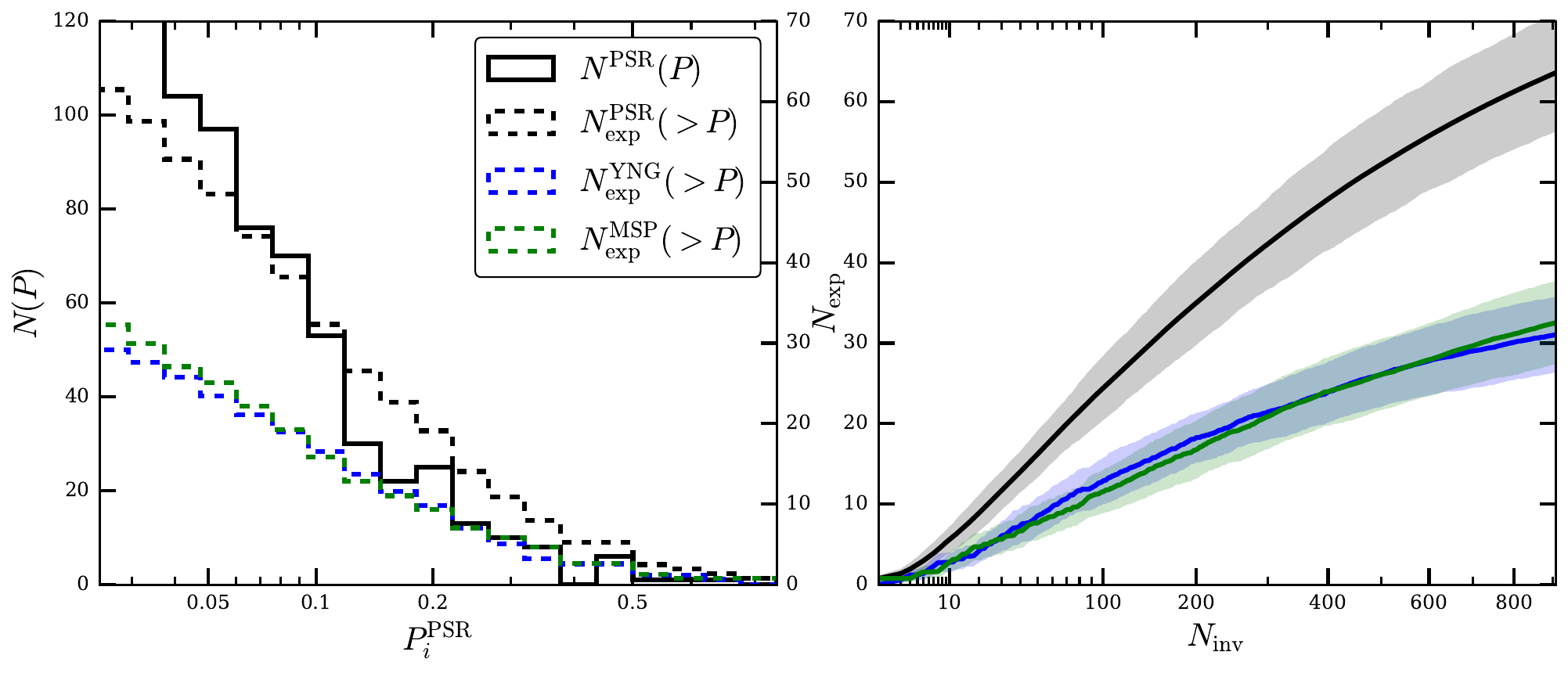} 
\caption{Overview of match probabilities. The left panel shows a histogram of pulsar-match probabilities $P_{i}^{\rm PSR}$ (black, solid line) and the expected number of pulsar detections $N_{\rm exp}$ above a certain threshold of $P_{i}^{\rm PSR}$ (black, dashed), after visual inspection. The green and blue dashed lines indicate the expected numbers of young pulsars and MSPs above the respective threshold. 
The right panel shows the dependence of the expected number of detections with the number of investigated X-ray sources $N_{\rm inv}$ for all pulsars and the two subsets in green and blue. The shaded regions indicate the expected statistical error on the predicted number. }
\label{MatchingPotential}
\end{figure*}

\begin{figure*}
\centering
\includegraphics[width=14cm]{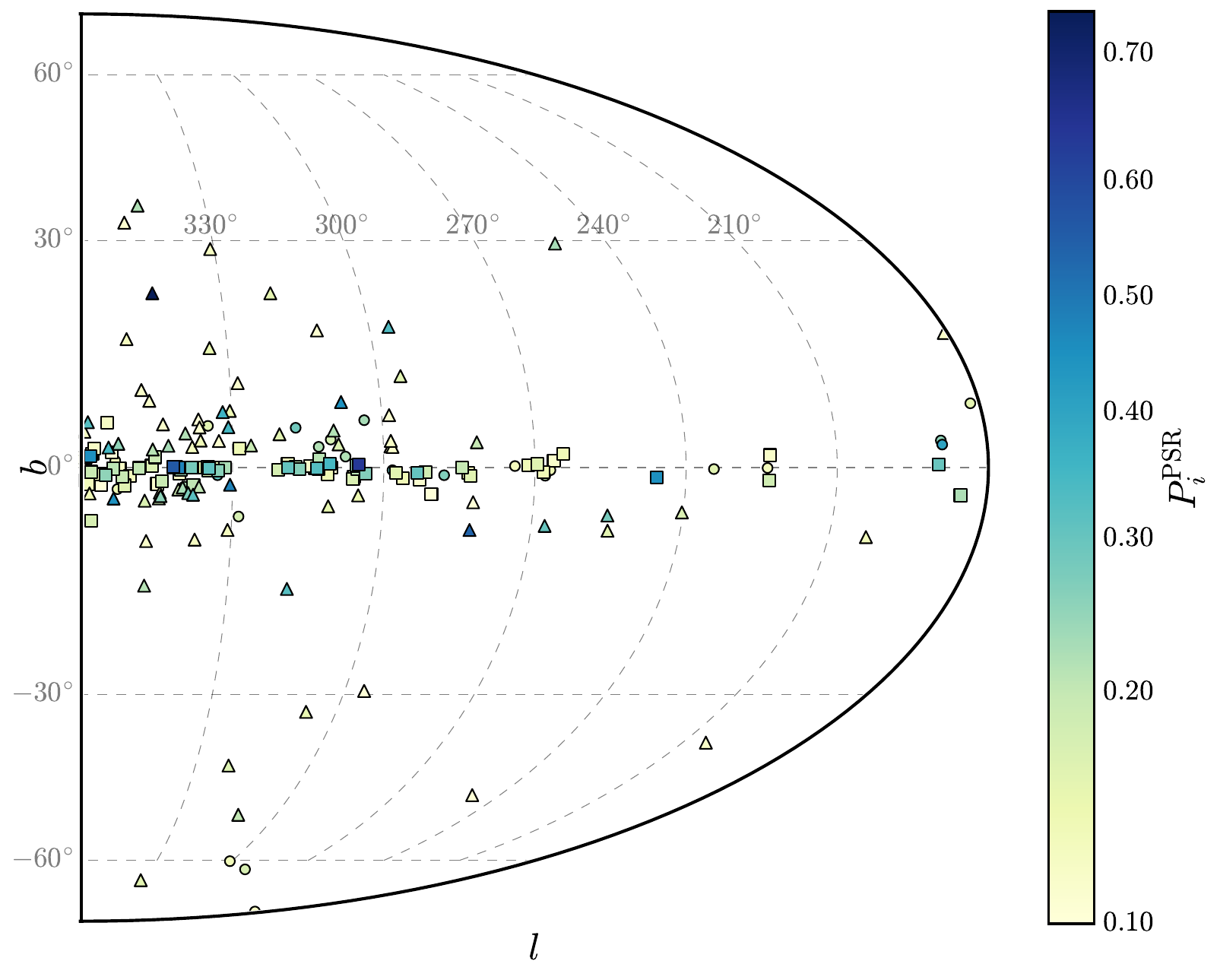} 
\caption{Distribution of the best pulsar-type match candidates to {\it Fermi} sources in the western Galactic hemisphere. The color bar indicates the maximum probability of having an X-ray-detected pulsar counterpart $P_{i}^{\rm PSR}$ for each source. 
Sources deemed likelier to be young pulsars are marked with squares and likelier MSPs with triangles. Sources that did not pass visual inspection are marked with circles.}
\label{MatchLonLatDistribution}
\end{figure*}

\subsection{Verifying our approach with previously associated 4FGL sources} \label{Calibration}
To test the reliability of our method and verify its probabilistic predictions, we also ran our algorithm on samples of previously associated sources in the 4FGL catalog. In two separate cross-matching runs, we applied our method to all previously identified or associated pulsar- and blazar-type $\gamma$-ray sources with {\tt CLASS1 == 'psr', 'msp'} and {\tt CLASS1 == 'bll', 'fsrq', 'bcu'}, respectively. Importantly, this approach provides us with a ground truth to test our method against, as the positions of the true multiwavelength counterparts of 4FGL sources are given in the catalog. 
To emulate the behavior of our machine-learning classifier on unassociated sources, we predicted the $\gamma$-ray-based likelihood for a source to be pulsar $P_{\gamma}^{\rm PSR}$, using our out-of-bag estimates created via 5-fold cross-validation (Sect.~\ref{Prior}). This means predictions for a particular source were made by a classifier that had not previously seen that source. 
Furthermore, we naturally used the full $\gamma$-ray-only error ellipses to model match probabilities rather than the known counterpart positions. 
By binning the resulting cross-matches by match probability $P_{i}^{\rm PSR}$ or $P_{i}^{\rm AGN}$, and evaluating the number of correctly and incorrectly identified counterparts in each bin, we are thus able to quantitatively test the performance and reliability of our algorithm. We considered a given X-ray source matched correctly if its position was less than $15\arcsec$ away from the counterpart given in 4FGL-DR4. Since pulsars frequently exhibit extended, not necessarily symmetric, pulsar wind nebulae, we used a more generous radius of $60\arcsec$ for extended X-ray sources.

The result of this exercise is displayed in Fig.~\ref{ProbCalibration}. In the cases of both pulsars and blazars, the sample of sources with high match probabilities is found to be very pure, as desired. Furthermore, the quantitative reliability of our output probabilities is illustrated by the close correspondence between predicted match probabilities $P_{i}^{\rm PSR}$/$P_{i}^{\rm AGN}$, and the observed fraction of correctly identified counterparts $f_{\rm det}$, despite a larger scatter for the pulsar class due to the much smaller test sample.

\section{Newly discovered X-ray counterparts to known rotation-powered pulsars \label{PSRs_X}}
In addition to the identification work done on the 4FGL catalog, we correlated the ATNF pulsar database \citep{ATNF} in version 1.71 with the eRASS:4 source catalog, as quite a significant number of pulsars found by Fermi were subsequently detected in the radio band. 
An earlier correlation by \cite{Prinz_Becker_2015} based on archival XMM-Newton and Chandra data identified the X-ray counterpart of 20 previously undetected rotation-powered pulsars. eROSITA with its unlimited field of view in survey mode, allows for searching for X-ray emission from all rotation-powered pulsars down to a homogeneous sensitivity limit. 
A statistical analysis of source positions in eRASS:1 showed that the astrometric accuracy varies between $\sim 4-15$ arcsec for sources detected with a likelihood below 10 \citep{Merloni23}. We thus restricted the search radius to $15$ arcsec, and, using the combined eRASS:4 data set, we found that 97 pulsars from the ATNF pulsar database in the western Galactic hemisphere match by position with an eROSITA source. 
Among the list of pulsars matching with an eROSITA source, 52 have a counterpart in the 4FGL-DR4 catalog (see Sect.~\ref{FluxRatio}).
For 15 investigated sources, we detected an X-ray counterpart for which we did not find a previous report in the literature. These sources are listed in Table \ref{PSR_counterpart} along with their basic pulsar parameters as adopted from the ATNF pulsar database \citep{ATNF}. Nine of these 15 pulsars 
have a $\gamma$-ray counterpart according to the 4FGL-DR4 catalog. 

Of all 15 pulsars in Table \ref{PSR_counterpart}, the X-ray counterpart of PSR J0837$-$2454 has the highest detection likelihood. This pulsar was discovered recently by \cite{PSR_J0837}. Its age and distance estimates are $28.6\,\si{kyr}$ and $0.2-0.9$ kpc \citep{PSR_J0837}, respectively. Its X-ray counterpart is detected only in the $0.2-0.7$ keV band, implying that the pulsar has a very soft X-ray spectrum. The 30 source counts detected in eRASS:4 do not support detailed spectral modeling due to the sparse photon statistics but approximately constrain the interstellar column absorption  towards PSR J0837$-$2454 to $N_{\rm H} \sim 10^{20}\, \si{cm^{-2}}$. This is in agreement with the column density through the Galaxy in the direction of PSR J0837$-$2454, which is $N_{\rm H} \approx 7.5 \times 10^{20}\, \si{cm^{-2}}$ \citep{HI4PI}. The low galactic column density further implies that the distance to PSR J0837$-$2454 is likely rather small. In any case, it is in strong contrast with the large dispersion measure toward the pulsar, which would imply a column density around $N_{\rm H} \approx 4 \times 10^{21}\, \si{cm^{-2}}$ \citep{PSR_J0837, He13}.
This may imply an unusual ionization state of the interstellar medium along this line of sight, with an increased abundance of free electrons, meaning an enhanced ionization fraction. 

\section{Results and discussion \label{Results}}

\subsection{Catalog of candidate pulsar counterparts to unassociated 4FGL sources}
Table \ref{GoodMatches} shows our 50 most likely pulsar candidates after visually inspecting all potential matches between eRASS:4 and 4FGL-DR4 catalogs. The full catalog of candidate matches, ranked by descending probability of pulsar-type matches $P_{i}^{\rm PSR}$ 
is available in electronic form at CDS.
This probability combines the probabilities of having a young and a recycled pulsar, meaning $P_{i}^{\rm PSR} = P_{i}^{\rm YNG} + P_{i}^{\rm MSP}$.
In addition, we also provide the visual inspection plots for each 4FGL source and a catalog of those sources that were removed from our final sample during visual inspections. 
Table \ref{ColumnDescriptions} gives an overview of the released catalog's columns. Apart from the posterior match probabilities for different source classes, this catalog also includes X-ray properties such as source positions, errors, and fluxes (derived from count rates), as well as the individual quantities constructed to compute the match probabilities, such as flux ratios or $\gamma$-ray-based source classifications. 
Fig.~\ref{MatchingPotential} summarizes the pulsar-match probabilities contained in our catalog and predicts how many new pulsars one may ideally expect to detect when observing a certain number of candidate counterparts, assuming our probabilities are quantitatively reliable (see Sect.~\ref{Calibration}). For instance, $6-11$ pulsar detections may be expected among the top 20 of our match candidates or around $30-40$ detections among the top 200. Among our candidates, an approximately equal number of MSPs and younger pulsars is predicted to be detected. While MSPs as high-latitude sources are usually in less crowded regions and, therefore, allow for more secure associations with their X-ray counterpart, unidentified sources at low latitudes, which are likelier to be young pulsars, are more numerous. 
This is substantiated by Fig.~\ref{MatchLonLatDistribution}, which depicts the spatial distribution and match probabilities of potential pulsar-type counterparts to 4FGL sources over the western Galactic hemisphere. 

To our knowledge, this work constitutes the first probabilistic cross-match between a {\it Fermi}-LAT source catalog and an X-ray catalog, and while only a few sources have pulsar-type match probabilities $>50\%$, the expected yield of possible follow-up campaigns can be accurately quantified.  
Nonetheless, even though around $92\%$ of all associated sources in 4FGL-DR4 are classified pulsars or blazars, the assumption that all unidentified sources belong to either of those source classes is most likely an oversimplification. This assumption neglects the existence of further $\gamma$-ray bright source types, such as supernova remnants, pulsar wind nebulae, or star-forming regions, which are particularly relevant for Galactic unassociated sources. 
Despite this limitation of our analysis, as Sect.~\ref{Calibration} shows, our algorithm can efficiently identify pulsar-type counterparts to 4FGL sources, as verified on the set of associated sources. Hence, we believe our final candidate list offers a reliable starting point for any follow-up campaign.  

\subsection{Remarks on similar multiwavelength studies}

\begin{figure}
\centering
\includegraphics[width=1.0\linewidth]{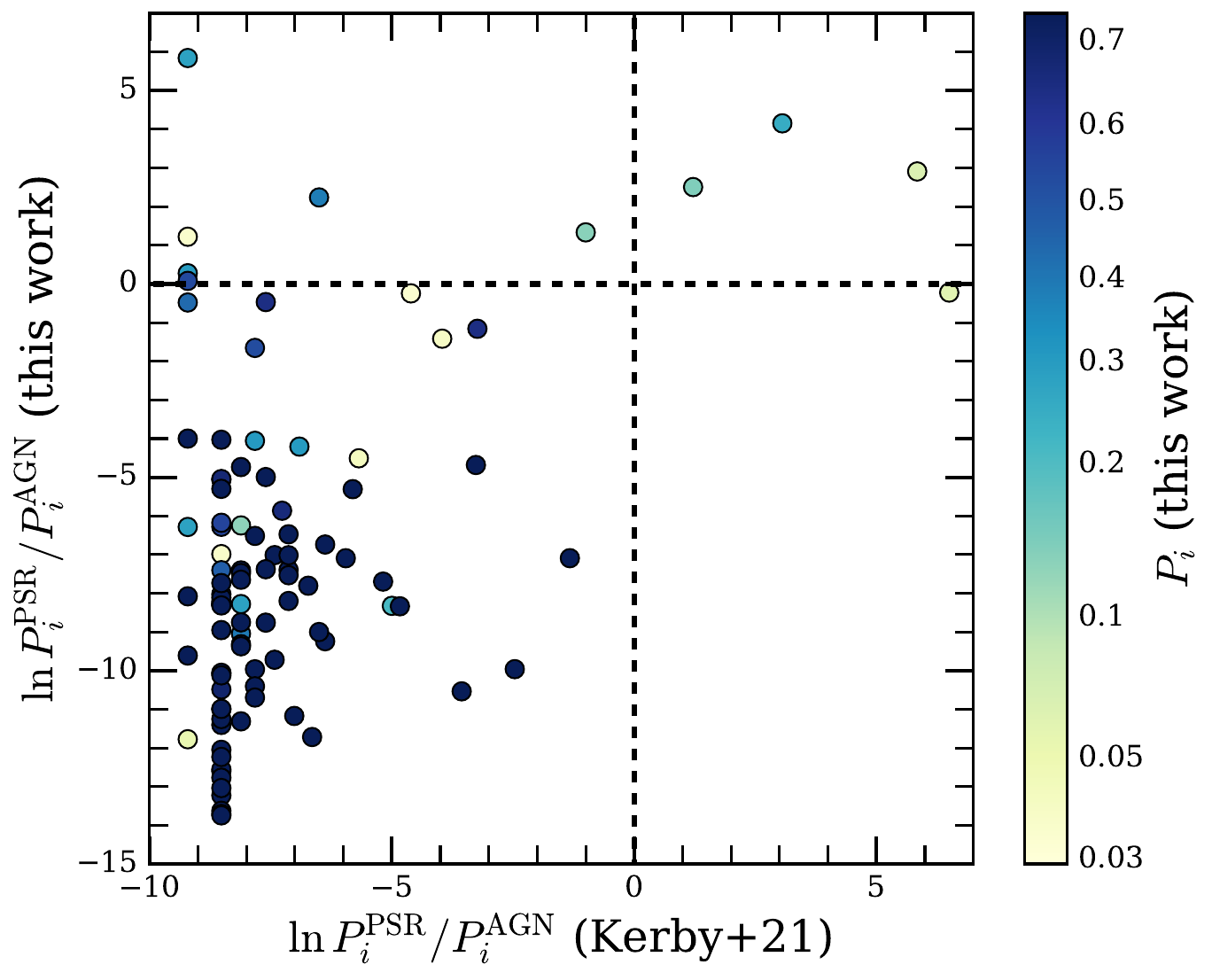} 
\caption{Comparison between our results and those of \citet{Kerby21} for the sample of X-ray sources common to both works. We plot the relative probability for a source to be a pulsar rather than a blazar, as determined here against the results from \citet{Kerby21}. The marker colors indicate the total probability of an X-ray source being the correct counterpart to the matched {\it Fermi} source $P_{i} = P_{i}^{\rm PSR} + P_{i}^{\rm AGN}$.}
\label{KerbyComparison}
\end{figure}

Systematic counterpart searches for unidentified {\it Fermi}-LAT sources have been carried out with multiple instruments and at multiple wavelengths \citep[see][]{3PC}. For instance, \citet{Bruzewski23} identified a sample of high-likelihood pulsar candidates by searching for steep-spectrum radio sources coincident with 4FGL sources. Unfortunately, through a brief positional cross-match, we found that none of their sources were likely to be a counterpart to any of our X-ray sources. 
Furthermore, a brief literature search yielded that, while a few of our X-ray sources have been known to be promising pulsar candidates for several years \citep{SazParkinson16, Dai17}, to our knowledge, none of our candidates have been independently confirmed to be pulsars yet. 
However, several likely candidates from a previous iteration of this project applied to the 4FGL-DR3 catalog \citep{4FGLDR3} have recently been confirmed to be very likely pulsars. This includes the redback system PSR J1803-6707 \citep{Clark23} or the MSP PSR J1259-8148 \citep{3PC} discovered by the TRAPUM collaboration, but also several sources now classified as ``binaries'', namely 4FGL J1702.7$-$5655, 4FGL J1910.7-5320, and 4FGL J1120.0-2204 \citep{Corbet22, Swihart22b, Au22}. No pulsations have been detected yet for the latter sources, but periodic optical variability on hour timescales is observed, as is typical for MSP binaries.  

Similar systematic studies to this work have been carried out using data from the {\it Swift}-XRT follow-up program of unassociated {\it Fermi} sources \citep{StrohFalcone}, for instance, the recent machine learning classification of X-ray counterparts to 4FGL sources by \citet{Kerby21}. 
To compare the results of their method to ours, we positionally cross-matched their output catalog to our set of X-ray sources (before applying the pulsar-specific cuts in Sect.~\ref{Combine}) with a radius of $15\arcsec$. In Fig.~\ref{KerbyComparison}, the output probabilities for matched sources are compared, considering the overall probability of a given match. 
Generally, secure matches of X-rays to $\gamma$-rays tend to agree between the two works, with all of these being classified as likely blazars. Similarly, all sources classified as more likely to be pulsars than blazars ($P_{i}^{\rm PSR} > P_{i}^{\rm AGN}$) by \citet{Kerby21} are recovered quite well here. In two cases, however, sources deemed likely blazars in their work are classified as much more likely to be pulsars by us. In both cases (4FGL J1357.3-6123 and 4FGL J1639.8-4642c), the {\it Fermi} source lies in the Galactic plane, which in principle makes a pulsar scenario preferable. However, on the other hand, the X-ray fluxes of the sources observed by {\it Swift} are several orders of magnitude larger than those observed by eROSITA, and the source positions are $6\arcsec$ and $13\arcsec$ apart, respectively. 
Hence, it seems questionable whether truly identical sources were truly detected by {\it Swift} and eROSITA. If the vastly differing fluxes indeed stemmed from a variable single astrophysical source, this source would be unlikely to be a pulsar, as stable emission is expected on long timescales.   

Importantly, the work of \citet{Kerby21} assumes that if an X-ray source is detected within an observed $\gamma$-ray error ellipse, it is in fact its likely counterpart. While this approach appears suitable for X-ray-bright sources such as blazars, in the case of $\gamma$-ray pulsars, it is important to consider the possibility of not having detected the correct counterpart at all (see Fig.~\ref{KerbyComparison}), as given the expected X-ray fluxes, in most cases an X-ray nondetection is a priori more likely than a detection (see Sect.~\ref{FluxRatio}).
Quantitatively estimating the match probability requires knowledge of the density and properties of the background X-ray source population, which underlines why an all-sky survey is an ideal tool to carry out such a task.

\subsection{Prospects for multiwavelength follow-up}
The results of our cross-matching effort do not constitute immediate identifications of pulsars, as the positional uncertainties of the 4FGL $\gamma$-ray sources are much too large to perform secure associations. However, we believe that our results are ideally suited as input to multiwavelength follow-up campaigns with the target of detecting new pulsars \citep[see][]{3PC}, especially since quantification of the expected success rate is possible with our output probabilities. 

The most natural choice for such a follow-up campaign would be a radio search for pulsed emission, similar to \citet{Clark23}. 
In particular, our X-ray positions, accurate to around $5\arcsec$, would vastly reduce the area, or the number of barycentric corrections, to be searched compared to the arcminute-sized error ellipses of 4FGL sources. Furthermore, a search for pulsar candidates in the imaging domain via deep pointed observations similar to \citet{Bruzewski23} could be a viable alternative. While this would not allow for directly detecting pulsations, the combination of X-ray emission and a steep-spectrum radio source coincident with a $\gamma$-ray source would make a rotation-powered pulsar appear very likely. 

For those X-ray sources that we classify as high-likelihood MSP candidates, optical observations with time-resolved photometry would likely constitute a similarly effective path, which may be somewhat less expensive than radio follow-up. Since, due to their evolutionary history, MSPs are expected to exist in binary systems primarily, optical emission is expected from their companions. Due to their short orbital periods, particularly for the spider subclass \citep{Roberts13}, modulations of the optical emission are expected on time scales of hours. While optical data do not allow for the detection of pulsations from the pulsar magnetosphere itself, the detection of such rapid periodic modulation from an X-ray bright source constitutes a ``smoking gun'' for a binary pulsar and is an ideal tool for the identification of (candidate) MSP binaries \citep[e.g.,][]{Clark23, Corbet22, Swihart22, Swihart22b, Romani12, Romani11}.

Finally, while the majority of MSPs are radio-loud, around half of the young $\gamma$-bright pulsars are radio-quiet \citep{3PC}. For this class of objects, neither optical nor radio searches would appear particularly promising. 
Hence, one possible, yet observationally expensive, avenue would be to carry out X-ray follow-up campaigns with the target of directly detecting pulsations. Given the X-ray faint nature of our typical pulsar candidates in this work, {\it XMM-Newton} \citep{Struder01, Turner01} appears likely to be the only current X-ray mission feasible for this purpose since a high number of source counts is as crucial for detecting pulsations as the ability to separate source from background emission, which makes the {\it Chandra} \citep{Weisskopf02} or {\it NICER} \citep{Gendreau16} telescopes appear less attractive. However, even with {\it XMM-Newton}, this effort would require quite long exposures. For instance, a typical eRASS:4 source flux of $F_{X}=2\times10^{-14}\,\si{erg.s^{-1}.cm^{-2}}$, corresponding to an {\it XMM} $0.2-12\,\si{keV}$ count rate of $1.8\times10^{-2}\,\si{ct.s^{-1}}$ in a typical aperture, together with a pulsed fraction of $20\%$, would require an exposure of around $35\,\si{ks}$ for a $5\sigma$ detection of a pulsed signal.  
The only viable alternative to X-ray follow-up would be to directly search the available $\gamma$-ray data from {\it Fermi}-LAT for pulsations at the candidate positions. Blind searches for $\gamma$-ray pulsations from unidentified {\it Fermi}-LAT sources were carried out successfully in the past via the distributed computing system {\tt Einstein@Home} \citep{Clark17}. These searches tend to suffer from having to search multiple possible positions due to the localization inaccuracy of {\it Fermi}-LAT, a problem which would be alleviated by a targeted search of candidate positions with arcsecond-level accuracy \citep[such as][]{Smith19}.

\section{Summary and conclusion \label{Summary}}

In this work, we have conducted a probabilistic cross-matching analysis 
between the 14-year catalog of {\it Fermi}-LAT $\gamma$-ray sources (4FGL-DR4), and the {\it SRG}/eROSITA X-ray catalog based on four all-sky surveys. The primary target of this effort was the identification of new high-energy rotation-powered pulsars by searching for likely X-ray counterparts to unassociated $\gamma$-ray sources. This was motivated by the fact that pulsars are the second most common class of identified 4FGL sources behind blazars. A dedicated multiwavelength study of the eROSITA-detected population of (candidate) blazars will be published elsewhere (S. H\"ammerich et al., in prep.).
Given the large positional uncertainty inherent to {\it Fermi}-LAT sources, the identification of an X-ray counterpart facilitates their identification, exploiting the characteristic thermal surface or magnetospheric X-ray emission from pulsars, even if little more than arcsecond-level positional information is provided.

For the identification of new pulsar candidates, we employed a Bayesian cross-matching scheme designed for astronomical catalogs, which combines positional overlap with prior information on the likelihood of a given match to produce a probabilistic statement on the quality of a given candidate match of pulsar- or blazar-type. 
More specifically, we constructed a random forest classifier, which, based on the $\gamma$-ray properties of the investigated {\it Fermi}-LAT sources, predicts the probability for it to be of blazar or (young or recycled) pulsar nature. While this classifier achieves similar accuracies as in previously published works, it predicts a much larger fraction of pulsars among unassociated sources. We attribute this to our emphasis on a balanced treatment of the source classes and the exclusion of explicit source fluxes from the approach. 

These ingredients were combined with a data-driven constraint on the expected X-ray flux for a match of either type based on the observed $\gamma$- to X-ray flux ratios for sources with known multiwavelength counterparts. This effort clearly confirmed previously established results, namely the higher relative X-ray brightness of blazars, and the large spread in X-ray fluxes of young pulsars.  
Finally, the X-ray images of our resulting candidates were visually screened in order to identify potential spurious detections. In addition, we investigated optical and infrared catalogs with the target of excluding sources classifiable as non-pulsar based on their multiwavelength counterparts.  

We verified our method of probabilistic cross-matching by applying it ``blindly'' to the set of previously associated sources in 4FGL-DR4, finding decent agreement between the predicted match probabilities and the fraction of counterparts that were in fact correctly identified. 
Our final list of candidate matches contains around 900 X-ray sources with probabilities of being the X-ray counterpart to a $\gamma$-ray pulsar hidden in an unassociated 4FGL source $P_{i}^{\rm PSR} \geq 0.02$. Around 50 sources exhibit probabilities $P_{i}^{\rm PSR} \geq 0.2$, most of which lie close to the Galactic plane. Based on our posterior match probabilities, we predict that a ``perfect pulsar detector'' should find 30 -- 40 new pulsars among our 200 best candidates, with about equal shares of young energetic and old millisecond pulsars.
Support for the reliability of our method comes from several sources now known to be pulsars, which were likely candidates in previous iterations of this project, and from its qualitative agreement with X-ray follow-up campaigns by {\it Swift}-XRT.

In addition to our counterpart search for {\it Fermi}-LAT sources, using our homogeneous survey coverage of the X-ray sky, we have performed a systematic search for X-ray counterparts of known rotation-powered pulsars, including all known radio and $\gamma$-ray pulsars. Thereby, we have identified 15 previously unknown X-ray counterparts of pulsars, including the soft emission from the recently discovered young and nearby PSR J0837$-$3754.  

We hope that our study will motivate new, or support existing, multiwavelength identification programs of pulsars in unassociated {\it Fermi}-LAT sources. Targeted searches for pulsations in the radio or $\gamma$-ray domain could profit from our candidate positions with arcsecond-level accuracy, as it would imply a large reduction in the area to search, relative to the $\gamma$-ray error ellipses. 
Similarly, optical searches for periodically variable counterparts, induced by the orbital modulation of the brightness of a potential binary companion to the pulsar, would be facilitated by only having to search the X-ray uncertainty region rather than the full $\gamma$-ray error ellipse, which may contain thousands of sources.

\begin{acknowledgements}
We are grateful to the anonymous referee for their helpful and fair criticism, which helped improve our paper. 
We would like to thank Dus\'an Tub\'in-Arenas for providing us with X-ray upper limits for all associated 4FGL sources. Further, we would like to thank Julien Wolf and Mara Salvato for useful discussions on multiwavelength counterparts in early phases of this project. WB acknowledges fruitful discussions with Luciano Nicastro. 
MGFM acknowledges support by the International Max-Planck Research School on Astrophysics at the Ludwig-Maximilians University (IMPRS).  \\
This work is based on data from eROSITA, the soft X-ray instrument aboard SRG, a joint Russian-German science mission supported by the Russian Space Agency (Roskosmos), in the interests of the Russian Academy of Sciences represented by its Space Research Institute (IKI), and the Deutsches Zentrum f\"ur Luft- und Raumfahrt (DLR). The SRG spacecraft was built by Lavochkin Association (NPOL) and its subcontractors and is operated by NPOL with support from the Max Planck Institute for Extraterrestrial Physics (MPE). The development and construction of the eROSITA X-ray instrument was led by MPE, with contributions from the Dr. Karl Remeis Observatory Bamberg \& ECAP (FAU Erlangen-Nuernberg), the University of Hamburg Observatory, the Leibniz Institute for Astrophysics Potsdam (AIP), and the Institute for Astronomy and Astrophysics of the University of T\"ubingen, with the support of DLR and the Max Planck Society. The Argelander Institute for Astronomy of the University of Bonn and the Ludwig Maximilians Universit\"at Munich also participated in the science preparation for eROSITA.
The eROSITA data shown here were processed using the eSASS software system developed by the German eROSITA consortium.
This work has made use of data from the European Space Agency (ESA) mission
{\it Gaia} (\url{https://www.cosmos.esa.int/gaia}), processed by the {\it Gaia} Data Processing and Analysis Consortium (DPAC, \url{https://www.cosmos.esa.int/web/gaia/dpac/consortium}). Funding for the DPAC has been provided by national institutions, in particular, the institutions participating in the {\it Gaia} Multilateral Agreement. \\
This research made use of Astropy,\footnote{\url{http://www.astropy.org}} a community-developed core Python package for Astronomy \citep{astropy:2013, astropy:2018}. Further, we acknowledge the use of the Python packages Matplotlib \citep{Hunter:2007}, SciPy \citep{SciPy}, and NumPy \citep{NumPy}. 
\end{acknowledgements}

\begingroup
\bibliographystyle{aa} 
\bibliography{Citations}

\begin{thebibliography}{92}
\expandafter\ifx\csname natexlab\endcsname\relax\def\natexlab#1{#1}\fi

\bibitem[{{Abdo} {et~al.}(2013){Abdo}, {Ajello}, {Allafort}, {Baldini},
  {Ballet}, {Barbiellini}, {Baring}, {Bastieri}, {Belfiore}, {Bellazzini},
  {Bhattacharyya}, {Bissaldi}, {Bloom}, {Bonamente}, {Bottacini}, {Brandt},
  {Bregeon}, {Brigida}, {Bruel}, {Buehler}, {Burgay}, {Burnett}, {Busetto},
  {Buson}, {Caliandro}, {Cameron}, {Camilo}, {Caraveo}, {Casandjian}, {Cecchi},
  {{\c{C}}elik}, {Charles}, {Chaty}, {Chaves}, {Chekhtman}, {Chen}, {Chiang},
  {Chiaro}, {Ciprini}, {Claus}, {Cognard}, {Cohen-Tanugi}, {Cominsky},
  {Conrad}, {Cutini}, {D'Ammando}, {de Angelis}, {DeCesar}, {De Luca}, {den
  Hartog}, {de Palma}, {Dermer}, {Desvignes}, {Digel}, {Di Venere}, {Drell},
  {Drlica-Wagner}, {Dubois}, {Dumora}, {Espinoza}, {Falletti}, {Favuzzi},
  {Ferrara}, {Focke}, {Franckowiak}, {Freire}, {Funk}, {Fusco}, {Gargano},
  {Gasparrini}, {Germani}, {Giglietto}, {Giommi}, {Giordano}, {Giroletti},
  {Glanzman}, {Godfrey}, {Gotthelf}, {Grenier}, {Grondin}, {Grove},
  {Guillemot}, {Guiriec}, {Hadasch}, {Hanabata}, {Harding}, {Hayashida},
  {Hays}, {Hessels}, {Hewitt}, {Hill}, {Horan}, {Hou}, {Hughes}, {Jackson},
  {Janssen}, {Jogler}, {J{\'o}hannesson}, {Johnson}, {Johnson}, {Johnson},
  {Johnson}, {Johnston}, {Kamae}, {Kataoka}, {Keith}, {Kerr}, {Kn{\"o}dlseder},
  {Kramer}, {Kuss}, {Lande}, {Larsson}, {Latronico}, {Lemoine-Goumard},
  {Longo}, {Loparco}, {Lovellette}, {Lubrano}, {Lyne}, {Manchester}, {Marelli},
  {Massaro}, {Mayer}, {Mazziotta}, {McEnery}, {McLaughlin}, {Mehault},
  {Michelson}, {Mignani}, {Mitthumsiri}, {Mizuno}, {Moiseev}, {Monzani},
  {Morselli}, {Moskalenko}, {Murgia}, {Nakamori}, {Nemmen}, {Nuss}, {Ohno},
  {Ohsugi}, {Orienti}, {Orlando}, {Ormes}, {Paneque}, {Panetta}, {Parent},
  {Perkins}, {Pesce-Rollins}, {Pierbattista}, {Piron}, {Pivato}, {Pletsch},
  {Porter}, {Possenti}, {Rain{\`o}}, {Rando}, {Ransom}, {Ray}, {Razzano},
  {Rea}, {Reimer}, {Reimer}, {Renault}, {Reposeur}, {Ritz}, {Romani}, {Roth},
  {Rousseau}, {Roy}, {Ruan}, {Sartori}, {Saz Parkinson}, {Scargle}, {Schulz},
  {Sgr{\`o}}, {Shannon}, {Siskind}, {Smith}, {Spandre}, {Spinelli}, {Stappers},
  {Strong}, {Suson}, {Takahashi}, {Thayer}, {Thayer}, {Theureau}, {Thompson},
  {Thorsett}, {Tibaldo}, {Tibolla}, {Tinivella}, {Torres}, {Tosti}, {Troja},
  {Uchiyama}, {Usher}, {Vandenbroucke}, {Vasileiou}, {Venter}, {Vianello},
  {Vitale}, {Wang}, {Weltevrede}, {Winer}, {Wolff}, {Wood}, {Wood}, {Wood}, \&
  {Yang}}]{2PC}
{Abdo}, A.~A., {Ajello}, M., {Allafort}, A., {et~al.} 2013, \apjs, 208, 17

\bibitem[{{Abdollahi} {et~al.}(2020){Abdollahi}, {Acero}, {Ackermann},
  {Ajello}, {Atwood}, {Axelsson}, {Baldini}, {Ballet}, {Barbiellini},
  {Bastieri}, {Becerra Gonzalez}, {Bellazzini}, {Berretta}, {Bissaldi},
  {Blandford}, {Bloom}, {Bonino}, {Bottacini}, {Brandt}, {Bregeon}, {Bruel},
  {Buehler}, {Burnett}, {Buson}, {Cameron}, {Caputo}, {Caraveo}, {Casandjian},
  {Castro}, {Cavazzuti}, {Charles}, {Chaty}, {Chen}, {Cheung}, {Chiaro},
  {Ciprini}, {Cohen-Tanugi}, {Cominsky}, {Coronado-Bl{\'a}zquez}, {Costantin},
  {Cuoco}, {Cutini}, {D'Ammando}, {DeKlotz}, {de la Torre Luque}, {de Palma},
  {Desai}, {Digel}, {Di Lalla}, {Di Mauro}, {Di Venere}, {Dom{\'\i}nguez},
  {Dumora}, {Fana Dirirsa}, {Fegan}, {Ferrara}, {Franckowiak}, {Fukazawa},
  {Funk}, {Fusco}, {Gargano}, {Gasparrini}, {Giglietto}, {Giommi}, {Giordano},
  {Giroletti}, {Glanzman}, {Green}, {Grenier}, {Griffin}, {Grondin}, {Grove},
  {Guiriec}, {Harding}, {Hayashi}, {Hays}, {Hewitt}, {Horan},
  {J{\'o}hannesson}, {Johnson}, {Kamae}, {Kerr}, {Kocevski}, {Kovac'evic'},
  {Kuss}, {Landriu}, {Larsson}, {Latronico}, {Lemoine-Goumard}, {Li},
  {Liodakis}, {Longo}, {Loparco}, {Lott}, {Lovellette}, {Lubrano}, {Madejski},
  {Maldera}, {Malyshev}, {Manfreda}, {Marchesini}, {Marcotulli},
  {Mart{\'\i}-Devesa}, {Martin}, {Massaro}, {Mazziotta}, {McEnery}, {Mereu},
  {Meyer}, {Michelson}, {Mirabal}, {Mizuno}, {Monzani}, {Morselli},
  {Moskalenko}, {Negro}, {Nuss}, {Ojha}, {Omodei}, {Orienti}, {Orlando},
  {Ormes}, {Palatiello}, {Paliya}, {Paneque}, {Pei}, {Pe{\~n}a-Herazo},
  {Perkins}, {Persic}, {Pesce-Rollins}, {Petrosian}, {Petrov}, {Piron}, {Poon},
  {Porter}, {Principe}, {Rain{\`o}}, {Rando}, {Razzano}, {Razzaque}, {Reimer},
  {Reimer}, {Remy}, {Reposeur}, {Romani}, {Saz Parkinson}, {Schinzel},
  {Serini}, {Sgr{\`o}}, {Siskind}, {Smith}, {Spandre}, {Spinelli}, {Strong},
  {Suson}, {Tajima}, {Takahashi}, {Tak}, {Thayer}, {Thompson}, {Tibaldo},
  {Torres}, {Torresi}, {Valverde}, {Van Klaveren}, {van Zyl}, {Wood},
  {Yassine}, \& {Zaharijas}}]{4FGL}
{Abdollahi}, S., {Acero}, F., {Ackermann}, M., {et~al.} 2020, \apjs, 247, 33

\bibitem[{{Abdollahi} {et~al.}(2022){Abdollahi}, {Acero}, {Baldini}, {Ballet},
  {Bastieri}, {Bellazzini}, {Berenji}, {Berretta}, {Bissaldi}, {Blandford},
  {Bloom}, {Bonino}, {Brill}, {Britto}, {Bruel}, {Burnett}, {Buson}, {Cameron},
  {Caputo}, {Caraveo}, {Castro}, {Chaty}, {Cheung}, {Chiaro}, {Cibrario},
  {Ciprini}, {Coronado-Bl{\'a}zquez}, {Crnogorcevic}, {Cutini}, {D'Ammando},
  {De Gaetano}, {Digel}, {Di Lalla}, {Dirirsa}, {Di Venere}, {Dom{\'\i}nguez},
  {Fallah Ramazani}, {Fegan}, {Ferrara}, {Fiori}, {Fleischhack}, {Franckowiak},
  {Fukazawa}, {Funk}, {Fusco}, {Galanti}, {Gammaldi}, {Gargano}, {Garrappa},
  {Gasparrini}, {Giacchino}, {Giglietto}, {Giordano}, {Giroletti}, {Glanzman},
  {Green}, {Grenier}, {Grondin}, {Guillemot}, {Guiriec}, {Gustafsson},
  {Harding}, {Hays}, {Hewitt}, {Horan}, {Hou}, {J{\'o}hannesson}, {Karwin},
  {Kayanoki}, {Kerr}, {Kuss}, {Landriu}, {Larsson}, {Latronico},
  {Lemoine-Goumard}, {Li}, {Liodakis}, {Longo}, {Loparco}, {Lott}, {Lubrano},
  {Maldera}, {Malyshev}, {Manfreda}, {Mart{\'\i}-Devesa}, {Mazziotta}, {Mereu},
  {Meyer}, {Michelson}, {Mirabal}, {Mitthumsiri}, {Mizuno}, {Moiseev},
  {Monzani}, {Morselli}, {Moskalenko}, {Negro}, {Nuss}, {Omodei}, {Orienti},
  {Orlando}, {Paneque}, {Pei}, {Perkins}, {Persic}, {Pesce-Rollins},
  {Petrosian}, {Pillera}, {Poon}, {Porter}, {Principe}, {Rain{\`o}}, {Rando},
  {Rani}, {Razzano}, {Razzaque}, {Reimer}, {Reimer}, {Reposeur},
  {S{\'a}nchez-Conde}, {Saz Parkinson}, {Scotton}, {Serini}, {Sgr{\`o}},
  {Siskind}, {Smith}, {Spandre}, {Spinelli}, {Sueoka}, {Suson}, {Tajima},
  {Tak}, {Thayer}, {Thompson}, {Torres}, {Troja}, {Valverde}, {Wood}, \&
  {Zaharijas}}]{4FGLDR3}
{Abdollahi}, S., {Acero}, F., {Baldini}, L., {et~al.} 2022, \apjs, 260, 53

\bibitem[{{Akaike}(1974)}]{Akaike74}
{Akaike}, H. 1974, IEEE Transactions on Automatic Control, 19, 716

\bibitem[{{Alpar} {et~al.}(1982){Alpar}, {Cheng}, {Ruderman}, \&
  {Shaham}}]{Alpar82}
{Alpar}, M.~A., {Cheng}, A.~F., {Ruderman}, M.~A., \& {Shaham}, J. 1982, \nat,
  300, 728

\bibitem[{{Astropy Collaboration} {et~al.}(2018){Astropy Collaboration},
  {Price-Whelan}, {Sip{H{o}}cz}, {G{"u}nther}, {Lim}, {Crawford}, {Conseil},
  {Shupe}, {Craig}, {Dencheva}, {Ginsburg}, {Vand erPlas}, {Bradley},
  {P{'e}rez-Su{'a}rez}, {de Val-Borro}, {Aldcroft}, {Cruz}, {Robitaille},
  {Tollerud}, {Ardelean}, {Babej}, {Bach}, {Bachetti}, {Bakanov}, {Bamford},
  {Barentsen}, {Barmby}, {Baumbach}, {Berry}, {Biscani}, {Boquien}, {Bostroem},
  {Bouma}, {Brammer}, {Bray}, {Breytenbach}, {Buddelmeijer}, {Burke},
  {Calderone}, {Cano Rodr{'i}guez}, {Cara}, {Cardoso}, {Cheedella}, {Copin},
  {Corrales}, {Crichton}, {D'Avella}, {Deil}, {Depagne}, {Dietrich}, {Donath},
  {Droettboom}, {Earl}, {Erben}, {Fabbro}, {Ferreira}, {Finethy}, {Fox},
  {Garrison}, {Gibbons}, {Goldstein}, {Gommers}, {Greco}, {Greenfield},
  {Groener}, {Grollier}, {Hagen}, {Hirst}, {Homeier}, {Horton}, {Hosseinzadeh},
  {Hu}, {Hunkeler}, {Ivezi{'c}}, {Jain}, {Jenness}, {Kanarek}, {Kendrew},
  {Kern}, {Kerzendorf}, {Khvalko}, {King}, {Kirkby}, {Kulkarni}, {Kumar},
  {Lee}, {Lenz}, {Littlefair}, {Ma}, {Macleod}, {Mastropietro}, {McCully},
  {Montagnac}, {Morris}, {Mueller}, {Mumford}, {Muna}, {Murphy}, {Nelson},
  {Nguyen}, {Ninan}, {N{"o}the}, {Ogaz}, {Oh}, {Parejko}, {Parley}, {Pascual},
  {Patil}, {Patil}, {Plunkett}, {Prochaska}, {Rastogi}, {Reddy Janga},
  {Sabater}, {Sakurikar}, {Seifert}, {Sherbert}, {Sherwood-Taylor}, {Shih},
  {Sick}, {Silbiger}, {Singanamalla}, {Singer}, {Sladen}, {Sooley},
  {Sornarajah}, {Streicher}, {Teuben}, {Thomas}, {Tremblay}, {Turner},
  {Terr{'o}n}, {van Kerkwijk}, {de la Vega}, {Watkins}, {Weaver}, {Whitmore},
  {Woillez}, {Zabalza}, \& {Astropy Contributors}}]{astropy:2018}
{Astropy Collaboration}, {Price-Whelan}, A.~M., {Sip{H{o}}cz}, B.~M., {et~al.}
  2018, aj, 156, 123

\bibitem[{{Astropy Collaboration} {et~al.}(2013){Astropy Collaboration},
  {Robitaille}, {Tollerud}, {Greenfield}, {Droettboom}, {Bray}, {Aldcroft},
  {Davis}, {Ginsburg}, {Price-Whelan}, {Kerzendorf}, {Conley}, {Crighton},
  {Barbary}, {Muna}, {Ferguson}, {Grollier}, {Parikh}, {Nair}, {Unther},
  {Deil}, {Woillez}, {Conseil}, {Kramer}, {Turner}, {Singer}, {Fox}, {Weaver},
  {Zabalza}, {Edwards}, {Azalee Bostroem}, {Burke}, {Casey}, {Crawford},
  {Dencheva}, {Ely}, {Jenness}, {Labrie}, {Lim}, {Pierfederici}, {Pontzen},
  {Ptak}, {Refsdal}, {Servillat}, \& {Streicher}}]{astropy:2013}
{Astropy Collaboration}, {Robitaille}, T.~P., {Tollerud}, E.~J., {et~al.} 2013,
  \aap, 558, A33

\bibitem[{{Atwood} {et~al.}(2009){Atwood}, {Abdo}, {Ackermann}, {Althouse},
  {Anderson}, {Axelsson}, {Baldini}, {Ballet}, {Band}, {Barbiellini},
  {Bartelt}, {Bastieri}, {Baughman}, {Bechtol}, {B{\'e}d{\'e}r{\`e}de},
  {Bellardi}, {Bellazzini}, {Berenji}, {Bignami}, {Bisello}, {Bissaldi},
  {Blandford}, {Bloom}, {Bogart}, {Bonamente}, {Bonnell}, {Borgland},
  {Bouvier}, {Bregeon}, {Brez}, {Brigida}, {Bruel}, {Burnett}, {Busetto},
  {Caliandro}, {Cameron}, {Caraveo}, {Carius}, {Carlson}, {Casandjian},
  {Cavazzuti}, {Ceccanti}, {Cecchi}, {Charles}, {Chekhtman}, {Cheung},
  {Chiang}, {Chipaux}, {Cillis}, {Ciprini}, {Claus}, {Cohen-Tanugi},
  {Condamoor}, {Conrad}, {Corbet}, {Corucci}, {Costamante}, {Cutini}, {Davis},
  {Decotigny}, {DeKlotz}, {Dermer}, {de Angelis}, {Digel}, {do Couto e Silva},
  {Drell}, {Dubois}, {Dumora}, {Edmonds}, {Fabiani}, {Farnier}, {Favuzzi},
  {Flath}, {Fleury}, {Focke}, {Funk}, {Fusco}, {Gargano}, {Gasparrini},
  {Gehrels}, {Gentit}, {Germani}, {Giebels}, {Giglietto}, {Giommi}, {Giordano},
  {Glanzman}, {Godfrey}, {Grenier}, {Grondin}, {Grove}, {Guillemot}, {Guiriec},
  {Haller}, {Harding}, {Hart}, {Hays}, {Healey}, {Hirayama}, {Hjalmarsdotter},
  {Horn}, {Hughes}, {J{\'o}hannesson}, {Johansson}, {Johnson}, {Johnson},
  {Johnson}, {Johnson}, {Kamae}, {Katagiri}, {Kataoka}, {Kavelaars}, {Kawai},
  {Kelly}, {Kerr}, {Klamra}, {Kn{\"o}dlseder}, {Kocian}, {Komin}, {Kuehn},
  {Kuss}, {Landriu}, {Latronico}, {Lee}, {Lee}, {Lemoine-Goumard}, {Lionetto},
  {Longo}, {Loparco}, {Lott}, {Lovellette}, {Lubrano}, {Madejski}, {Makeev},
  {Marangelli}, {Massai}, {Mazziotta}, {McEnery}, {Menon}, {Meurer},
  {Michelson}, {Minuti}, {Mirizzi}, {Mitthumsiri}, {Mizuno}, {Moiseev},
  {Monte}, {Monzani}, {Moretti}, {Morselli}, {Moskalenko}, {Murgia},
  {Nakamori}, {Nishino}, {Nolan}, {Norris}, {Nuss}, {Ohno}, {Ohsugi}, {Omodei},
  {Orlando}, {Ormes}, {Paccagnella}, {Paneque}, {Panetta}, {Parent}, {Pearce},
  {Pepe}, {Perazzo}, {Pesce-Rollins}, {Picozza}, {Pieri}, {Pinchera}, {Piron},
  {Porter}, {Poupard}, {Rain{\`o}}, {Rando}, {Rapposelli}, {Razzano}, {Reimer},
  {Reimer}, {Reposeur}, {Reyes}, {Ritz}, {Rochester}, {Rodriguez}, {Romani},
  {Roth}, {Russell}, {Ryde}, {Sabatini}, {Sadrozinski}, {Sanchez}, {Sander},
  {Sapozhnikov}, {Parkinson}, {Scargle}, {Schalk}, {Scolieri}, {Sgr{\`o}},
  {Share}, {Shaw}, {Shimokawabe}, {Shrader}, {Sierpowska-Bartosik}, {Siskind},
  {Smith}, {Smith}, {Spandre}, {Spinelli}, {Starck}, {Stephens}, {Strickman},
  {Strong}, {Suson}, {Tajima}, {Takahashi}, {Takahashi}, {Tanaka}, {Tenze},
  {Tether}, {Thayer}, {Thayer}, {Thompson}, {Tibaldo}, {Tibolla}, {Torres},
  {Tosti}, {Tramacere}, {Turri}, {Usher}, {Vilchez}, {Vitale}, {Wang},
  {Watters}, {Winer}, {Wood}, {Ylinen}, \& {Ziegler}}]{Atwood09}
{Atwood}, W.~B., {Abdo}, A.~A., {Ackermann}, M., {et~al.} 2009, \apj, 697, 1071

\bibitem[{{Au} {et~al.}(2023){Au}, {Strader}, {Swihart}, {Lin}, {Kong},
  {Takata}, {Hui}, {Panurach}, {Molina}, {Aydi}, {Sokolovsky}, \& {Li}}]{Au22}
{Au}, K.-Y., {Strader}, J., {Swihart}, S.~J., {et~al.} 2023, \apj, 943, 103

\bibitem[{{Ballet} {et~al.}(2023){Ballet}, {Bruel}, {Burnett}, {Lott}, \& {The
  Fermi-LAT collaboration}}]{4FGLDR4}
{Ballet}, J., {Bruel}, P., {Burnett}, T.~H., {Lott}, B., \& {The Fermi-LAT
  collaboration}. 2023, arXiv e-prints, arXiv:2307.12546

\bibitem[{{Ballet} {et~al.}(2020){Ballet}, {Burnett}, {Digel}, \&
  {Lott}}]{4FGL_DR2}
{Ballet}, J., {Burnett}, T.~H., {Digel}, S.~W., \& {Lott}, B. 2020, arXiv
  e-prints, arXiv:2005.11208

\bibitem[{{Becker}(2009)}]{Becker09}
{Becker}, W. 2009, in Astrophysics and Space Science Library, Vol. 357,
  Astrophysics and Space Science Library, ed. W.~{Becker}, 91

\bibitem[{{Becker} \& {Tr\"umper}(1997)}]{Becker97}
{Becker}, W. \& {Tr\"umper}, J. 1997, \aap, 326, 682

\bibitem[{{Bhat} \& {Malyshev}(2022)}]{Bhat22}
{Bhat}, A. \& {Malyshev}, D. 2022, \aap, 660, A87

\bibitem[{{Braglia} {et~al.}(2020){Braglia}, {Mignani}, {Belfiore}, {Marelli},
  {Israel}, {Novara}, {De Luca}, {Tiengo}, \& {Saz Parkinson}}]{Braglia20}
{Braglia}, C., {Mignani}, R.~P., {Belfiore}, A., {et~al.} 2020, \mnras, 497,
  5364

\bibitem[{{Brunner} {et~al.}(2022){Brunner}, {Liu}, {Lamer}, {Georgakakis},
  {Merloni}, {Brusa}, {Bulbul}, {Dennerl}, {Friedrich}, {Liu}, {Maitra},
  {Nandra}, {Ramos-Ceja}, {Sanders}, {Stewart}, {Boller}, {Buchner}, {Clerc},
  {Comparat}, {Dwelly}, {Eckert}, {Finoguenov}, {Freyberg}, {Ghirardini},
  {Gueguen}, {Haberl}, {Kreykenbohm}, {Krumpe}, {Osterhage}, {Pacaud},
  {Predehl}, {Reiprich}, {Robrade}, {Salvato}, {Santangelo}, {Schrabback},
  {Schwope}, \& {Wilms}}]{Brunner22}
{Brunner}, H., {Liu}, T., {Lamer}, G., {et~al.} 2022, \aap, 661, A1

\bibitem[{{Bruzewski} {et~al.}(2023){Bruzewski}, {Schinzel}, \&
  {Taylor}}]{Bruzewski23}
{Bruzewski}, S., {Schinzel}, F.~K., \& {Taylor}, G.~B. 2023, \apj, 943, 51

\bibitem[{{Budav{\'a}ri} \& {Szalay}(2008)}]{Budavari08}
{Budav{\'a}ri}, T. \& {Szalay}, A.~S. 2008, \apj, 679, 301

\bibitem[{Chawla {et~al.}(2002)Chawla, Bowyer, Hall, \& Kegelmeyer}]{SMOTE}
Chawla, N.~V., Bowyer, K.~W., Hall, L.~O., \& Kegelmeyer, W.~P. 2002, J. Artif.
  Int. Res., 16, 321–357

\bibitem[{Chen \& Breiman(2004)}]{Chen04}
Chen, C. \& Breiman, L. 2004, University of California, Berkeley

\bibitem[{{Clark} {et~al.}(2023){Clark}, {Breton}, {Barr}, {Burgay},
  {Thongmeearkom}, {Nieder}, {Buchner}, {Stappers}, {Kramer}, {Becker},
  {Mayer}, {Phosrisom}, {Ashok}, {Bezuidenhout}, {Calore}, {Cognard}, {Freire},
  {Geyer}, {Grie{\ss}meier}, {Karuppusamy}, {Levin}, {Padmanabh}, {Possenti},
  {Ransom}, {Serylak}, {Venkatraman Krishnan}, {Vleeschower}, {Behrend},
  {Champion}, {Chen}, {Horn}, {Keane}, {K{\"u}nkel}, {Men}, {Ridolfi},
  {Dhillon}, {Marsh}, \& {Papa}}]{Clark23}
{Clark}, C.~J., {Breton}, R.~P., {Barr}, E.~D., {et~al.} 2023, \mnras, 519,
  5590

\bibitem[{{Clark} {et~al.}(2017){Clark}, {Wu}, {Pletsch}, {Guillemot}, {Allen},
  {Aulbert}, {Beer}, {Bock}, {Cu{\'e}llar}, {Eggenstein}, {Fehrmann}, {Kramer},
  {Machenschalk}, \& {Nieder}}]{Clark17}
{Clark}, C.~J., {Wu}, J., {Pletsch}, H.~J., {et~al.} 2017, \apj, 834, 106

\bibitem[{{Condon} {et~al.}(1998){Condon}, {Cotton}, {Greisen}, {Yin},
  {Perley}, {Taylor}, \& {Broderick}}]{NVSS}
{Condon}, J.~J., {Cotton}, W.~D., {Greisen}, E.~W., {et~al.} 1998, \aj, 115,
  1693

\bibitem[{{Corbet} {et~al.}(2022){Corbet}, {Chomiuk}, {Coley}, {Dubus},
  {Edwards}, {Islam}, {McBride}, {Stevens}, {Strader}, {Swihart}, \&
  {Townsend}}]{Corbet22}
{Corbet}, R.~H.~D., {Chomiuk}, L., {Coley}, J.~B., {et~al.} 2022, \apj, 935, 2

\bibitem[{{Coronado-Bl{\'a}zquez}(2022)}]{Coronado-Blazquez22}
{Coronado-Bl{\'a}zquez}, J. 2022, \mnras, 515, 1807

\bibitem[{{Dai} {et~al.}(2017){Dai}, {Wang}, {Vadakkumthani}, \&
  {Xing}}]{Dai17}
{Dai}, X.-J., {Wang}, Z.-X., {Vadakkumthani}, J., \& {Xing}, Y. 2017, Research
  in Astronomy and Astrophysics, 17, 072

\bibitem[{{De Breuck} {et~al.}(2002){De Breuck}, {Tang}, {de Bruyn},
  {R{\"o}ttgering}, \& {van Breugel}}]{WISH}
{De Breuck}, C., {Tang}, Y., {de Bruyn}, A.~G., {R{\"o}ttgering}, H., \& {van
  Breugel}, W. 2002, \aap, 394, 59

\bibitem[{{Fruchter} {et~al.}(1988){Fruchter}, {Stinebring}, \&
  {Taylor}}]{Fruchter88}
{Fruchter}, A.~S., {Stinebring}, D.~R., \& {Taylor}, J.~H. 1988, \nat, 333, 237

\bibitem[{{Gaia Collaboration} {et~al.}(2016){Gaia Collaboration}, {Prusti},
  {de Bruijne}, {Brown}, {Vallenari}, {Babusiaux}, {Bailer-Jones}, {Bastian},
  {Biermann}, {Evans}, {Eyer}, {Jansen}, {Jordi}, {Klioner}, {Lammers},
  {Lindegren}, {Luri}, {Mignard}, {Milligan}, {Panem}, {Poinsignon},
  {Pourbaix}, {Randich}, {Sarri}, {Sartoretti}, {Siddiqui}, {Soubiran},
  {Valette}, {van Leeuwen}, {Walton}, {Aerts}, {Arenou}, {Cropper}, {Drimmel},
  {H{\o}g}, {Katz}, {Lattanzi}, {O'Mullane}, {Grebel}, {Holland}, {Huc},
  {Passot}, {Bramante}, {Cacciari}, {Casta{\~n}eda}, {Chaoul}, {Cheek}, {De
  Angeli}, {Fabricius}, {Guerra}, {Hern{\'a}ndez}, {Jean-Antoine-Piccolo},
  {Masana}, {Messineo}, {Mowlavi}, {Nienartowicz}, {Ord{\'o}{\~n}ez-Blanco},
  {Panuzzo}, {Portell}, {Richards}, {Riello}, {Seabroke}, {Tanga},
  {Th{\'e}venin}, {Torra}, {Els}, {Gracia-Abril}, {Comoretto},
  {Garcia-Reinaldos}, {Lock}, {Mercier}, {Altmann}, {Andrae}, {Astraatmadja},
  {Bellas-Velidis}, {Benson}, {Berthier}, {Blomme}, {Busso}, {Carry},
  {Cellino}, {Clementini}, {Cowell}, {Creevey}, {Cuypers}, {Davidson}, {De
  Ridder}, {de Torres}, {Delchambre}, {Dell'Oro}, {Ducourant}, {Fr{\'e}mat},
  {Garc{\'\i}a-Torres}, {Gosset}, {Halbwachs}, {Hambly}, {Harrison}, {Hauser},
  {Hestroffer}, {Hodgkin}, {Huckle}, {Hutton}, {Jasniewicz}, {Jordan},
  {Kontizas}, {Korn}, {Lanzafame}, {Manteiga}, {Moitinho}, {Muinonen},
  {Osinde}, {Pancino}, {Pauwels}, {Petit}, {Recio-Blanco}, {Robin}, {Sarro},
  {Siopis}, {Smith}, {Smith}, {Sozzetti}, {Thuillot}, {van Reeven}, {Viala},
  {Abbas}, {Abreu Aramburu}, {Accart}, {Aguado}, {Allan}, {Allasia},
  {Altavilla}, {{\'A}lvarez}, {Alves}, {Anderson}, {Andrei}, {Anglada Varela},
  {Antiche}, {Antoja}, {Ant{\'o}n}, {Arcay}, {Atzei}, {Ayache}, {Bach},
  {Baker}, {Balaguer-N{\'u}{\~n}ez}, {Barache}, {Barata}, {Barbier}, {Barblan},
  {Baroni}, {Barrado y Navascu{\'e}s}, {Barros}, {Barstow}, {Becciani},
  {Bellazzini}, {Bellei}, {Bello Garc{\'\i}a}, {Belokurov}, {Bendjoya},
  {Berihuete}, {Bianchi}, {Bienaym{\'e}}, {Billebaud}, {Blagorodnova},
  {Blanco-Cuaresma}, {Boch}, {Bombrun}, {Borrachero}, {Bouquillon}, {Bourda},
  {Bouy}, {Bragaglia}, {Breddels}, {Brouillet}, {Br{\"u}semeister},
  {Bucciarelli}, {Budnik}, {Burgess}, {Burgon}, {Burlacu}, {Busonero}, {Buzzi},
  {Caffau}, {Cambras}, {Campbell}, {Cancelliere}, {Cantat-Gaudin}, {Carlucci},
  {Carrasco}, {Castellani}, {Charlot}, {Charnas}, {Charvet}, {Chassat},
  {Chiavassa}, {Clotet}, {Cocozza}, {Collins}, {Collins}, {Costigan}, {Crifo},
  {Cross}, {Crosta}, {Crowley}, {Dafonte}, {Damerdji}, {Dapergolas}, {David},
  {David}, {De Cat}, {de Felice}, {de Laverny}, {De Luise}, {De March}, {de
  Martino}, {de Souza}, {Debosscher}, {del Pozo}, {Delbo}, {Delgado},
  {Delgado}, {di Marco}, {Di Matteo}, {Diakite}, {Distefano}, {Dolding}, {Dos
  Anjos}, {Drazinos}, {Dur{\'a}n}, {Dzigan}, {Ecale}, {Edvardsson}, {Enke},
  {Erdmann}, {Escolar}, {Espina}, {Evans}, {Eynard Bontemps}, {Fabre},
  {Fabrizio}, {Faigler}, {Falc{\~a}o}, {Farr{\`a}s Casas}, {Faye}, {Federici},
  {Fedorets}, {Fern{\'a}ndez-Hern{\'a}ndez}, {Fernique}, {Fienga}, {Figueras},
  {Filippi}, {Findeisen}, {Fonti}, {Fouesneau}, {Fraile}, {Fraser}, {Fuchs},
  {Furnell}, {Gai}, {Galleti}, {Galluccio}, {Garabato}, {Garc{\'\i}a-Sedano},
  {Gar{\'e}}, {Garofalo}, {Garralda}, {Gavras}, {Gerssen}, {Geyer}, {Gilmore},
  {Girona}, {Giuffrida}, {Gomes}, {Gonz{\'a}lez-Marcos},
  {Gonz{\'a}lez-N{\'u}{\~n}ez}, {Gonz{\'a}lez-Vidal}, {Granvik}, {Guerrier},
  {Guillout}, {Guiraud}, {G{\'u}rpide}, {Guti{\'e}rrez-S{\'a}nchez}, {Guy},
  {Haigron}, {Hatzidimitriou}, {Haywood}, {Heiter}, {Helmi}, {Hobbs},
  {Hofmann}, {Holl}, {Holland}, {Hunt}, {Hypki}, {Icardi}, {Irwin}, {Jevardat
  de Fombelle}, {Jofr{\'e}}, {Jonker}, {Jorissen}, {Julbe}, {Karampelas},
  {Kochoska}, {Kohley}, {Kolenberg}, {Kontizas}, {Koposov}, {Kordopatis},
  {Koubsky}, {Kowalczyk}, {Krone-Martins}, {Kudryashova}, {Kull}, {Bachchan},
  {Lacoste-Seris}, {Lanza}, {Lavigne}, {Le Poncin-Lafitte}, {Lebreton},
  {Lebzelter}, {Leccia}, {Leclerc}, {Lecoeur-Taibi}, {Lemaitre}, {Lenhardt},
  {Leroux}, {Liao}, {Licata}, {Lindstr{\o}m}, {Lister}, {Livanou}, {Lobel},
  {L{\"o}ffler}, {L{\'o}pez}, {Lopez-Lozano}, {Lorenz}, {Loureiro},
  {MacDonald}, {Magalh{\~a}es Fernandes}, {Managau}, {Mann}, {Mantelet},
  {Marchal}, {Marchant}, {Marconi}, {Marie}, {Marinoni}, {Marrese},
  {Marschalk{\'o}}, {Marshall}, {Mart{\'\i}n-Fleitas}, {Martino}, {Mary},
  {Matijevi{\v{c}}}, {Mazeh}, {McMillan}, {Messina}, {Mestre}, {Michalik},
  {Millar}, {Miranda}, {Molina}, {Molinaro}, {Molinaro}, {Moln{\'a}r},
  {Moniez}, {Montegriffo}, {Monteiro}, {Mor}, {Mora}, {Morbidelli}, {Morel},
  {Morgenthaler}, {Morley}, {Morris}, {Mulone}, {Muraveva}, {Musella},
  {Narbonne}, {Nelemans}, {Nicastro}, {Noval}, {Ord{\'e}novic},
  {Ordieres-Mer{\'e}}, {Osborne}, {Pagani}, {Pagano}, {Pailler}, {Palacin},
  {Palaversa}, {Parsons}, {Paulsen}, {Pecoraro}, {Pedrosa}, {Pentik{\"a}inen},
  {Pereira}, {Pichon}, {Piersimoni}, {Pineau}, {Plachy}, {Plum}, {Poujoulet},
  {Pr{\v{s}}a}, {Pulone}, {Ragaini}, {Rago}, {Rambaux}, {Ramos-Lerate},
  {Ranalli}, {Rauw}, {Read}, {Regibo}, {Renk}, {Reyl{\'e}}, {Ribeiro},
  {Rimoldini}, {Ripepi}, {Riva}, {Rixon}, {Roelens}, {Romero-G{\'o}mez},
  {Rowell}, {Royer}, {Rudolph}, {Ruiz-Dern}, {Sadowski}, {Sagrist{\`a}
  Sell{\'e}s}, {Sahlmann}, {Salgado}, {Salguero}, {Sarasso}, {Savietto},
  {Schnorhk}, {Schultheis}, {Sciacca}, {Segol}, {Segovia}, {Segransan},
  {Serpell}, {Shih}, {Smareglia}, {Smart}, {Smith}, {Solano}, {Solitro},
  {Sordo}, {Soria Nieto}, {Souchay}, {Spagna}, {Spoto}, {Stampa}, {Steele},
  {Steidelm{\"u}ller}, {Stephenson}, {Stoev}, {Suess}, {S{\"u}veges}, {Surdej},
  {Szabados}, {Szegedi-Elek}, {Tapiador}, {Taris}, {Tauran}, {Taylor},
  {Teixeira}, {Terrett}, {Tingley}, {Trager}, {Turon}, {Ulla}, {Utrilla},
  {Valentini}, {van Elteren}, {Van Hemelryck}, {van Leeuwen}, {Varadi},
  {Vecchiato}, {Veljanoski}, {Via}, {Vicente}, {Vogt}, {Voss}, {Votruba},
  {Voutsinas}, {Walmsley}, {Weiler}, {Weingrill}, {Werner}, {Wevers},
  {Whitehead}, {Wyrzykowski}, {Yoldas}, {{\v{Z}}erjal}, {Zucker}, {Zurbach},
  {Zwitter}, {Alecu}, {Allen}, {Allende Prieto}, {Amorim},
  {Anglada-Escud{\'e}}, {Arsenijevic}, {Azaz}, {Balm}, {Beck}, {Bernstein},
  {Bigot}, {Bijaoui}, {Blasco}, {Bonfigli}, {Bono}, {Boudreault}, {Bressan},
  {Brown}, {Brunet}, {Bunclark}, {Buonanno}, {Butkevich}, {Carret}, {Carrion},
  {Chemin}, {Ch{\'e}reau}, {Corcione}, {Darmigny}, {de Boer}, {de Teodoro}, {de
  Zeeuw}, {Delle Luche}, {Domingues}, {Dubath}, {Fodor}, {Fr{\'e}zouls},
  {Fries}, {Fustes}, {Fyfe}, {Gallardo}, {Gallegos}, {Gardiol}, {Gebran},
  {Gomboc}, {G{\'o}mez}, {Grux}, {Gueguen}, {Heyrovsky}, {Hoar}, {Iannicola},
  {Isasi Parache}, {Janotto}, {Joliet}, {Jonckheere}, {Keil}, {Kim},
  {Klagyivik}, {Klar}, {Knude}, {Kochukhov}, {Kolka}, {Kos}, {Kutka}, {Lainey},
  {LeBouquin}, {Liu}, {Loreggia}, {Makarov}, {Marseille}, {Martayan},
  {Martinez-Rubi}, {Massart}, {Meynadier}, {Mignot}, {Munari}, {Nguyen},
  {Nordlander}, {Ocvirk}, {O'Flaherty}, {Olias Sanz}, {Ortiz}, {Osorio},
  {Oszkiewicz}, {Ouzounis}, {Palmer}, {Park}, {Pasquato}, {Peltzer}, {Peralta},
  {P{\'e}turaud}, {Pieniluoma}, {Pigozzi}, {Poels}, {Prat}, {Prod'homme},
  {Raison}, {Rebordao}, {Risquez}, {Rocca-Volmerange}, {Rosen}, {Ruiz-Fuertes},
  {Russo}, {Sembay}, {Serraller Vizcaino}, {Short}, {Siebert}, {Silva},
  {Sinachopoulos}, {Slezak}, {Soffel}, {Sosnowska}, {Strai{\v{z}}ys}, {ter
  Linden}, {Terrell}, {Theil}, {Tiede}, {Troisi}, {Tsalmantza}, {Tur},
  {Vaccari}, {Vachier}, {Valles}, {Van Hamme}, {Veltz}, {Virtanen}, {Wallut},
  {Wichmann}, {Wilkinson}, {Ziaeepour}, \& {Zschocke}}]{Gaia}
{Gaia Collaboration}, {Prusti}, T., {de Bruijne}, J.~H.~J., {et~al.} 2016,
  \aap, 595, A1

\bibitem[{{Gaia Collaboration} {et~al.}(2023){Gaia Collaboration}, {Vallenari},
  {Brown}, {Prusti}, {de Bruijne}, {Arenou}, {Babusiaux}, {Biermann},
  {Creevey}, {Ducourant}, {Evans}, {Eyer}, {Guerra}, {Hutton}, {Jordi},
  {Klioner}, {Lammers}, {Lindegren}, {Luri}, {Mignard}, {Panem}, {Pourbaix},
  {Randich}, {Sartoretti}, {Soubiran}, {Tanga}, {Walton}, {Bailer-Jones},
  {Bastian}, {Drimmel}, {Jansen}, {Katz}, {Lattanzi}, {van Leeuwen}, {Bakker},
  {Cacciari}, {Casta{\~n}eda}, {De Angeli}, {Fabricius}, {Fouesneau},
  {Fr{\'e}mat}, {Galluccio}, {Guerrier}, {Heiter}, {Masana}, {Messineo},
  {Mowlavi}, {Nicolas}, {Nienartowicz}, {Pailler}, {Panuzzo}, {Riclet}, {Roux},
  {Seabroke}, {Sordo}, {Th{\'e}venin}, {Gracia-Abril}, {Portell}, {Teyssier},
  {Altmann}, {Andrae}, {Audard}, {Bellas-Velidis}, {Benson}, {Berthier},
  {Blomme}, {Burgess}, {Busonero}, {Busso}, {C{\'a}novas}, {Carry}, {Cellino},
  {Cheek}, {Clementini}, {Damerdji}, {Davidson}, {de Teodoro}, {Nu{\~n}ez
  Campos}, {Delchambre}, {Dell'Oro}, {Esquej}, {Fern{\'a}ndez-Hern{\'a}ndez},
  {Fraile}, {Garabato}, {Garc{\'\i}a-Lario}, {Gosset}, {Haigron}, {Halbwachs},
  {Hambly}, {Harrison}, {Hern{\'a}ndez}, {Hestroffer}, {Hodgkin}, {Holl},
  {Jan{\ss}en}, {Jevardat de Fombelle}, {Jordan}, {Krone-Martins}, {Lanzafame},
  {L{\"o}ffler}, {Marchal}, {Marrese}, {Moitinho}, {Muinonen}, {Osborne},
  {Pancino}, {Pauwels}, {Recio-Blanco}, {Reyl{\'e}}, {Riello}, {Rimoldini},
  {Roegiers}, {Rybizki}, {Sarro}, {Siopis}, {Smith}, {Sozzetti}, {Utrilla},
  {van Leeuwen}, {Abbas}, {{\'A}brah{\'a}m}, {Abreu Aramburu}, {Aerts},
  {Aguado}, {Ajaj}, {Aldea-Montero}, {Altavilla}, {{\'A}lvarez}, {Alves},
  {Anders}, {Anderson}, {Anglada Varela}, {Antoja}, {Baines}, {Baker},
  {Balaguer-N{\'u}{\~n}ez}, {Balbinot}, {Balog}, {Barache}, {Barbato},
  {Barros}, {Barstow}, {Bartolom{\'e}}, {Bassilana}, {Bauchet}, {Becciani},
  {Bellazzini}, {Berihuete}, {Bernet}, {Bertone}, {Bianchi}, {Binnenfeld},
  {Blanco-Cuaresma}, {Blazere}, {Boch}, {Bombrun}, {Bossini}, {Bouquillon},
  {Bragaglia}, {Bramante}, {Breedt}, {Bressan}, {Brouillet}, {Brugaletta},
  {Bucciarelli}, {Burlacu}, {Butkevich}, {Buzzi}, {Caffau}, {Cancelliere},
  {Cantat-Gaudin}, {Carballo}, {Carlucci}, {Carnerero}, {Carrasco},
  {Casamiquela}, {Castellani}, {Castro-Ginard}, {Chaoul}, {Charlot}, {Chemin},
  {Chiaramida}, {Chiavassa}, {Chornay}, {Comoretto}, {Contursi}, {Cooper},
  {Cornez}, {Cowell}, {Crifo}, {Cropper}, {Crosta}, {Crowley}, {Dafonte},
  {Dapergolas}, {David}, {David}, {de Laverny}, {De Luise}, {De March}, {De
  Ridder}, {de Souza}, {de Torres}, {del Peloso}, {del Pozo}, {Delbo},
  {Delgado}, {Delisle}, {Demouchy}, {Dharmawardena}, {Di Matteo}, {Diakite},
  {Diener}, {Distefano}, {Dolding}, {Edvardsson}, {Enke}, {Fabre}, {Fabrizio},
  {Faigler}, {Fedorets}, {Fernique}, {Fienga}, {Figueras}, {Fournier},
  {Fouron}, {Fragkoudi}, {Gai}, {Garcia-Gutierrez}, {Garcia-Reinaldos},
  {Garc{\'\i}a-Torres}, {Garofalo}, {Gavel}, {Gavras}, {Gerlach}, {Geyer},
  {Giacobbe}, {Gilmore}, {Girona}, {Giuffrida}, {Gomel}, {Gomez},
  {Gonz{\'a}lez-N{\'u}{\~n}ez}, {Gonz{\'a}lez-Santamar{\'\i}a},
  {Gonz{\'a}lez-Vidal}, {Granvik}, {Guillout}, {Guiraud},
  {Guti{\'e}rrez-S{\'a}nchez}, {Guy}, {Hatzidimitriou}, {Hauser}, {Haywood},
  {Helmer}, {Helmi}, {Sarmiento}, {Hidalgo}, {Hilger}, {H{\l}adczuk}, {Hobbs},
  {Holland}, {Huckle}, {Jardine}, {Jasniewicz}, {Jean-Antoine Piccolo},
  {Jim{\'e}nez-Arranz}, {Jorissen}, {Juaristi Campillo}, {Julbe}, {Karbevska},
  {Kervella}, {Khanna}, {Kontizas}, {Kordopatis}, {Korn}, {K{\'o}sp{\'a}l},
  {Kostrzewa-Rutkowska}, {Kruszy{\'n}ska}, {Kun}, {Laizeau}, {Lambert},
  {Lanza}, {Lasne}, {Le Campion}, {Lebreton}, {Lebzelter}, {Leccia}, {Leclerc},
  {Lecoeur-Taibi}, {Liao}, {Licata}, {Lindstr{\o}m}, {Lister}, {Livanou},
  {Lobel}, {Lorca}, {Loup}, {Madrero Pardo}, {Magdaleno Romeo}, {Managau},
  {Mann}, {Manteiga}, {Marchant}, {Marconi}, {Marcos}, {Marcos Santos},
  {Mar{\'\i}n Pina}, {Marinoni}, {Marocco}, {Marshall}, {Martin Polo},
  {Mart{\'\i}n-Fleitas}, {Marton}, {Mary}, {Masip}, {Massari},
  {Mastrobuono-Battisti}, {Mazeh}, {McMillan}, {Messina}, {Michalik}, {Millar},
  {Mints}, {Molina}, {Molinaro}, {Moln{\'a}r}, {Monari}, {Mongui{\'o}},
  {Montegriffo}, {Montero}, {Mor}, {Mora}, {Morbidelli}, {Morel}, {Morris},
  {Muraveva}, {Murphy}, {Musella}, {Nagy}, {Noval}, {Oca{\~n}a}, {Ogden},
  {Ordenovic}, {Osinde}, {Pagani}, {Pagano}, {Palaversa}, {Palicio},
  {Pallas-Quintela}, {Panahi}, {Payne-Wardenaar}, {Pe{\~n}alosa Esteller},
  {Penttil{\"a}}, {Pichon}, {Piersimoni}, {Pineau}, {Plachy}, {Plum}, {Poggio},
  {Pr{\v{s}}a}, {Pulone}, {Racero}, {Ragaini}, {Rainer}, {Raiteri}, {Rambaux},
  {Ramos}, {Ramos-Lerate}, {Re Fiorentin}, {Regibo}, {Richards}, {Rios Diaz},
  {Ripepi}, {Riva}, {Rix}, {Rixon}, {Robichon}, {Robin}, {Robin}, {Roelens},
  {Rogues}, {Rohrbasser}, {Romero-G{\'o}mez}, {Rowell}, {Royer}, {Ruz Mieres},
  {Rybicki}, {Sadowski}, {S{\'a}ez N{\'u}{\~n}ez}, {Sagrist{\`a} Sell{\'e}s},
  {Sahlmann}, {Salguero}, {Samaras}, {Sanchez Gimenez}, {Sanna},
  {Santove{\~n}a}, {Sarasso}, {Schultheis}, {Sciacca}, {Segol}, {Segovia},
  {S{\'e}gransan}, {Semeux}, {Shahaf}, {Siddiqui}, {Siebert}, {Siltala},
  {Silvelo}, {Slezak}, {Slezak}, {Smart}, {Snaith}, {Solano}, {Solitro},
  {Souami}, {Souchay}, {Spagna}, {Spina}, {Spoto}, {Steele},
  {Steidelm{\"u}ller}, {Stephenson}, {S{\"u}veges}, {Surdej}, {Szabados},
  {Szegedi-Elek}, {Taris}, {Taylor}, {Teixeira}, {Tolomei}, {Tonello}, {Torra},
  {Torra}, {Torralba Elipe}, {Trabucchi}, {Tsounis}, {Turon}, {Ulla}, {Unger},
  {Vaillant}, {van Dillen}, {van Reeven}, {Vanel}, {Vecchiato}, {Viala},
  {Vicente}, {Voutsinas}, {Weiler}, {Wevers}, {Wyrzykowski}, {Yoldas}, {Yvard},
  {Zhao}, {Zorec}, {Zucker}, \& {Zwitter}}]{GaiaDR3}
{Gaia Collaboration}, {Vallenari}, A., {Brown}, A.~G.~A., {et~al.} 2023, \aap,
  674, A1

\bibitem[{{Ge} {et~al.}(2022){Ge}, {Paltani}, {Eckert}, \& {Salvato}}]{Ge21}
{Ge}, L., {Paltani}, S., {Eckert}, D., \& {Salvato}, M. 2022, \aap, 667, A153

\bibitem[{{Gendreau} {et~al.}(2016){Gendreau}, {Arzoumanian}, {Adkins},
  {Albert}, {Anders}, {Aylward}, {Baker}, {Balsamo}, {Bamford}, {Benegalrao},
  {Berry}, {Bhalwani}, {Black}, {Blaurock}, {Bronke}, {Brown}, {Budinoff},
  {Cantwell}, {Cazeau}, {Chen}, {Clement}, {Colangelo}, {Coleman},
  {Coopersmith}, {Dehaven}, {Doty}, {Egan}, {Enoto}, {Fan}, {Ferro}, {Foster},
  {Galassi}, {Gallo}, {Green}, {Grosh}, {Ha}, {Hasouneh}, {Heefner}, {Hestnes},
  {Hoge}, {Jacobs}, {J{\o}rgensen}, {Kaiser}, {Kellogg}, {Kenyon}, {Koenecke},
  {Kozon}, {LaMarr}, {Lambertson}, {Larson}, {Lentine}, {Lewis}, {Lilly},
  {Liu}, {Malonis}, {Manthripragada}, {Markwardt}, {Matonak}, {Mcginnis},
  {Miller}, {Mitchell}, {Mitchell}, {Mohammed}, {Monroe}, {Montt de Garcia},
  {Mul{\'e}}, {Nagao}, {Ngo}, {Norris}, {Norwood}, {Novotka}, {Okajima},
  {Olsen}, {Onyeachu}, {Orosco}, {Peterson}, {Pevear}, {Pham}, {Pollard},
  {Pope}, {Powers}, {Powers}, {Price}, {Prigozhin}, {Ramirez}, {Reid},
  {Remillard}, {Rogstad}, {Rosecrans}, {Rowe}, {Sager}, {Sanders}, {Savadkin},
  {Saylor}, {Schaeffer}, {Schweiss}, {Semper}, {Serlemitsos}, {Shackelford},
  {Soong}, {Struebel}, {Vezie}, {Villasenor}, {Winternitz}, {Wofford},
  {Wright}, {Yang}, \& {Yu}}]{Gendreau16}
{Gendreau}, K.~C., {Arzoumanian}, Z., {Adkins}, P.~W., {et~al.} 2016, in
  Society of Photo-Optical Instrumentation Engineers (SPIE) Conference Series,
  Vol. 9905, Space Telescopes and Instrumentation 2016: Ultraviolet to Gamma
  Ray, ed. J.-W.~A. {den Herder}, T.~{Takahashi}, \& M.~{Bautz}, 99051H

\bibitem[{{Gentile} {et~al.}(2014){Gentile}, {Roberts}, {McLaughlin}, {Camilo},
  {Hessels}, {Kerr}, {Ransom}, {Ray}, \& {Stairs}}]{Gentile14}
{Gentile}, P.~A., {Roberts}, M.~S.~E., {McLaughlin}, M.~A., {et~al.} 2014,
  \apj, 783, 69

\bibitem[{{Germani} {et~al.}(2021){Germani}, {Tosti}, {Lubrano}, {Cutini},
  {Mereu}, \& {Berretta}}]{Germani21}
{Germani}, S., {Tosti}, G., {Lubrano}, P., {et~al.} 2021, \mnras, 505, 5853

\bibitem[{{Hare} {et~al.}(2022){Hare}, {Kargaltsev}, {Younes}, {Volkov}, \&
  {Rangelov}}]{Hare22}
{Hare}, J., {Kargaltsev}, O., {Younes}, G., {Volkov}, I., \& {Rangelov}, B.
  2022, in AAS/High Energy Astrophysics Division, Vol.~54, AAS/High Energy
  Astrophysics Division, 110.11

\bibitem[{Harris {et~al.}(2020)Harris, Millman, van~der Walt, Gommers,
  Virtanen, Cournapeau, Wieser, Taylor, Berg, Smith, Kern, Picus, Hoyer, van
  Kerkwijk, Brett, Haldane, Fernández~del Río, Wiebe, Peterson,
  Gérard-Marchant, Sheppard, Reddy, Weckesser, Abbasi, Gohlke, \&
  Oliphant}]{NumPy}
Harris, C.~R., Millman, K.~J., van~der Walt, S.~J., {et~al.} 2020, Nature, 585,
  357–362

\bibitem[{{He} {et~al.}(2013){He}, {Ng}, \& {Kaspi}}]{He13}
{He}, C., {Ng}, C.~Y., \& {Kaspi}, V.~M. 2013, \apj, 768, 64

\bibitem[{{Helfand} {et~al.}(2015){Helfand}, {White}, \& {Becker}}]{FIRST}
{Helfand}, D.~J., {White}, R.~L., \& {Becker}, R.~H. 2015, \apj, 801, 26

\bibitem[{{HI4PI Collaboration} {et~al.}(2016){HI4PI Collaboration}, {Ben
  Bekhti}, {Fl{\"o}er}, {Keller}, {Kerp}, {Lenz}, {Winkel}, {Bailin},
  {Calabretta}, {Dedes}, {Ford}, {Gibson}, {Haud}, {Janowiecki}, {Kalberla},
  {Lockman}, {McClure-Griffiths}, {Murphy}, {Nakanishi}, {Pisano}, \&
  {Staveley-Smith}}]{HI4PI}
{HI4PI Collaboration}, {Ben Bekhti}, N., {Fl{\"o}er}, L., {et~al.} 2016, \aap,
  594, A116

\bibitem[{{Hui} \& {Li}(2019)}]{Hui19}
{Hui}, C.~Y. \& {Li}, K.~L. 2019, Galaxies, 7, 93

\bibitem[{Hunter(2007)}]{Hunter:2007}
Hunter, J.~D. 2007, Computing in Science \& Engineering, 9, 90

\bibitem[{{Hurley-Walker} {et~al.}(2017){Hurley-Walker}, {Callingham},
  {Hancock}, {Franzen}, {Hindson}, {Kapi{\'n}ska}, {Morgan}, {Offringa},
  {Wayth}, {Wu}, {Zheng}, {Murphy}, {Bell}, {Dwarakanath}, {For}, {Gaensler},
  {Johnston-Hollitt}, {Lenc}, {Procopio}, {Staveley-Smith}, {Ekers}, {Bowman},
  {Briggs}, {Cappallo}, {Deshpande}, {Greenhill}, {Hazelton}, {Kaplan},
  {Lonsdale}, {McWhirter}, {Mitchell}, {Morales}, {Morgan}, {Oberoi}, {Ord},
  {Prabu}, {Shankar}, {Srivani}, {Subrahmanyan}, {Tingay}, {Webster},
  {Williams}, \& {Williams}}]{GLEAM}
{Hurley-Walker}, N., {Callingham}, J.~R., {Hancock}, P.~J., {et~al.} 2017,
  \mnras, 464, 1146

\bibitem[{{Intema} {et~al.}(2017){Intema}, {Jagannathan}, {Mooley}, \&
  {Frail}}]{TGSS}
{Intema}, H.~T., {Jagannathan}, P., {Mooley}, K.~P., \& {Frail}, D.~A. 2017,
  \aap, 598, A78

\bibitem[{{Kaur} {et~al.}(2019){Kaur}, {Falcone}, {Stroh}, {Kennea}, \&
  {Ferrara}}]{Kaur19}
{Kaur}, A., {Falcone}, A.~D., {Stroh}, M.~D., {Kennea}, J.~A., \& {Ferrara},
  E.~C. 2019, \apj, 887, 18

\bibitem[{{Kerby} {et~al.}(2021{\natexlab{a}}){Kerby}, {Kaur}, {Falcone},
  {Eskenasy}, {Hancock}, {Stroh}, {Ferrara}, {Ray}, {Kennea}, \&
  {Grove}}]{Kerby21}
{Kerby}, S., {Kaur}, A., {Falcone}, A.~D., {et~al.} 2021{\natexlab{a}}, \apj,
  923, 75

\bibitem[{{Kerby} {et~al.}(2021{\natexlab{b}}){Kerby}, {Kaur}, {Falcone},
  {Stroh}, {Ferrara}, {Kennea}, \& {Colosimo}}]{Kerby21b}
{Kerby}, S., {Kaur}, A., {Falcone}, A.~D., {et~al.} 2021{\natexlab{b}}, \aj,
  161, 154

\bibitem[{{Koljonen} \& {Linares}(2023)}]{Koljonen23}
{Koljonen}, K. I.~I. \& {Linares}, M. 2023, \mnras, 525, 3963

\bibitem[{{Lane} {et~al.}(2014){Lane}, {Cotton}, {van Velzen}, {Clarke},
  {Kassim}, {Helmboldt}, {Lazio}, \& {Cohen}}]{VLSSR}
{Lane}, W.~M., {Cotton}, W.~D., {van Velzen}, S., {et~al.} 2014, \mnras, 440,
  327

\bibitem[{Lema{{\^i}}tre {et~al.}(2017)Lema{{\^i}}tre, Nogueira, \&
  Aridas}]{Lemaitre17}
Lema{{\^i}}tre, G., Nogueira, F., \& Aridas, C.~K. 2017, Journal of Machine
  Learning Research, 18, 1

\bibitem[{{Luo} {et~al.}(2020){Luo}, {Leung}, {Hui}, \& {Li}}]{Luo20}
{Luo}, S., {Leung}, A.~P., {Hui}, C.~Y., \& {Li}, K.~L. 2020, \mnras, 492, 5377

\bibitem[{{Malyshev}(2023)}]{Malyshev23}
{Malyshev}, D.~V. 2023, RAS Techniques and Instruments, 2, 735

\bibitem[{{Manchester} {et~al.}(2005){Manchester}, {Hobbs}, {Teoh}, \&
  {Hobbs}}]{ATNF}
{Manchester}, R.~N., {Hobbs}, G.~B., {Teoh}, A., \& {Hobbs}, M. 2005, \aj, 129,
  1993

\bibitem[{{Marelli} {et~al.}(2011){Marelli}, {De Luca}, \&
  {Caraveo}}]{Marelli11}
{Marelli}, M., {De Luca}, A., \& {Caraveo}, P.~A. 2011, \apj, 733, 82

\bibitem[{{Marelli} {et~al.}(2015){Marelli}, {Mignani}, {De Luca}, {Saz
  Parkinson}, {Salvetti}, {Den Hartog}, \& {Wolff}}]{Marelli15}
{Marelli}, M., {Mignani}, R.~P., {De Luca}, A., {et~al.} 2015, \apj, 802, 78

\bibitem[{{Marocco} {et~al.}(2021){Marocco}, {Eisenhardt}, {Fowler},
  {Kirkpatrick}, {Meisner}, {Schlafly}, {Stanford}, {Garcia}, {Caselden},
  {Cushing}, {Cutri}, {Faherty}, {Gelino}, {Gonzalez}, {Jarrett}, {Koontz},
  {Mainzer}, {Marchese}, {Mobasher}, {Schlegel}, {Stern}, {Teplitz}, \&
  {Wright}}]{CatWISE2020}
{Marocco}, F., {Eisenhardt}, P. R.~M., {Fowler}, J.~W., {et~al.} 2021, \apjs,
  253, 8

\bibitem[{{Mauch} {et~al.}(2003){Mauch}, {Murphy}, {Buttery}, {Curran},
  {Hunstead}, {Piestrzynski}, {Robertson}, \& {Sadler}}]{SUMSS}
{Mauch}, T., {Murphy}, T., {Buttery}, H.~J., {et~al.} 2003, \mnras, 342, 1117

\bibitem[{{Merloni} {et~al.}(2024){Merloni}, {Lamer}, {Liu}, {Ramos-Ceja},
  {Brunner}, \& et~al.}]{Merloni23}
{Merloni}, A., {Lamer}, G., {Liu}, T., {et~al.} 2024, \aap, 682, A34

\bibitem[{{Merloni} {et~al.}(2012){Merloni}, {Predehl}, {Becker},
  {B{\"o}hringer}, {Boller}, {Brunner}, {Brusa}, {Dennerl}, {Freyberg},
  {Friedrich}, {Georgakakis}, {Haberl}, {Hasinger}, {Meidinger}, {Mohr},
  {Nandra}, {Rau}, {Reiprich}, {Robrade}, {Salvato}, {Santangelo}, {Sasaki},
  {Schwope}, {Wilms}, \& {German eROSITA Consortium}}]{Merloni12}
{Merloni}, A., {Predehl}, P., {Becker}, W., {et~al.} 2012, arXiv e-prints,
  arXiv:1209.3114

\bibitem[{{Mignani}(2011)}]{Mignani11}
{Mignani}, R.~P. 2011, Advances in Space Research, 47, 1281

\bibitem[{{Pineau} {et~al.}(2017){Pineau}, {Derriere}, {Motch}, {Carrera},
  {Genova}, {Michel}, {Mingo}, {Mints}, {Nebot G{\'o}mez-Mor{\'a}n}, {Rosen},
  \& {Ruiz Camu{\~n}as}}]{Pineau17}
{Pineau}, F.~X., {Derriere}, S., {Motch}, C., {et~al.} 2017, \aap, 597, A89

\bibitem[{{Pizzolato} {et~al.}(2003){Pizzolato}, {Maggio}, {Micela},
  {Sciortino}, \& {Ventura}}]{Pizzolato03}
{Pizzolato}, N., {Maggio}, A., {Micela}, G., {Sciortino}, S., \& {Ventura}, P.
  2003, \aap, 397, 147

\bibitem[{{Pol} {et~al.}(2021){Pol}, {Burke-Spolaor}, {Hurley-Walker},
  {Blumer}, {Johnston}, {Keith}, {Keane}, {Burgay}, {Possenti}, {Petroff}, \&
  {Bhat}}]{PSR_J0837}
{Pol}, N., {Burke-Spolaor}, S., {Hurley-Walker}, N., {et~al.} 2021, \apj, 911,
  121

\bibitem[{{Predehl} {et~al.}(2021){Predehl}, {Andritschke}, {Arefiev},
  {Babyshkin}, {Batanov}, {Becker}, {B{\"o}hringer}, {Bogomolov}, {Boller},
  {Borm}, {Bornemann}, {Br{\"a}uninger}, {Br{\"u}ggen}, {Brunner}, {Brusa},
  {Bulbul}, {Buntov}, {Burwitz}, {Burkert}, {Clerc}, {Churazov}, {Coutinho},
  {Dauser}, {Dennerl}, {Doroshenko}, {Eder}, {Emberger}, {Eraerds},
  {Finoguenov}, {Freyberg}, {Friedrich}, {Friedrich}, {F{\"u}rmetz},
  {Georgakakis}, {Gilfanov}, {Granato}, {Grossberger}, {Gueguen}, {Gureev},
  {Haberl}, {H{\"a}lker}, {Hartner}, {Hasinger}, {Huber}, {Ji}, {Kienlin},
  {Kink}, {Korotkov}, {Kreykenbohm}, {Lamer}, {Lomakin}, {Lapshov}, {Liu},
  {Maitra}, {Meidinger}, {Menz}, {Merloni}, {Mernik}, {Mican}, {Mohr},
  {M{\"u}ller}, {Nandra}, {Nazarov}, {Pacaud}, {Pavlinsky}, {Perinati},
  {Pfeffermann}, {Pietschner}, {Ramos-Ceja}, {Rau}, {Reiffers}, {Reiprich},
  {Robrade}, {Salvato}, {Sanders}, {Santangelo}, {Sasaki}, {Scheuerle},
  {Schmid}, {Schmitt}, {Schwope}, {Shirshakov}, {Steinmetz}, {Stewart},
  {Str{\"u}der}, {Sunyaev}, {Tenzer}, {Tiedemann}, {Tr{\"u}mper}, {Voron},
  {Weber}, {Wilms}, \& {Yaroshenko}}]{Predehl21}
{Predehl}, P., {Andritschke}, R., {Arefiev}, V., {et~al.} 2021, \aap, 647, A1

\bibitem[{{Prinz} \& {Becker}(2015)}]{Prinz_Becker_2015}
{Prinz}, T. \& {Becker}, W. 2015, arXiv e-prints, arXiv:1511.07713

\bibitem[{{Radhakrishnan} \& {Srinivasan}(1982)}]{Radhakrishnan82}
{Radhakrishnan}, V. \& {Srinivasan}, G. 1982, Current Science, 51, 1096

\bibitem[{{Rengelink} {et~al.}(1997){Rengelink}, {Tang}, {de Bruyn}, {Miley},
  {Bremer}, {Roettgering}, \& {Bremer}}]{WENSS}
{Rengelink}, R.~B., {Tang}, Y., {de Bruyn}, A.~G., {et~al.} 1997, \aaps, 124,
  259

\bibitem[{{Roberts}(2013)}]{Roberts13}
{Roberts}, M. S.~E. 2013, in Neutron Stars and Pulsars: Challenges and
  Opportunities after 80 years, ed. J.~{van Leeuwen}, Vol. 291, 127--132

\bibitem[{{Romani} {et~al.}(2012){Romani}, {Filippenko}, {Silverman}, {Cenko},
  {Greiner}, {Rau}, {Elliott}, \& {Pletsch}}]{Romani12}
{Romani}, R.~W., {Filippenko}, A.~V., {Silverman}, J.~M., {et~al.} 2012, \apjl,
  760, L36

\bibitem[{{Romani} \& {Shaw}(2011)}]{Romani11}
{Romani}, R.~W. \& {Shaw}, M.~S. 2011, \apjl, 743, L26

\bibitem[{{Salvato} {et~al.}(2018){Salvato}, {Buchner}, {Budav{\'a}ri},
  {Dwelly}, {Merloni}, {Brusa}, {Rau}, {Fotopoulou}, \& {Nandra}}]{Salvato18}
{Salvato}, M., {Buchner}, J., {Budav{\'a}ri}, T., {et~al.} 2018, \mnras, 473,
  4937

\bibitem[{{Salvato} {et~al.}(2022){Salvato}, {Wolf}, {Dwelly}, {Georgakakis},
  {Brusa}, {Merloni}, {Liu}, {Toba}, {Nandra}, {Lamer}, {Buchner}, {Schneider},
  {Freund}, {Rau}, {Schwope}, {Nishizawa}, {Klein}, {Arcodia}, {Comparat},
  {Musiimenta}, {Nagao}, {Brunner}, {Malyali}, {Finoguenov}, {Anderson},
  {Shen}, {Ibarra-Medel}, {Trump}, {Brandt}, {Urry}, {Rivera}, {Krumpe},
  {Urrutia}, {Miyaji}, {Ichikawa}, {Schneider}, {Fresco}, {Boller}, {Haase},
  {Brownstein}, {Lane}, {Bizyaev}, \& {Nitschelm}}]{Salvato22}
{Salvato}, M., {Wolf}, J., {Dwelly}, T., {et~al.} 2022, \aap, 661, A3

\bibitem[{Salvetti(2014)}]{Salvetti16}
Salvetti, D. 2014, PhD thesis, Pavia U.

\bibitem[{{Saz Parkinson} {et~al.}(2016){Saz Parkinson}, {Xu}, {Yu},
  {Salvetti}, {Marelli}, \& {Falcone}}]{SazParkinson16}
{Saz Parkinson}, P.~M., {Xu}, H., {Yu}, P.~L.~H., {et~al.} 2016, \apj, 820, 8

\bibitem[{{Schneider} {et~al.}(2022){Schneider}, {Freund}, {Czesla}, {Robrade},
  {Salvato}, \& {Schmitt}}]{Schneider22}
{Schneider}, P.~C., {Freund}, S., {Czesla}, S., {et~al.} 2022, \aap, 661, A6

\bibitem[{{Shibata} {et~al.}(2016){Shibata}, {Watanabe}, {Yatsu}, {Enoto}, \&
  {Bamba}}]{Shibata16}
{Shibata}, S., {Watanabe}, E., {Yatsu}, Y., {Enoto}, T., \& {Bamba}, A. 2016,
  \apj, 833, 59

\bibitem[{{Shimwell} {et~al.}(2022){Shimwell}, {Hardcastle}, {Tasse}, {Best},
  {R{\"o}ttgering}, {Williams}, {Botteon}, {Drabent}, {Mechev}, {Shulevski},
  {van Weeren}, {Bester}, {Br{\"u}ggen}, {Brunetti}, {Callingham}, {Chy{\.z}y},
  {Conway}, {Dijkema}, {Duncan}, {de Gasperin}, {Hale}, {Haverkorn}, {Hugo},
  {Jackson}, {Mevius}, {Miley}, {Morabito}, {Morganti}, {Offringa}, {Oonk},
  {Rafferty}, {Sabater}, {Smith}, {Schwarz}, {Smirnov}, {O'Sullivan},
  {Vedantham}, {White}, {Albert}, {Alegre}, {Asabere}, {Bacon}, {Bonafede},
  {Bonnassieux}, {Brienza}, {Bilicki}, {Bonato}, {Calistro Rivera}, {Cassano},
  {Cochrane}, {Croston}, {Cuciti}, {Dallacasa}, {Danezi}, {Dettmar}, {Di
  Gennaro}, {Edler}, {En{\ss}lin}, {Emig}, {Franzen}, {Garc{\'\i}a-Vergara},
  {Grange}, {G{\"u}rkan}, {Hajduk}, {Heald}, {Heesen}, {Hoang}, {Hoeft},
  {Horellou}, {Iacobelli}, {Jamrozy}, {Jeli{\'c}}, {Kondapally}, {Kukreti},
  {Kunert-Bajraszewska}, {Magliocchetti}, {Mahatma}, {Ma{\l}ek}, {Mandal},
  {Massaro}, {Meyer-Zhao}, {Mingo}, {Mostert}, {Nair}, {Nakoneczny},
  {Nikiel-Wroczy{\'n}ski}, {Orr{\'u}}, {Pajdosz-{\'S}mierciak}, {Pasini},
  {Prandoni}, {van Piggelen}, {Rajpurohit}, {Retana-Montenegro}, {Riseley},
  {Rowlinson}, {Saxena}, {Schrijvers}, {Sweijen}, {Siewert}, {Timmerman},
  {Vaccari}, {Vink}, {West}, {Wo{\l}owska}, {Zhang}, \& {Zheng}}]{LOTSS}
{Shimwell}, T.~W., {Hardcastle}, M.~J., {Tasse}, C., {et~al.} 2022, \aap, 659,
  A1

\bibitem[{{Smith} {et~al.}(2023){Smith}, {Abdollahi}, {Ajello}, {Bailes},
  {Baldini}, {Ballet}, {Baring}, {Bassa}, {Gonzalez}, {Bellazzini}, {Berretta},
  {Bhattacharyya}, {Bissaldi}, {Bonino}, {Bottacini}, {Bregeon}, {Bruel},
  {Burgay}, {Burnett}, {Cameron}, {Camilo}, {Caputo}, {Caraveo}, {Cavazzuti},
  {Chiaro}, {Ciprini}, {Clark}, {Cognard}, {Corongiu}, {Orestano},
  {Crnogorcevic}, {Cuoco}, {Cutini}, {D'Ammando}, {de Angelis}, {DeCesar}, {De
  Gaetano}, {de Menezes}, {Deneva}, {de Palma}, {Di Lalla}, {Dirirsa}, {Di
  Venere}, {Dom{\'\i}nguez}, {Dumora}, {Fegan}, {Ferrara}, {Fiori},
  {Fleischhack}, {Flynn}, {Franckowiak}, {Freire}, {Fukazawa}, {Fusco},
  {Galanti}, {Gammaldi}, {Gargano}, {Gasparrini}, {Giacchino}, {Giglietto},
  {Giordano}, {Giroletti}, {Green}, {Grenier}, {Guillemot}, {Guiriec},
  {Gustafsson}, {Harding}, {Hays}, {Hewitt}, {Horan}, {Hou}, {Jankowski},
  {Johnson}, {Johnson}, {Johnston}, {Kataoka}, {Keith}, {Kerr}, {Kramer},
  {Kuss}, {Latronico}, {Lee}, {Li}, {Li}, {Limyansky}, {Longo}, {Loparco},
  {Lorusso}, {Lovellette}, {Lower}, {Lubrano}, {Lyne}, {Maan}, {Maldera},
  {Manchester}, {Manfreda}, {Marelli}, {Mart{\'\i}-Devesa}, {Mazziotta},
  {McEnery}, {Mereu}, {Michelson}, {Mickaliger}, {Mitthumsiri}, {Mizuno},
  {Moiseev}, {Monzani}, {Morselli}, {Negro}, {Nemmen}, {Nieder}, {Nuss},
  {Omodei}, {Orienti}, {Orlando}, {Ormes}, {Palatiello}, {Paneque},
  {Panzarini}, {Parthasarathy}, {Persic}, {Pesce-Rollins}, {Pillera}, {Poon},
  {Porter}, {Possenti}, {Principe}, {Rain{\`o}}, {Rando}, {Ransom}, {Ray},
  {Razzano}, {Razzaque}, {Reimer}, {Reimer}, {Renault-Tinacci}, {Romani},
  {S{\'a}nchez-Conde}, {Parkinson}, {Scotton}, {Serini}, {Sgr{\`o}}, {Shannon},
  {Sharma}, {Shen}, {Siskind}, {Spandre}, {Spinelli}, {Stappers}, {Stephens},
  {Suson}, {Tabassum}, {Tajima}, {Tak}, {Theureau}, {Thompson}, {Tibolla},
  {Torres}, {Valverde}, {Venter}, {Wadiasingh}, {Wang}, {Wang}, {Wang},
  {Weltevrede}, {Wood}, {Yan}, {Zaharijas}, {Zhang}, \& {Zhu}}]{3PC}
{Smith}, D.~A., {Abdollahi}, S., {Ajello}, M., {et~al.} 2023, \apj, 958, 191

\bibitem[{{Smith} {et~al.}(2019){Smith}, {Bruel}, {Cognard}, {Cameron},
  {Camilo}, {Dai}, {Guillemot}, {Johnson}, {Johnston}, {Keith}, {Kerr},
  {Kramer}, {Lyne}, {Manchester}, {Shannon}, {Sobey}, {Stappers}, \&
  {Weltevrede}}]{Smith19}
{Smith}, D.~A., {Bruel}, P., {Cognard}, I., {et~al.} 2019, \apj, 871, 78

\bibitem[{{Strader} {et~al.}(2019){Strader}, {Swihart}, {Chomiuk}, {Bahramian},
  {Britt}, {Cheung}, {Dage}, {Halpern}, {Li}, {Mignani}, {Orosz}, {Peacock},
  {Salinas}, {Shishkovsky}, \& {Tremou}}]{Strader19}
{Strader}, J., {Swihart}, S., {Chomiuk}, L., {et~al.} 2019, \apj, 872, 42

\bibitem[{{Stroh} \& {Falcone}(2013)}]{StrohFalcone}
{Stroh}, M.~C. \& {Falcone}, A.~D. 2013, \apjs, 207, 28

\bibitem[{{Str{\"u}der} {et~al.}(2001){Str{\"u}der}, {Briel}, {Dennerl},
  {Hartmann}, {Kendziorra}, {Meidinger}, {Pfeffermann}, {Reppin}, {Aschenbach},
  {Bornemann}, {Br{\"a}uninger}, {Burkert}, {Elender}, {Freyberg}, {Haberl},
  {Hartner}, {Heuschmann}, {Hippmann}, {Kastelic}, {Kemmer}, {Kettenring},
  {Kink}, {Krause}, {M{\"u}ller}, {Oppitz}, {Pietsch}, {Popp}, {Predehl},
  {Read}, {Stephan}, {St{\"o}tter}, {Tr{\"u}mper}, {Holl}, {Kemmer}, {Soltau},
  {St{\"o}tter}, {Weber}, {Weichert}, {von Zanthier}, {Carathanassis}, {Lutz},
  {Richter}, {Solc}, {B{\"o}ttcher}, {Kuster}, {Staubert}, {Abbey}, {Holland},
  {Turner}, {Balasini}, {Bignami}, {La Palombara}, {Villa}, {Buttler},
  {Gianini}, {Lain{\'e}}, {Lumb}, \& {Dhez}}]{Struder01}
{Str{\"u}der}, L., {Briel}, U., {Dennerl}, K., {et~al.} 2001, \aap, 365, L18

\bibitem[{{Sunyaev} {et~al.}(2021){Sunyaev}, {Arefiev}, {Babyshkin},
  {Bogomolov}, {Borisov}, {Buntov}, {Brunner}, {Burenin}, {Churazov},
  {Coutinho}, {Eder}, {Eismont}, {Freyberg}, {Gilfanov}, {Gureyev}, {Hasinger},
  {Khabibullin}, {Kolmykov}, {Komovkin}, {Krivonos}, {Lapshov}, {Levin},
  {Lomakin}, {Lutovinov}, {Medvedev}, {Merloni}, {Mernik}, {Mikhailov},
  {Molodtsov}, {Mzhelsky}, {M{\"u}ller}, {Nandra}, {Nazarov}, {Pavlinsky},
  {Poghodin}, {Predehl}, {Robrade}, {Sazonov}, {Scheuerle}, {Shirshakov},
  {Tkachenko}, \& {Voron}}]{Sunyaev21}
{Sunyaev}, R., {Arefiev}, V., {Babyshkin}, V., {et~al.} 2021, \aap, 656, A132

\bibitem[{{Swihart} {et~al.}(2022{\natexlab{a}}){Swihart}, {Strader}, {Aydi},
  {Chomiuk}, {Dage}, {Kawash}, {Sokolovsky}, \& {Ferrara}}]{Swihart22b}
{Swihart}, S.~J., {Strader}, J., {Aydi}, E., {et~al.} 2022{\natexlab{a}}, \apj,
  926, 201

\bibitem[{{Swihart} {et~al.}(2022{\natexlab{b}}){Swihart}, {Strader},
  {Chomiuk}, {Aydi}, {Sokolovsky}, {Ray}, \& {Kerr}}]{Swihart22}
{Swihart}, S.~J., {Strader}, J., {Chomiuk}, L., {et~al.} 2022{\natexlab{b}},
  \apj, 941, 199

\bibitem[{{Tranin} {et~al.}(2022){Tranin}, {Godet}, {Webb}, \&
  {Primorac}}]{Tranin22}
{Tranin}, H., {Godet}, O., {Webb}, N., \& {Primorac}, D. 2022, \aap, 657, A138

\bibitem[{{Truemper}(1982)}]{Truemper82}
{Truemper}, J. 1982, Advances in Space Research, 2, 241

\bibitem[{{Tub\'in-Arenas} {et~al.}(2023){Tub\'in-Arenas}, {Krumpe}, {Lamer},
  {Brunner}, {Haase}, \& et~al.}]{Tubin23}
{Tub\'in-Arenas}, D., {Krumpe}, M., {Lamer}, G., {et~al.} 2023, \aap, submitted

\bibitem[{{Turner} {et~al.}(2001){Turner}, {Abbey}, {Arnaud}, {Balasini},
  {Barbera}, {Belsole}, {Bennie}, {Bernard}, {Bignami}, {Boer}, {Briel},
  {Butler}, {Cara}, {Chabaud}, {Cole}, {Collura}, {Conte}, {Cros}, {Denby},
  {Dhez}, {Di Coco}, {Dowson}, {Ferrando}, {Ghizzardi}, {Gianotti}, {Goodall},
  {Gretton}, {Griffiths}, {Hainaut}, {Hochedez}, {Holland}, {Jourdain},
  {Kendziorra}, {Lagostina}, {Laine}, {La Palombara}, {Lortholary}, {Lumb},
  {Marty}, {Molendi}, {Pigot}, {Poindron}, {Pounds}, {Reeves}, {Reppin},
  {Rothenflug}, {Salvetat}, {Sauvageot}, {Schmitt}, {Sembay}, {Short},
  {Spragg}, {Stephen}, {Str{\"u}der}, {Tiengo}, {Trifoglio}, {Tr{\"u}mper},
  {Vercellone}, {Vigroux}, {Villa}, {Ward}, {Whitehead}, \& {Zonca}}]{Turner01}
{Turner}, M.~J.~L., {Abbey}, A., {Arnaud}, M., {et~al.} 2001, \aap, 365, L27

\bibitem[{Virtanen {et~al.}(2020)Virtanen, Gommers, Oliphant, Haberland, Reddy,
  Cournapeau, Burovski, Peterson, Weckesser, Bright, {van der Walt}, Brett,
  Wilson, Millman, Mayorov, Nelson, Jones, Kern, Larson, Carey, Polat, Feng,
  Moore, {VanderPlas}, Laxalde, Perktold, Cimrman, Henriksen, Quintero, Harris,
  Archibald, Ribeiro, Pedregosa, {van Mulbregt}, \& {SciPy 1.0
  Contributors}}]{SciPy}
Virtanen, P., Gommers, R., Oliphant, T.~E., {et~al.} 2020, Nature Methods, 17,
  261

\bibitem[{{Weisskopf} {et~al.}(2002){Weisskopf}, {Brinkman}, {Canizares},
  {Garmire}, {Murray}, \& {Van Speybroeck}}]{Weisskopf02}
{Weisskopf}, M.~C., {Brinkman}, B., {Canizares}, C., {et~al.} 2002, \pasp, 114,
  1

\bibitem[{{Wenger} {et~al.}(2000){Wenger}, {Ochsenbein}, {Egret}, {Dubois},
  {Bonnarel}, {Borde}, {Genova}, {Jasniewicz}, {Lalo{\"e}}, {Lesteven}, \&
  {Monier}}]{Wenger00}
{Wenger}, M., {Ochsenbein}, F., {Egret}, D., {et~al.} 2000, \aaps, 143, 9

\bibitem[{{Wright} {et~al.}(2011){Wright}, {Drake}, {Mamajek}, \&
  {Henry}}]{Wright11}
{Wright}, N.~J., {Drake}, J.~J., {Mamajek}, E.~E., \& {Henry}, G.~W. 2011,
  \apj, 743, 48

\end{thebibliography}

\begin{appendix}

\section{Tuning of random forest classifiers} \label{ML_Tuning}

\begin{figure*}[t!]
\centering
\resizebox{18.5cm}{!}{
\includegraphics[width=6.0cm]{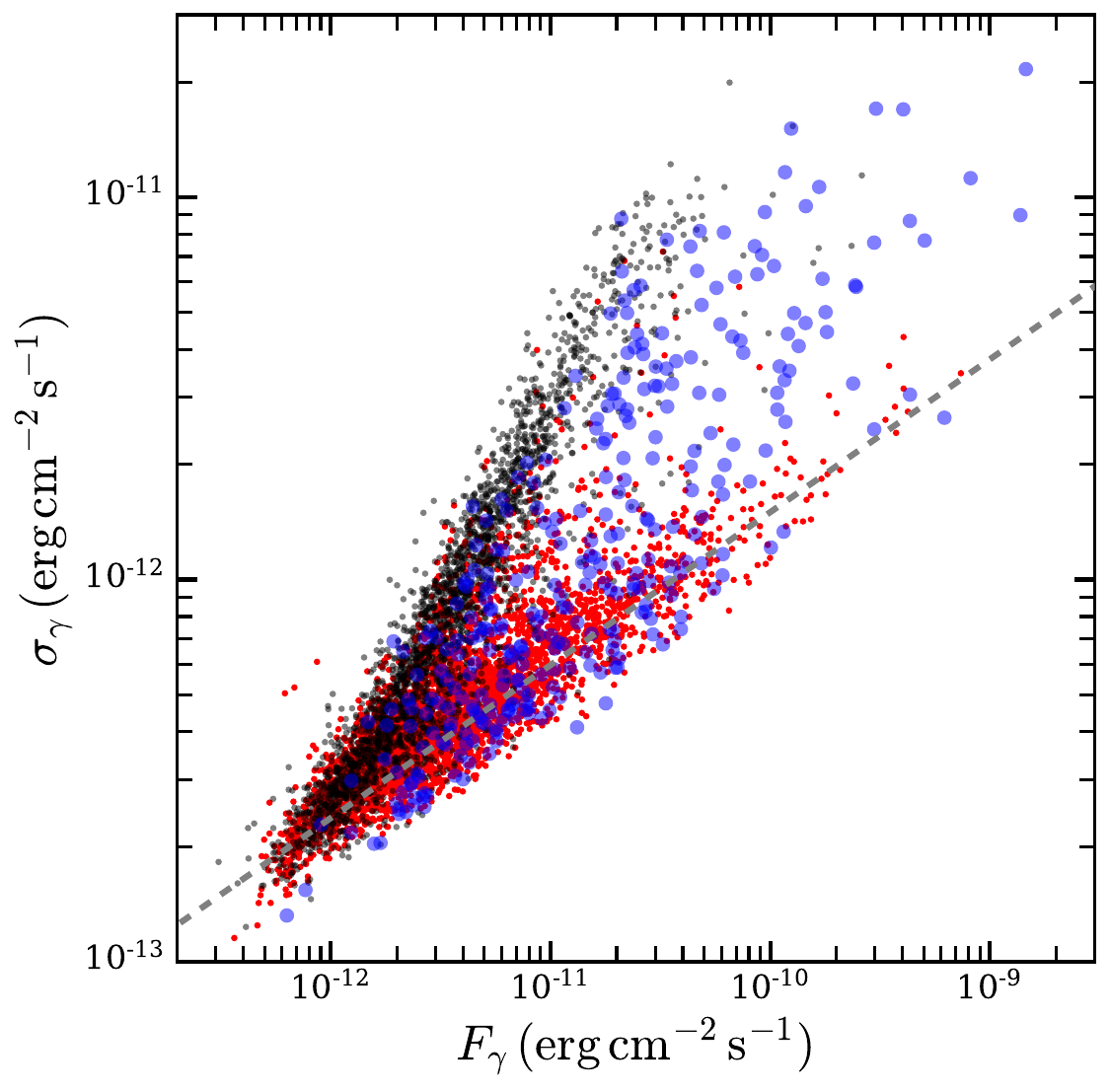} 
\includegraphics[width=6.1cm]{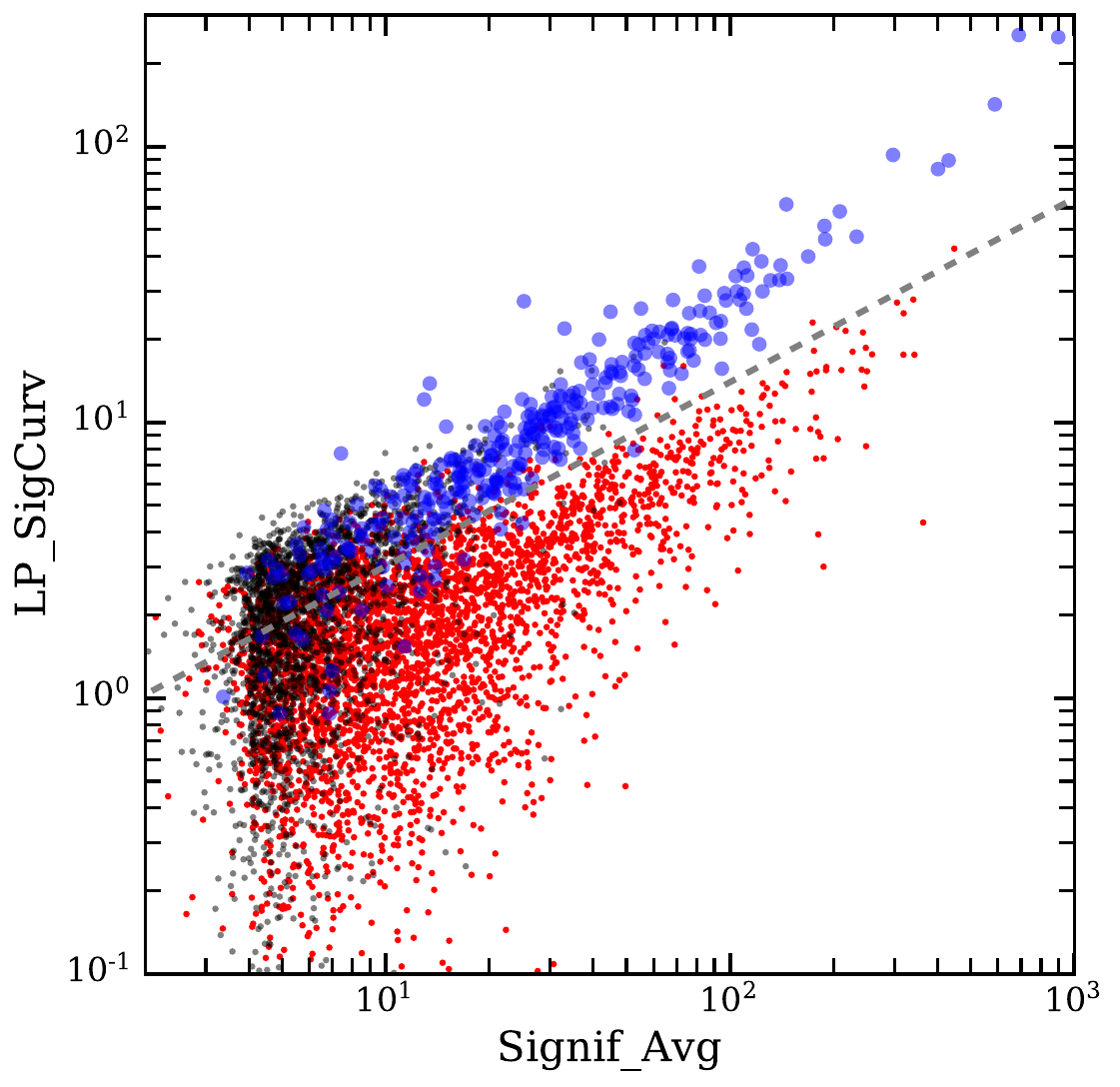} 
\includegraphics[width=6.1cm]{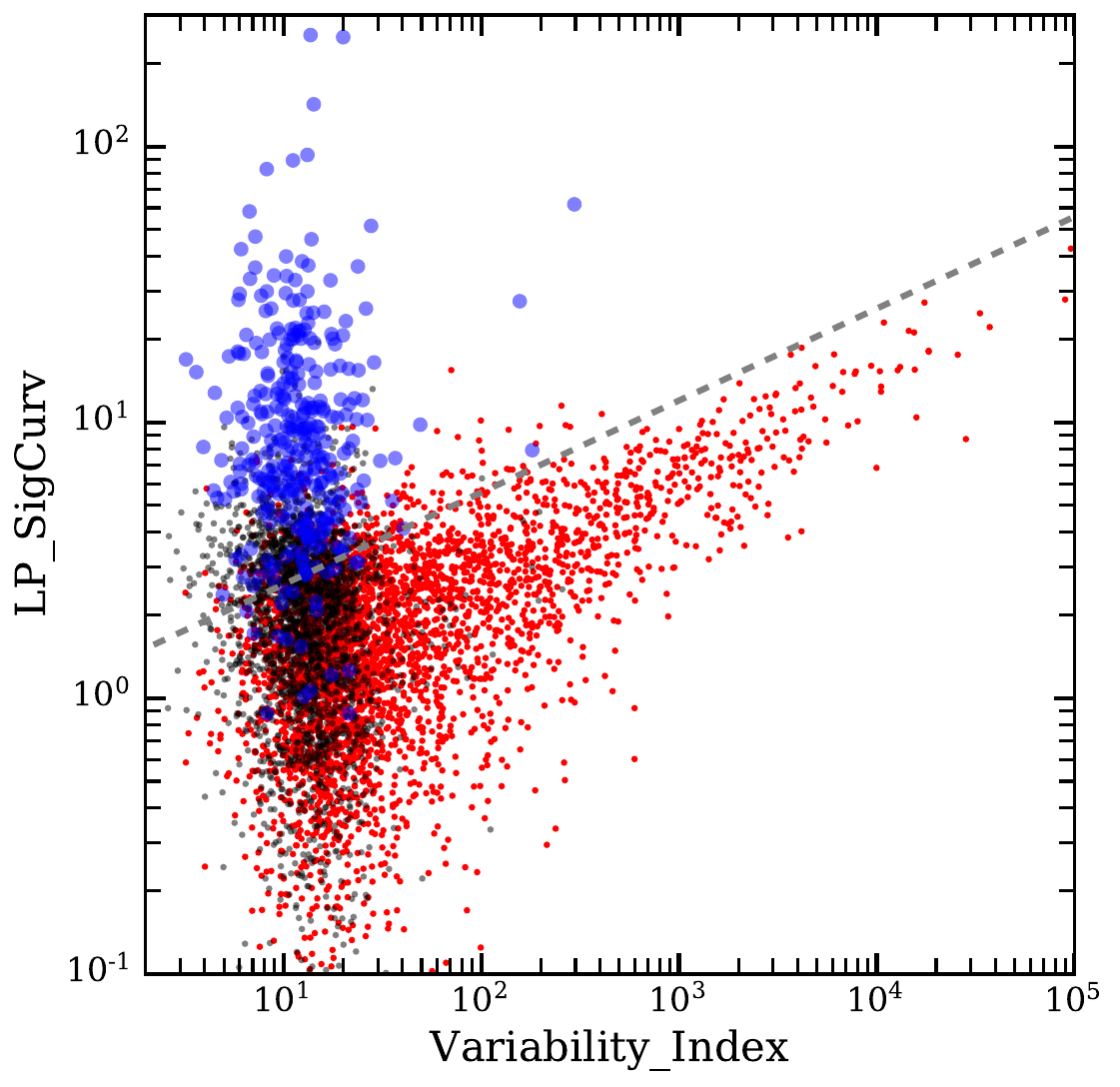}
}
\caption{Distribution of 4FGL-DR4 sources in feature space. In each panel, we show the distribution of pulsars (blue), blazars (red), and unassociated sources (black) in different parameters. The left panel compares the observed energy flux $F_{\gamma}$ in the $100 \,\si{MeV}-100\,\si{GeV}$ band to its estimated error $\sigma_{\gamma}$. The central panel compares the significance of detection with that of spectral curvature for a log-parabola model. In the right panel, this is compared with the measure of variability on monthly to yearly scales. In each panel, the dashed gray line indicates the direction perpendicular to which we defined our composite parameters to be able to better separate the source types. }
\label{FeatureEngineering}
\end{figure*}

\begin{table*}
    \centering
    \caption{List of the twenty candidate features used for our classification task. }
    \begin{tabular}{cc}
    \hline\hline
        Feature & Description \\ \hline
        {\tt SymLat} & Symmetric Galactic latitude, i.e. $\lvert b\rvert$ \\
        {\tt SymLon} & Symmetric Galactic longitude, i.e. $\lvert 180^{\circ}-l\rvert$ \\
        {\tt SigCombined} & $3 \log {\tt LP\_SigCurv}\tablefootmark{a} - \log {\tt Variability\_Index}\tablefootmark{a}$ \\
        {\tt Frac\_Variability}\tablefootmark{a} & \\
        {\tt Variability\_Index}\tablefootmark{a} & \\   
        {\tt ModSigCurv} & $3\log {\tt LP\_SigCurv}\tablefootmark{a} - 2\log {\tt Signif\_Avg}\tablefootmark{a}$ \\
        {\tt LP\_SigCurv}\tablefootmark{a} & \\
        {\tt LP\_beta}\tablefootmark{a} & \\
        {\tt LP\_Index}\tablefootmark{a} &  \\
        {\tt PL\_Index}\tablefootmark{a} & \\
        {\tt DeltaInd} & ${\tt LP\_Index}\tablefootmark{a}-{\tt PL\_Index}\tablefootmark{a}$ \\[0.3ex]        
        {\tt Pivot\_Energy}\tablefootmark{a} & \\
        {\tt EFluxErr} & $\log {\tt Unc\_Energy\_Flux100}\tablefootmark{a} - 0.4\log {\tt Energy\_Flux100}\tablefootmark{a}$ \\
        {\tt HR12} & $\frac{F(0.3-1.0\,\si{GeV}) \,-\, F(0.05-0.3\,\si{GeV}) }{F(0.3-1.0\,\si{GeV}) \,+\, F(0.05-0.3\,\si{GeV}) }$ \\[0.3ex]
        {\tt HR23} & $\frac{F(1.0-3.0\,\si{GeV}) \,-\, F(0.3-1.0\,\si{GeV}) }{F(1.0-3.0\,\si{GeV}) \,+\, F(0.3-1.0\,\si{GeV}) }$ \\[0.3ex]
        {\tt HR34} & $\frac{F(3-30\,\si{GeV}) \,-\, F(1.0-3.0\,\si{GeV}) }{F(3-30\,\si{GeV}) \,+\, F(1.0-3.0\,\si{GeV}) }$ \\[0.3ex]
        {\tt HR45} & $\frac{F(30-1000\,\si{GeV}) \,-\, F(3-30\,\si{GeV}) }{F(30-1000\,\si{GeV}) \,+\, F(3-30\,\si{GeV}) }$ \\[0.3ex]        
        {\tt K13} & $\frac{2 F(0.3-1.0\,\si{GeV}) \,-\, F(0.05-0.3\,\si{GeV}) \,- \,F(1.0-3.0\,\si{GeV})}{2 F(0.3-1.0\,\si{GeV}) \,+\, F(0.05-0.3\,\si{GeV}) \,+\, F(1.0-3.0\,\si{GeV})}$ \\[0.3ex]
        {\tt K24} & $\frac{2 F(1.0-3.0\,\si{GeV}) \,-\, F(0.3-1.0\,\si{GeV}) \,- \,F(3-30\,\si{GeV})}{2 F(1.0-3.0\,\si{GeV}) \,+\, F(0.3-1.0\,\si{GeV}) \,+\, F(3-30\,\si{GeV})}$ \\[0.3ex]
        {\tt K35} & $\frac{2 F(3-30\,\si{GeV}) \,-\, F(1.0-3.0\,\si{GeV}) \,- \,F(30-1000\,\si{GeV})}{2 F(3-30\,\si{GeV}) \,+\, F(1.0-3.0\,\si{GeV}) \,+\, F(30-1000\,\si{GeV})}$ \\[0.3ex]
        \hline
    \end{tabular}
    \tablefoot{\tablefoottext{a}{Catalog column \citep[see][for a detailed description]{4FGLDR3}}}
    \label{CandidateFeatures}
\end{table*}

\begin{figure}
\centering
\includegraphics[width=\linewidth]{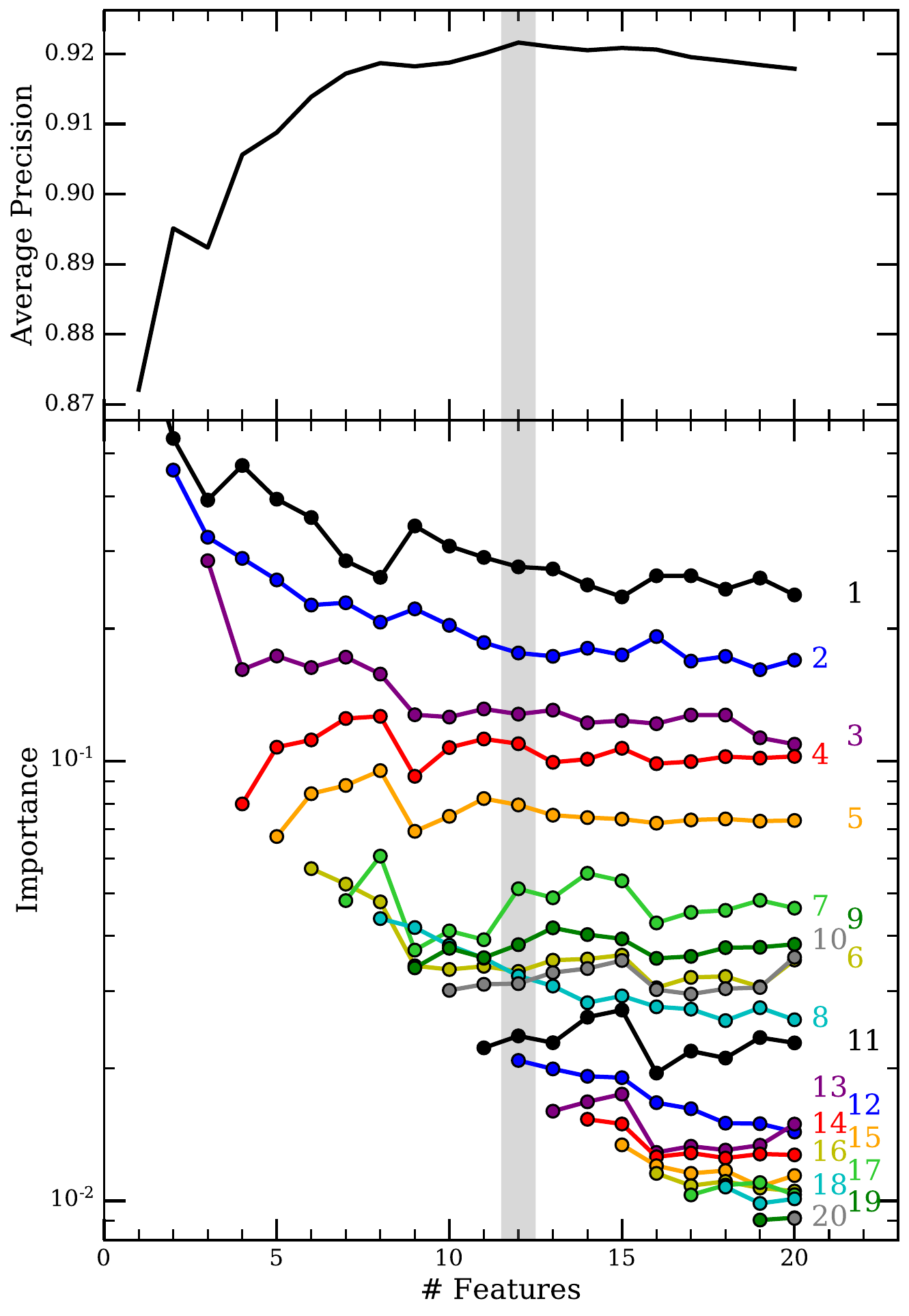} 
\caption{Illustration of recursive feature elimination. The top panel illustrates the evolution of the average precision of our classifier with the number of features. The bottom panel illustrates the evolution of the importance of individual features, as the least important feature is recursively removed. The shaded bar indicates the set of features we found to be optimal. \\ 
The curves indicate the parameters numbered by their reverse order of elimination: (1) {\tt SigCombined}, (2) {\tt ModSigCurv}, (3) {\tt LP\_beta}, (4) {\tt LP\_SigCurv}, (5) {\tt K24}, (6) {\tt SymLat}, (7) {\tt Frac\_Variability}, (8) {\tt EFluxErr}, (9) {\tt HR34}, (10) {\tt Variability\_Index}, (11) {\tt HR45}, (12) {\tt PL\_Index}, (13) {\tt HR12}, (14) {\tt DeltaInd}, (15) {\tt SymLon}, (16) {\tt K13}, (17) {\tt Pivot\_Energy}, (18) {\tt LP\_Index}, (19) {\tt K35}, (20) {\tt HR23}.
}
\label{FeatureElimination}
\end{figure}

The goal of classifying the 4FGL-DR4 catalog into likely blazars and pulsars based on machine learning suffers from a strong inherent imbalance: since about ten times more identified blazars can be used for training the classifier, a standard random forest would tend to learn a bias against the minority class, which happens to be the class of interest, here. Hence, in order to assign equal ``prior'' probabilities to the AGN and pulsar scenarios, samples of similar size need to be used for training. 
To achieve this, one could, on one hand, try to oversample the minority class, either by using sources multiple times or by producing artificial samples with the synthetic minority oversampling technique  \citep[SMOTE;][]{SMOTE}. 
A somewhat more natural choice is to undersample the majority class, meaning to draw only a random subsample for each individual decision tree, so that equal sample sizes are effectively given \citep{Chen04}. As we found that, even for low recall, the SMOTE method struggled to achieve a pure pulsar sample in our tests, we decided to employ the latter method, which is implemented in the {\tt BalancedRandomForest} classifier in the {\tt imbalanced-learn} Python package \citep{Lemaitre17}. 

Generally, we found that the hyperparameters of our classifier had a smaller impact on its performance than the choice and definition of the input features. The classifiers discussed in the following employed a maximum tree depth of 10 and a maximum number of features per node of 2, which was one of the setups found to yield optimal performance. In order to allow for reliable probabilistic statements in our output, we trained a large number of trees, i.e.~10\,000. 

Several possible biases may impact the performance of our binary classifier, for instance, a possible covariate shift, where the distribution of input features is different between training and target samples \citep{Malyshev23}. A natural example of this is the source flux, since the sample of unassociated sources is systematically fainter in $\gamma$-rays than the set of associated sources, in particular pulsars (see Fig.~\ref{FeatureEngineering}). This is because the direct detection of $\gamma$-ray pulsations requires a comparatively large number of photons, hence a sufficiently bright $\gamma$-ray source \citep{SazParkinson16}.
In order to counteract biases in our classification against the pulsar class introduced by this effect, we did not include any absolute fluxes in our set of candidate features. Further, we constructed composite features, in which we attempted to remove the dependence on source flux or significance. 

The three panels in Fig.~\ref{FeatureEngineering} illustrate this interdependence between some features which have been shown to be important for classification \citep[e.g.,][]{Salvetti16, SazParkinson16}. Clearly, the flux error strongly scales with the measured energy flux, even though the lower envelope exhibits a slightly shallower scaling than expected in a purely Poissonian regime with around $\sigma_{\gamma}\propto F_{\gamma}^{0.4}$. The scatter relative to this relation is probably related to varying background levels in the vicinity of the sources. 
Similarly, the curvature significance for the log-parabola model {\tt LP\_SigCurv} scales with the source detection significance, but pulsars clearly tend to show more significant curvature in a relative sense \citep{4FGLDR3}. Similarly, due to their dependence on source flux, {\tt LP\_SigCurv} exhibits a weak scaling with the measure of variability, most clearly for bright blazars. 
We constructed several composite features based on the observed scalings. Outside the modified flux error measure (see left panel in Fig.~\ref{FeatureEngineering}), the primary goal of these was to optimally separate the two source classes in the displayed parameter spaces, rather than being strictly flux-independent. Furthermore, similarly to previous works \citep[e.g.,]{Bhat22, Germani21, SazParkinson16}, we constructed several empirical hardness ratios as well curvature measures based from the band fluxes given in the catalog. 
Our full list of candidate features is given in Table \ref{CandidateFeatures}. 

We followed the established approach of recursive feature elimination \citep{Luo20} for identifying the optimal set of features to use for our classification task: Starting from a classifier with all candidate features, we repeatedly determined its performance, characterized by its average precision measured with repeated five-fold cross-validation, and successively removed the feature with the lowest importance, as illustrated in Fig.~\ref{FeatureElimination}. The average precision measures the ability of the classifier to produce a pure and complete sample of pulsars in the presence of a class imbalance, hence we chose the version of our classifier with 12 features (see Table \ref{ML_Features}), which maximizes this quantity. Many of the features with the highest importance for the classifier are directly related to the spectral curvature, and in a few cases also to the relative variability of the sources. This is an unsurprising behavior, as it has been known for a while that pulsars show less variability and have spectra which are more curved than blazars \citep{Salvetti16, SazParkinson16}.     
 
\begin{figure}[t!]
\centering
\includegraphics[width=1.0\linewidth]{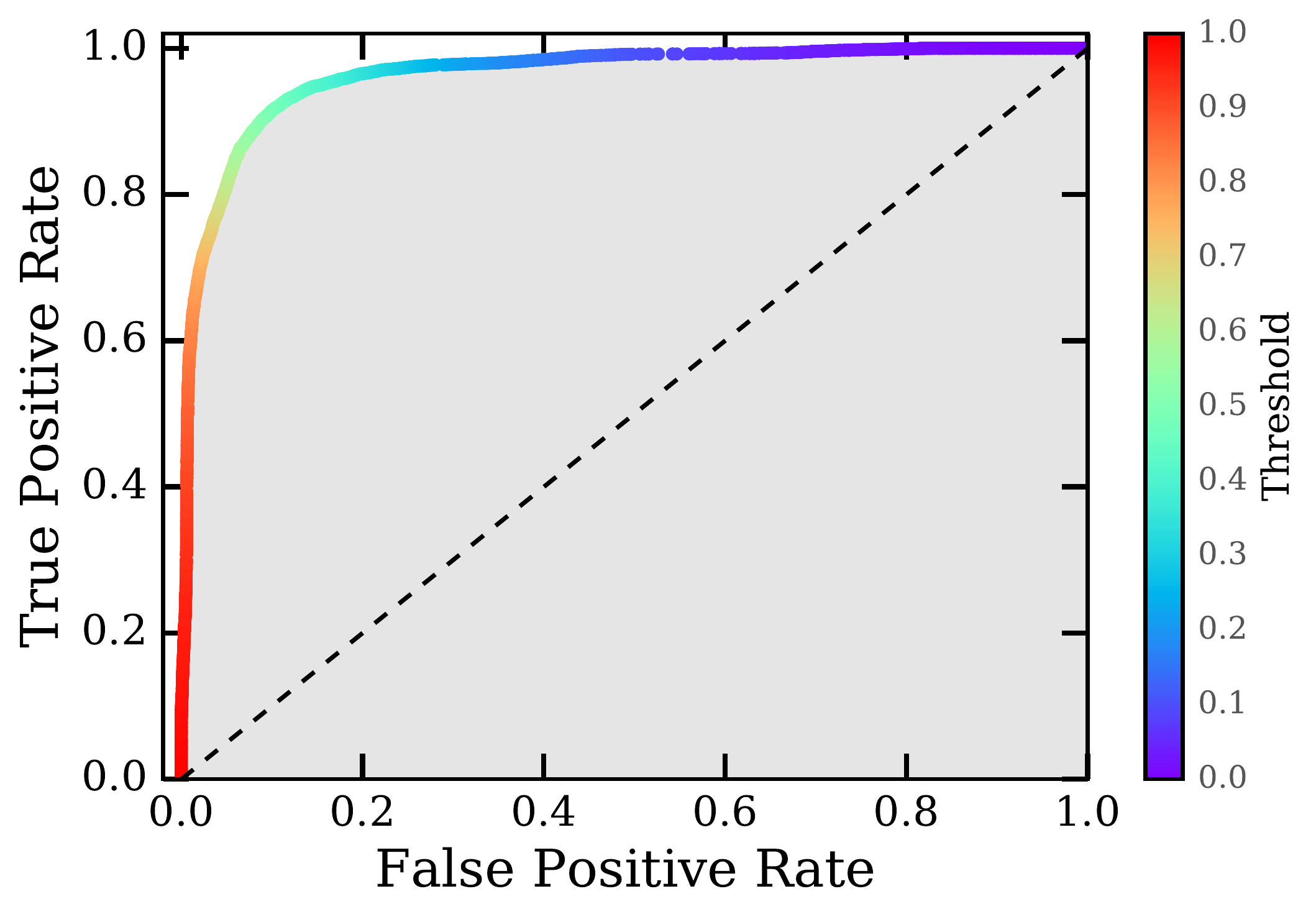} 
\caption{Receiver operating characteristic of the binary classifier separating young pulsars and MSPs. This graph compares the true positive rate, meaning the fraction of young pulsars that are correctly identified, to the false positive rate, meaning the fraction of MSPs wrongly identified as young pulsars, dependent on the decision threshold.    
}
\label{ROC_youngold}
\end{figure}

\begin{table}[t!]
    \centering
    \caption{Optimal set of features and their importance for the classifier separating young pulsars and MSPs among $\gamma$-ray sources.}
    \begin{tabular}{cc}
    \hline\hline
        Feature & Importance \\ \hline
        {\tt SymLat} & $0.263$ \\
        {\tt EFluxErr} & $0.157$ \\
        {\tt LP\_Index} & $0.096$ \\
        {\tt ModSigCurv} & $0.074$ \\
        {\tt PL\_Index} & $0.069$ \\
        {\tt SigCombined} & $0.065$ \\
        {\tt DeltaInd} & $0.063$ \\
        {\tt LP\_SigCurv} & $0.063$ \\
        {\tt Pivot\_Energy} & $0.043$ \\
        {\tt HR23} & $0.039$ \\
        {\tt HR34} & $0.036$ \\
        {\tt SymLon} & $0.031$ \\
        \hline
    \end{tabular}
    \label{ML_Features_Youngold}
\end{table}

As a final step, we converted the prediction of our random forest into a quantitative probability $P_{\gamma}^{\rm PSR}$, specifying the chance for a given $\gamma$-ray source to be a pulsar, assuming both scenarios are a priori equally likely. To do this, we used the output probabilities of our cross-validation and created calibration curves. This means we computed the fraction of pulsars $f_{\rm PSR}$ in the output sample dependent on the probability $p$ assigned by the classifier. Since the number of blazars $N_{\rm AGN}$ is much larger than the number of pulsars $N_{\rm PSR}$ in the test sample, corrected this value to obtain the ``balanced'' fractions which would be observed for equally abundant classes:
\begin{equation}
    \frac{f^{\rm bal}_{\rm PSR}}{1-f^{\rm bal}_{\rm PSR}} = \frac{N_{\rm AGN}}{N_{\rm PSR}}\frac{f_{\rm PSR}}{1-f_{\rm PSR}}.
\end{equation}
Given the binary nature of our classifier, a logistic curve of the form 
\begin{equation}
    f(x\,|\,a,b) = \frac{1}{1+\exp\left(-ax + b\right)},
\end{equation}
where $x$ is related to the classifier output $p$ as $x = \ln \, p/\left(1-p\right)$, is a natural choice for its calibration. By fitting this function to the observed dependence of pulsar fraction $f^{\rm bal}_{\rm PSR}$ on classifier prediction $p$, we obtained a quantitatively calibrated estimator $P_{\gamma}^{\rm PSR}=f(p)$ for the pulsar probability of any given $\gamma$-ray source.

Analogously to our classification of pulsars versus blazars, we created a binary classifier for estimating the probability for a given source to be a young pulsar, rather than a millisecond pulsar, assuming its pulsar nature. We followed the same approach as for our primary classifier, with the only exception that we used the receiver operating characteristic (ROC), to measure the performance of our classifier, as neither class is of larger interest for our study and source fractions are about balanced, here. The optimal ROC curve, with an area under the curve of $0.964$, is displayed in Fig.~\ref{ROC_youngold}. 
The optimal set of features for this classifier are shown in Table \ref{ML_Features_Youngold}, and are fundamentally different from the pulsar-blazar case. The highest importance was ascribed to the Galactic latitude of the source, as young pulsars tend to be much closer to their birth site, meaning in the Galactic plane. Further important features include the relative flux errors, which is probably related to the crowdedness of the field, and the overall slope of the $\gamma$-ray spectrum, as MSPs typically have somewhat harder spectra \citep{2PC}.   

Finally, we note that we also tested whether the two classifiers used in our work could be combined into a single three-way classifier separating blazar-like sources, young pulsars, and MSPs. However, we found that the performance of this single classifier fell slightly short of that combining the two individual ones. Quantitatively, the three-way classifier achieved an average precision of 0.916 in the separation of pulsars from AGN, and a ROC AUC of 0.959 in the subclassification of pulsars into young and recycled pulsars, compared to scores of 0.920 and 0.964 by the two separate classifiers, respectively.

\section{Supplementary tables}
Here, we provide additional information for readers who wish to reproduce the analysis performed in this work, or employ our results for their own research. 
Table \ref{ColumnDescriptions} describes the columns of our full catalog of candidate pulsar counterparts to 4FGL sources, which is provided in electronic form at CDS. Table \ref{XDetPulsarTable} lists all X-ray detections of pulsars in the 4FGL catalog, along with the X-ray fluxes derived from their count rates, as displayed in Fig.~\ref{XGFluxes}.   

\begin{table*}
  \centering
  \caption{Description of the columns of our cross-match catalogs between eRASS:4 and 4FGL-DR4.}
  \begin{tabular}{p{0.12\textwidth} p{0.15\textwidth} p{0.63\textwidth}}
\hline\hline
Name & Units & Description \\
   \hline
\tt \# &  & Rank of X-ray source by match probability \\
\tt Name\_ero &  & eROSITA source ID following IAU naming conventions \\
\tt RA\_ero & $\rm deg$ & Right ascension (ICRS) of X-ray source \\
\tt DEC\_ero & $\rm deg$ & Declination (ICRS) of X-ray source \\
\tt sigma\_ero & $\rm arcsec$ & One-sigma\tablefootmark{(a)} positional error of X-ray source\\
\tt CR\_ero & $\rm ct\,s^{-1}$ & X-ray count rate in $0.2-2.3 \,\si{keV}$ \\
\tt F\_ero & $\rm erg\,cm^{-2}\,s^{-1}$ & X-ray source flux in $0.2-2.3 \,\si{keV}$, determined via an energy conversion factor \citep{Merloni23} from the count rate \\
\tt DET\_LIKE\_ero &  & X-ray detection likelihood \citep{Brunner22, Merloni23}\\
\tt EXT\_LIKE\_ero &  & X-ray extent likelihood \citep{Brunner22, Merloni23}\\
\hline
\tt Name\_4FGL &  & Name of $\gamma$-ray source in 4FGL-DR4 catalog \citep{4FGLDR4}\\
\tt Class2\_4FGL &  & Class of low-probability association\tablefootmark{(b)} of $\gamma$-ray source in 4FGL-DR4 catalog \citep{4FGL} \\
\tt Assoc2\_4FGL &  & Low-probability association\tablefootmark{(b)} of $\gamma$-ray source in 4FGL-DR4 catalog \citep{4FGL} \\
\tt RA\_4FGL & $\rm deg$ & Right ascension (ICRS) of $\gamma$-ray source \\
\tt DEC\_4FGL & $\rm deg$ & Declination (ICRS) of $\gamma$-ray source \\
\tt aerr\_4FGL & $\rm arcmin$ & Semi-major axis of $\gamma$-ray $95\%$ error ellipse\\
\tt berr\_4FGL & $\rm arcmin$ & Semi-minor axis of $\gamma$-ray $95\%$ error ellipse\\
\tt phierr\_4FGL & $\rm deg$ (east of north) & Position angle of $\gamma$-ray  error ellipse on the sky\\
\tt nu\_4FGL & $\rm deg^{-2}$ & Density of eRASS:4 X-ray sources within a $2^{\circ}$ radius around $\gamma$-ray source\\
\tt exp\_Field &  & Expectation value for the number of X-ray chance alignments (denominator in Eq.~\ref{PositionalPost})\\
\tt F\_4FGL & $\rm erg\,cm^{-2}\,s^{-1}$ & $\gamma$-ray source flux in the range $100\,\si{MeV}-100\,\si{GeV}$\\
\tt Prior\_PSR &  & Prior probability $P_{\gamma}^{\rm PSR}$ for a source to be a pulsar, rather than blazar, based on its $\gamma$-ray properties\\
\tt Prior\_youngold &  & Prior probability $P_{\gamma}^{\rm YNG}$ for a source to be a young pulsar, rather than a millisecond pulsar, based on its $\gamma$-ray properties\\
\hline
\tt d\_mahalanobis & & Mahalanobis distance, i.e.~the distance between X-ray and $\gamma$-ray source positions, normalized by the total positional error in the given direction \\
\tt BF\_pos & & Purely positional Bayes factor of the association, i.e.~value of Eq.~\ref{PositionalPost} with $c=1$. \\
\tt log\_flux\_ratio & & Logarithmic $\gamma$-to-X-ray flux ratio, i.e. $\log \, F_{\gamma}/F_{x} $\\
\tt Prior\_fr\_AGN & & Likelihood ratio, assuming AGN nature, of having a match rather than a background X-ray source based on only X-ray and $\gamma$-ray fluxes, i.e.~$\phi_{t}(\log F_{X}\,|\,\log F_{\gamma})/\phi_{B}(\log F_{X})$ (Eqs.~\ref{FR_Back} and \ref{FR_Assoc})\\
\tt Prior\_fr\_YNG & & Same as above, but assuming young pulsar nature\\
\tt Prior\_fr\_MSP & & Same as above, but assuming millisecond pulsar nature\\
\tt P\_i\_PSR & & Combined posterior probability $P_{i}^{\rm PSR} = P_{i}^{\rm YNG}+P_{i}^{\rm MSP}$ (Eq.~\ref{FullPosterior}) for the given X-ray and $\gamma$-ray source to be matched, and to be of pulsar nature   \\
\tt P\_i\_YNG & & Same as above, but only for young pulsar nature \\
\tt P\_i\_MSP & & Same as above, but only for millisecond pulsar nature \\
\tt P\_i\_AGN & & Same as above, but for AGN (blazar) nature \\
\tt P\_i &  & Same as above, for any nature, i.e.~$P_{i} = P_{i}^{\rm PSR}+P_{i}^{\rm AGN}$\\
\tt Comment &  & Comment on the match, for instance, on possible alternative nature of the X-ray source or possible spurious origin. \\
\hline	
  \end{tabular}
  \tablefoot{The top segment describes the properties of matched X-ray sources, the middle segment the properties of the $\gamma$-ray source, and the final segment characterizes the quality of the match. 
  \\ \tablefoottext{a}{``One sigma'' implies that the true two-dimensional source position lies within a circle of this radius with a probability of $39\%$.}\tablefoottext{b}{We note that very likely associations ({\tt ASSOC1 $\neq$ `'}) are by definition not present, since our targets are unassociated sources.}}
      \label{ColumnDescriptions}
\end{table*}

\begin{table*}[t!]
    \centering
    \caption{Basic X-ray properties of 4FGL-DR4 pulsar entries detected in the eRASS:4 catalog.}
    \begin{tabular}{llllll}
    \hline\hline
        4FGL source & Pulsar name & Pulsar type & X-ray flux ($0.2-2.3\,\si{keV})$ & Det. likelihood\tablefootmark{a} & Separation \\ 
         &  &  & $\si{erg.s^{-1}.cm^{-2}}$ & & $\si{arcsec}$ \\ \hline
         4FGL J0514.6$-$4408 & PSR J0514$-$4408 &  YNG & $(1.2 \pm 0.4) \times\, 10^{-14}$ & 17.2 & 2.0 \\
        4FGL J0540.3$-$6920 & PSR J0540$-$6919 &  YNG & $(1.09 \pm 0.05) \times\, 10^{-11}$ & 292552.5 & 2.2 \\
        4FGL J0633.9+1746 & PSR J0633+1746 &  YNG & $(9.5 \pm 0.5) \times\, 10^{-13}$ & 1432.2 & 5.3 \\
        4FGL J0659.7+1416 & PSR J0659+1414 &  YNG & $(8.05 \pm 0.15) \times\, 10^{-12}$ & 17351.6 & 2.5 \\
        4FGL J0835.3$-$4510 & PSR J0835$-$4510 &  YNG & $(3.887 \pm 0.027) \times\, 10^{-11}$ & 83429.2 & 2.5 \\
        4FGL J1015.5$-$6030 & CXOU J101546.0$-$602939 &  YNG & $(4.31 \pm 0.22) \times\, 10^{-13}$ & 1142.6 & 3.0 \\
        4FGL J1028.5$-$5819 & PSR J1028$-$5819 &  YNG & $(7.1 \pm 1.1) \times\, 10^{-14}$ & 95.4 & 3.1 \\
        4FGL J1044.4$-$5737 & PSR J1044$-$5737 &  YNG & $(4.5 \pm 0.8) \times\, 10^{-14}$ & 52.6 & 2.6 \\
        4FGL J1056.9$-$5852 & PSR J1057$-$5851 &  YNG & $(1.0 \pm 0.4) \times\, 10^{-14}$ & 8.6 & 7.4 \\
        4FGL J1057.9$-$5227 & PSR J1057$-$5226 &  YNG & $(1.55 \pm 0.05) \times\, 10^{-12}$ & 4951.5 & 2.3 \\
        4FGL J1104.9$-$6037 & PSR J1105$-$6037 &  YNG & $(3.4 \pm 0.7) \times\, 10^{-14}$ & 33.9 & 3.0 \\
        4FGL J1111.8$-$6039 & PSR J1111$-$6039 &  YNG & $(4.7 \pm 0.4) \times\, 10^{-13}$ & 256.1 & 5.8 \\
        4FGL J1112.1$-$6108 & PSR J1112$-$6103 &  YNG & $(1.7 \pm 0.6) \times\, 10^{-14}$ & 10.2 & 9.7 \\
        4FGL J1119.1$-$6127 & PSR J1119$-$6127 &  YNG & $(1.4 \pm 0.5) \times\, 10^{-14}$ & 5.8 & 9.0 \\
        4FGL J1135.2$-$6055 & PSR J1135$-$6055 &  YNG & $(7.8 \pm 1.6) \times\, 10^{-14}$ & 17.5 & 12.3 \\
        4FGL J1203.9$-$6242 & PSR J1203$-$6242 &  YNG & $(2.1 \pm 0.6) \times\, 10^{-14}$ & 19.4 & 11.1 \\
        4FGL J1356.9$-$6432 & PSR J1357$-$6429 &  YNG & $(1.15 \pm 0.16) \times\, 10^{-13}$ & 60.9 & 1.3 \\
        4FGL J1709.7$-$4429 & PSR J1709$-$4429 &  YNG & $(2.17 \pm 0.23) \times\, 10^{-13}$ & 192.9 & 2.5 \\
        4FGL J1730.5$-$3352 & PSR J1730$-$3350 &  YNG & $(1.7 \pm 0.8) \times\, 10^{-14}$ & 5.3 & 14.1 \\
        4FGL J1732.5$-$3131 & PSR J1732$-$3131 &  YNG & $(1.5 \pm 1.0) \times\, 10^{-14}$ & 5.2 & 2.3 \\
        4FGL J1747.2$-$2957 & PSR J1747$-$2958 &  YNG & $(3.8 \pm 0.3) \times\, 10^{-13}$ & 249.1 & 3.2 \\
        4FGL J0101.1$-$6422 & PSR J0101$-$6422 &  MSP & $(5.6 \pm 1.0) \times\, 10^{-14}$ & 67.2 & 2.4 \\
        4FGL J0437.2$-$4715 & PSR J0437$-$4715 &  MSP & $(1.030 \pm 0.027) \times\, 10^{-12}$ & 6195.6 & 2.5 \\
        4FGL J0610.2$-$2100 & PSR J0610$-$2100 &  MSP & $(2.2 \pm 0.7) \times\, 10^{-14}$ & 15.6 & 5.7 \\
        4FGL J0614.1$-$3329 & PSR J0614$-$3329 &  MSP & $(6.9 \pm 1.1) \times\, 10^{-14}$ & 83.7 & 0.5 \\
        4FGL J0751.2+1808 & PSR J0751+1807 &  MSP & $(2.8 \pm 1.0) \times\, 10^{-14}$ & 10.7 & 10.8 \\
        4FGL J0952.1$-$0607 & PSR J0952$-$0607 &  MSP & $(3.0 \pm 1.1) \times\, 10^{-14}$ & 12.2 & 10.1 \\
        4FGL J1024.5$-$0719 & PSR J1024$-$0719 &  MSP & $(2.8 \pm 1.1) \times\, 10^{-14}$ & 12.8 & 5.0 \\
        4FGL J1035.4$-$6720 & PSR J1035$-$6720 &  MSP & $(1.2 \pm 0.4) \times\, 10^{-14}$ & 14.6 & 4.2 \\
        4FGL J1036.6$-$4349 & PSR J1036$-$4353 &  MSP & $(1.9 \pm 0.7) \times\, 10^{-14}$ & 10.1 & 11.9 \\
        4FGL J1124.0$-$3653 & PSR J1124$-$3653 &  MSP & $(1.7 \pm 0.6) \times\, 10^{-14}$ & 9.8 & 5.2 \\
        4FGL J1126.4$-$6011 & PSR J1125$-$6014 &  MSP & $(1.4 \pm 0.5) \times\, 10^{-14}$ & 9.6 & 2.3 \\
        4FGL J1207.4$-$5050 & PSR J1207$-$5050 &  MSP & $(1.4 \pm 0.5) \times\, 10^{-14}$ & 10.5 & 4.4 \\
        4FGL J1228.0$-$4853 & PSR J1227$-$4853 &  MSP & $(8.9 \pm 1.3) \times\, 10^{-14}$ & 95.5 & 3.1 \\
        4FGL J1231.1$-$1412 & PSR J1231$-$1411 &  MSP & $(1.34 \pm 0.18) \times\, 10^{-13}$ & 130.4 & 1.8 \\
        4FGL J1301.6+0834 & PSR J1301+0833 &  MSP & $(2.9 \pm 1.0) \times\, 10^{-14}$ & 12.0 & 4.2 \\
        4FGL J1306.8$-$4035 & PSR J1306$-$4035 &  MSP & $(1.28 \pm 0.15) \times\, 10^{-13}$ & 178.6 & 1.6 \\
        4FGL J1311.7$-$3430 & PSR J1311$-$3430 &  MSP & $(6.8 \pm 1.2) \times\, 10^{-14}$ & 51.0 & 4.5 \\
        4FGL J1312.7+0050 & PSR J1312+0051 &  MSP & $(3.2 \pm 0.9) \times\, 10^{-14}$ & 19.2 & 0.9 \\
        4FGL J1400.6$-$1432 & PSR J1400$-$1431 &  MSP & $(1.7 \pm 0.7) \times\, 10^{-14}$ & 9.5 & 4.6 \\
        4FGL J1417.6$-$4403 & PSR J1417$-$4402 &  MSP & $(2.62 \pm 0.21) \times\, 10^{-13}$ & 387.0 & 1.4 \\
        4FGL J1440.2$-$5505 & PSR J1439$-$5501 &  MSP & $(2.0 \pm 0.8) \times\, 10^{-14}$ & 13.9 & 6.4 \\
        4FGL J1614.5$-$2230 & PSR J1614$-$2230 &  MSP & $(3.5 \pm 1.0) \times\, 10^{-14}$ & 17.8 & 3.4 \\
        4FGL J1628.1$-$3204 & PSR J1628$-$3205 &  MSP & $(4.2 \pm 1.1) \times\, 10^{-14}$ & 23.3 & 4.2 \\
        4FGL J1803.1$-$6708 & PSR J1803$-$6707 &  MSP & $(3.2 \pm 0.9) \times\, 10^{-14}$ & 17.9 & 2.1 \\
        4FGL J1858.3$-$5424 & PSR J1858$-$5422 &  MSP & $(2.3 \pm 0.9) \times\, 10^{-14}$ & 7.3 & 1.6 \\
        4FGL J1902.0$-$5105 & PSR J1902$-$5105 &  MSP & $(1.8 \pm 0.8) \times\, 10^{-14}$ & 5.3 & 3.3 \\
        4FGL J1909.7$-$3744 & PSR J1909$-$3744 &  MSP & $(1.8 \pm 0.9) \times\, 10^{-14}$ & 5.1 & 5.7 \\
        4FGL J2029.5$-$4237 & PSR J2029$-$4239 &  MSP & $(4.0 \pm 1.3) \times\, 10^{-14}$ & 12.5 & 5.7 \\
        4FGL J2039.5$-$5617 & PSR J2039$-$5617 &  MSP & $(2.1 \pm 0.8) \times\, 10^{-14}$ & 9.4 & 7.1 \\
        4FGL J2241.7$-$5236 & PSR J2241$-$5236 &  MSP & $(4.9 \pm 1.2) \times\, 10^{-14}$ & 28.0 & 3.0 \\
        4FGL J2333.1$-$5527 & PSR J2333$-$5526 &  MSP & $(6.0 \pm 1.7) \times\, 10^{-14}$ & 12.6 & 10.0 \\
        \hline
    \end{tabular}
    \tablefoot{\tablefoottext{a}{\citet{Brunner22}}}
    \label{XDetPulsarTable}
\end{table*}

\end{appendix}
\endgroup

\end{document}